\documentclass[useAMS,usenatbib,usegraphicx]{mn2e}

\usepackage{rotating}
\usepackage{aas_macros}
\usepackage{times}

\title[Mass distribution and structural parameters of SMC clusters]
{Mass distribution and structural parameters of Small Magellanic Cloud star clusters}
\author[F.F.S. Maia, A. E. Piatti \& J.F.C. Santos Jr.]
{F. F. S. Maia$^{1}$\thanks{E-mail: ffsmaia@ufmg.br}, A. E. Piatti$^{2}$ and 
J. F. C. Santos Jr.$^{1}$\\
$^{1}$Instituto de Ci\^{e}ncias Exatas, UFMG, Av. Ant\^{o}nio Carlos 6627, Belo Horizonte, 
Brazil\\
$^{2}$Observatorio Astron\'omico, Universidad Nacional de C\'ordoba, Laprida 854, X5000BGR, 
C\'ordoba, Argentina}

\begin{document}

\date{\today}

\pagerange{\pageref{firstpage}--\pageref{lastpage}} \pubyear{2013}

\maketitle

\label{firstpage}

\begin{abstract}

In this work we estimate, for the first time, the total masses and mass function slopes of a 
sample of 29 young and intermediate-age SMC clusters from CCD Washington photometry. 
We also derive age, interstellar reddening and structural parameters for most of the studied 
clusters by employing a statistical method to 
remove the unavoidable field star contamination. Only these 29 clusters out of 68 originally 
analysed cluster candidates present stellar overdensities and coherent distribution in their 
colour-magnitude diagrams 
compatible with the existence of a genuine star cluster. 
We employed simple stellar population models to derive general equations for estimating the 
cluster mass based 
only on its age and integrated light in the $B$, $V$, $I$, $C$ and $T_1$ filter. 
These equations were tested against mass values computed from luminosity functions, 
showing an excellent agreement. 
The sample contains clusters with ages between 60 Myr and 3 Gyr and masses between 300 
and 3000 M$_\odot$ distributed between $\sim$0.5$\degr$ and $\sim$2$\degr$ from the SMC 
optical centre.
We determined mass function slopes for 24 clusters, of which 19 have slopes compatible 
with that of Kroupa's IMF ($\alpha = 2.3\pm0.7$), considering the uncertainties. The remaining 
clusters -- H86-188, H86-190, K47, K63 and NGC242 -- showed flatter MFs. Additionally, only 
clusters with masses lower than $\sim$1000 M$_\odot$ and flatter MF were found within 
$\sim$0.6$\degr$ from the SMC rotational centre.

\end{abstract}


\section{Introduction}

Since the proximity of the Magellanic Clouds (MCs) allows us to spatially resolve their stellar 
populations from ground 
based telescopes, stellar clusters have been largely used to investigate their formation and 
chemical evolution, namely: the star formation history 
\citep{Glatt:2010}, the age-metallicity relationship \citep{Piatti:2011b} and the age distribution of 
their clusters \citep{Chiosi:2006,Piatti:2011c}, among others.

Unlike the well known cluster age gap in the Large Magellanic Cloud (LMC), the formation of 
stellar clusters in the Small Magellanic Cloud (SMC) 
seems to have occurred continuously over the last 10.5 Gyr, with some periods of 
enhancement possibly due to the close gravitational interaction with either the Milky Way (MW)
or the LMC \citep{Glatt:2008}.
Particularly, young and intermediate-age stellar clusters provide important information of the 
recent ($<$ 1 Gyr) interactions of the LMC-SMC-MW system. Investigations in this context have 
shown that the location of the young stellar populations of the SMC are biased towards the East 
side of the galaxy, while the bulk of the old population ($\sim$10 Gyr) presents a much 
more spherical and smooth distribution \citep{Gardiner:1992,Zaritsky:2000}. This trend is 
best noticed for stellar associations, HII regions and stellar clusters \citep{Bica:1995}.

As far as we are aware, \citet[][hereafter B08]{Bica:2008} presented the most complete 
compilation of extended objects in the MCs, containing over 7000 clusters and 
associations. However, the characterisation of such a large number of targets is beyond the 
scope of any single work. Therefore, many objects from that catalogue still lack their 
fundamental parameters while others do not correspond to physical systems at all 
\citep{Piatti:2012}. Particularly, the sparse, poorly populated clusters near the central regions of 
these galaxies have been neglected due to 
the difficulties imposed by the large field contamination and source crowding. 

Previous photometric studies of SMC clusters include the works of \citet{Pietrzynski:1999}, 
\citet{Chiosi:2006}, \citet{Glatt:2010}, \citet{Piatti:2011} and \citet{Piatti:2012} that used 
colour-magnitude diagrams (CMDs) and isochrone fittings to age-date stellar clusters. 
Hubble Space Telescope ($HST$) data have also been used to derive the age of a few 
SMC clusters via CMDs \citep[e.g.][]{Mighell:1998, Rich:2000,Rochau:2007,Chiosi:2007}.

As important as the age estimate for large samples of SMC clusters, the knowledge 
of their formation and evolution would benefit greatly from 
studies on their mass and initial mass function (IMF), despite all difficulties inherent 
to their derivation. 
Mass and IMF, combined with age information, are fundamental properties 
which allow a connection between  dynamics and stellar evolution within a cluster. 
Diagnostics of mass segregation, evaporation and cluster evolutionary state become 
feasible if these properties are known. 

\citet{Kontizas:1982} have performed star counts on photographic plates obtained for 
20 SMC clusters to derive their structural parameters (core and tidal radius) from 
\citet{King:1962} profile fittings. They estimated the cluster masses ($M_\mathrm{clu}$) 
using the \citet{King:1962} approximation for the cluster tidal radius:

\begin{equation}
R_t = d [M_\mathrm{clu}/(3.5 M_\mathrm{smc})]^{1/3}, 
\label{eqtr}
\end{equation}

\noindent
where $M_\mathrm{smc}$ is the 
SMC mass \citep[3$\times$10$^9$ M$_\odot$;][]{de-Vaucouleurs:1972} and $d$ is the 
distance between the cluster and the SMC dynamical centre, adopted as 
the rotation centre at RA$(1950.0)=1^\rmn{h} 03^\rmn{m}$ and 
Dec.$(1950.0)=-72\degr 45\arcmin$ 
\citep{Hindman:1967,Westerlund:1997}. The above expression gave an upper 
mass limit because the distances used were the projected ones. 

An often pursued approach, suitable to deal with large cluster samples, is to estimate 
masses from absolute magnitude and evolutionary models. In this context, 
\citet{Hunter:2003} compared $UBVR$ integrated colours of 939 SMC clusters with 
evolutionary models to obtain their ages and masses. They further investigated the effects
of fading and size-of-sample effects in star cluster analysis. \citet{Dias:2010} has derived the 
ages and metallicities of 14 SMC clusters, using integrated spectra fitted to theoretical models.
They showed that these parameters are not critically affected neither by the SSP model used 
nor by the fitting technique employed.
\citet{Mackey:2003} compiled ages from the literature and determined masses 
for 10 populous SMC clusters from their surface brightness profiles 
measured from $HST$ images. 
\citet{Carvalho:2008} fitted \citet{Elson:1987}
models to surface brightness profiles of 23 SMC clusters and calculated 
their masses, total cluster luminosities and the mass-to-light ratios as in 
\citet{Mackey:2003}.

On a more direct approach a cluster luminosity function (LF) can be converted into a 
mass distribution by using an isochrone mass-luminosity relation. It has the advantage that, 
in addition to the total mass, mass function (MF) slopes can also be derived. In this context
$HST$ data has been largely used to the investigation of SMC clusters, mostly due to its 
photometric depth and spatial resolution. \citet{Glatt:2011} estimated the present-day 
mass function (PDMF) and the total masses of 6 clusters older than 1 Gyr, detecting the 
occurrence of mass segregation.
\citet{Rochau:2007} derived a PDMF similar to the \citet{Salpeter:1955} IMF for BS90, a 4.5 Gyr 
old cluster, also detecting mass segregation as a result of the cluster dynamical evolution since 
its age is larger than its relaxation time. 

\citet{Chiosi:2007} estimated the IMF of the younger clusters NGC265, K29 and NGC290 
(age $\sim$ 100 Myr), founding that it is compatible 
with that of \citet{Kroupa:2001} for masses between 0.7 and 4 M$_\odot$. Additional young,
well-studied clusters are NGC330 \citep{Sirianni:2002,Gouliermis:2004}, NGC346 
\citep{Massey:1995,Sabbi:2008} and NGC602 \citep{Schmalzl:2008,Cignoni:2009}; 
all presenting MF slopes consistent with the Salpeter value. 
Mass segregation was also detected for these clusters, and suggested to be 
primordial as their age is smaller than their relaxation times. 

We aim at increasing the number of clusters with estimated mass in the SMC, thus 
helping to better understand the cluster dynamical evolution in this galaxy. 
We made use of a recently developed method \citep{Maia:2010} to account for the field star 
contamination by sampling the field population from a neighbouring region and statistically 
removing it from the cluster CMD. Then, we estimated for the first time the total mass and mass 
function slopes for a sample of 29 SMC clusters. 
This information can be used to infer the evolutionary state of these targets and 
also provide additional constraints on the environmental conditions of the clusters evolution in
the galaxy. For this purpose 
we made use of deep Washington photometry acquired at the CTIO 4m Blanco telescope, drawn 
from the NOAO Science Data Management 
Archives\footnote{http://www.noao.edu/sdm/archives.php}.

In Sect.~2 we describe the collected data, their reduction and the initial cluster sample. 
Sect.~3 deals with the methods used to select the cluster sample analysed in this work. 
Sect.~4 describes the field decontamination method and isochrone fitting. Mass functions of our 
targets are derived in Sect~5, where cluster mass is also estimated. In Sect~6 we discuss our
results and in Sect.~7 we draw our concluding remarks.

\section{Data handling and cluster selection}

The photometric data used in this work were taken from the NOAO Science Data 
Management Archives and included several star clusters inside 11 fields distributed throughout 
the SMC. The images were obtained in December, 2008
at the CTIO 4m Blanco telescope with the Mosaic II camera attached ($36 \times 36$ 
arcmin$^2$ field with a 0.27 arcsec.pixel$^{-1}$ plate scale) and the $C$ and $T_1$ 
Washington photometric filters. Exposure times in these filters were 1500s and 300s,
respectively.

The reduction and the calibration of the frames were carried out using standard {\sc iraf} routines 
from the mosaic data reduction package ({\sc mscred}) and the photometry was performed using 
the star finding and point spread functions fitting routines from the {\sc daphot/allstar} packages. 
The average seeing values measured in the images were 1.2\arcsec and 1.0\arcsec, in the
$C$ and $T_1$ filters respectively. The 50\% completeness level were reached at $C \sim 
23-24.5$ and $T_1 \sim 22.5-24.0$, depending on the crowding of the field, corresponding to 
a mass limit of $\sim 1.2\, \mathrm{M}_\odot$ if reddening is neglected.
Further details of the data processing and of the photometry of the images are described in
\citet{Piatti:2011,Piatti:2012a}.

The images contained a total of 152 clusters from the B08 catalogue. 
Recent works on this target list have already reported 20 intermediate-age 
or old clusters \citep{Piatti:2011, Piatti:2011b}, 4 moderately young clusters and 17 possible 
asterisms \citep{Piatti:2012}. The present work focuses on a subset of the remaining sample of 
68 candidate clusters, which includes potentially younger objects and asterisms. 

\section{Star count analysis}
\subsection{Density maps}

A stellar density enhancement over the surrounding field is the most basic condition to identify
a star cluster. However, discerning such an enhancement from field density fluctuations can 
be difficult, specially in dense regions.  

To address this issue we constructed stellar density maps for each target in our sample. Each 
map was constructed by: (i) selecting stars brighter/fainter than a magnitude threshold value 
(see below) inside a $2 \times 2$ arcmin box around the target literature coordinates; 
(ii) calculating the stellar density value on a circular region of 5 arcsec radius around 
each star; (iii) interpolating these stellar density values to a uniform grid with a resolution of 
$\approx 5$ arcsec and; (iv) plotting these values as a contour map.

\begin{figure}
\centering
\includegraphics[width=8cm]{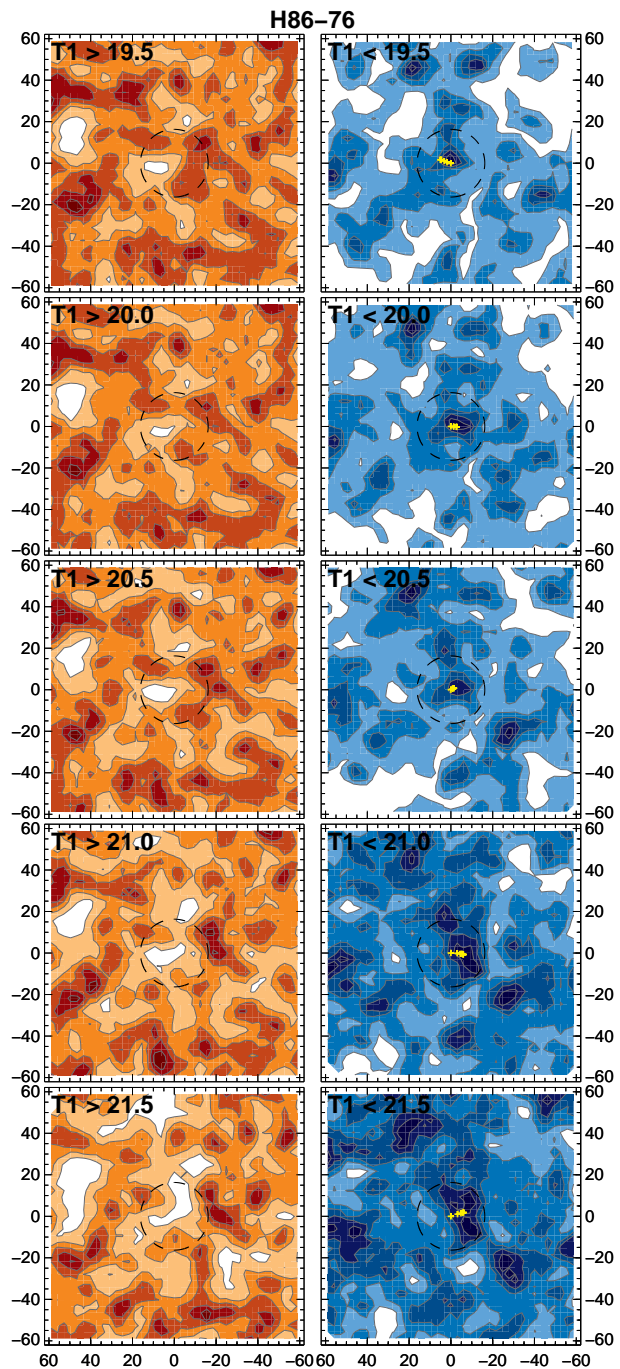}
\caption{"Red" (left) and "blue" (right) stellar density maps for the cluster H86-76. In each 
map, darker colours represent relatively higher stelar density levels. The density levels between
the maps are not related. The magnitude threshold for the top panel maps ($T_1 = 19.5$) 
ensures that this "blue" map contains at least 10 per cent of the total stellar population. The 
dashed line represents the cluster visual radius and the yellow symbols indicate iterations of the
centre finding algorithm.}
\label{fig1}
\end{figure}

In a tentative to probe the cluster stars distribution, we constructed a map considering 
only stars brighter than a magnitude threshold in the $T_1$ band. This threshold was defined 
so as to include the brightest 10 per cent of the total number of selected stars. 
In addition, a complementary map was also created using the stars fainter than the 
defined threshold to probe the field population. These maps are hereafter referred as the "blue" 
and "red" stellar density maps, respectively. Finally, additional "blue" and  "red" density maps 
were also created by shifting this magnitude threshold towards fainter magnitudes at increments 
of 0.5 mag in order to find the magnitude limit that gives the best contrast between the cluster 
and field population. Five "blue" and "red" maps were built for each cluster candidate, following 
the above procedure. Fig.~\ref{fig1} shows the resulting density maps for the cluster H86-76.

It can be seen that as the magnitude threshold is increased from its initial value, the cluster 
becomes better defined in the "blue" map up to a magnitude limit ($T_1 \approx 20.5$),
where the field population starts to dominate. This magnitude limit varies among our target
sample, as it depends on both the cluster age and the relative stellar overdensity 
over the nearby field. It can be more objectively defined by using the target cumulative 
luminosity function (see section~\ref{lfsect}).

\subsection{Centre calculation}

The next step in characterising any cluster candidate consists in determining its centre 
coordinates. For this purpose, we devised an algorithm to iteratively search for a stellar 
density peak by calculating the density weighted average position of the stars. 
Given a cluster visual radius and an initial centre coordinates (B08), the algorithm: (i) selects the
stars inside the cluster radius around the initial centre; (ii) calculates new centre coordinates as
the mean of the selected stars position weighted by the calculated stellar density around each
star; (iii) checks for convergence; (iv) either starts a new iteration by replacing the initial 
coordinates by the calculated coordinates or stops adopting the last calculated coordinates as 
the final centre coordinates.

The algorithm converges if the distance between the initial centre and the last derived centre is 
less than 0.5'' and the stellar density at the latter is 1-$\sigma$ above the sky density 
fluctuations. The algorithm aborts if the maximum number of 5 iterations is reached.

This centre finding algorithm was applied to the five blue density maps of each candidate 
cluster. Even 
though any young cluster would certainly appear prominent in the "blue" maps, the fainter stellar 
population of the cluster should also allow an overdensity to be detectable as the maps start 
sampling fainter stars. Therefore, genuine clusters should show a density enhancement in  
"blue" density maps.

To be classified as a possible cluster, the centre finding algorithm must have converged on at 
least 3 of the 5 "blue" maps of the candidate. Only 37 of our 68 objects met this criterion, most of 
them converging on all 5 density maps. Therefore, we concentrated our subsequent analysis on
this selected sample of 37 cluster candidates. Fig.~\ref{fig1} shows the results of the centre 
finding algorithm applied to the "blue" maps of the cluster H86-76 (yellow crosses).

The possible clusters selected had their centre coordinates calculated as the mean of the 
coordinates found in each density map. Clusters that have survived all our selection criteria 
(see below) have their centre coordinates (RA, Dec.) shown in Table~\ref{tab1}.

\subsection{Radial density profile}
\label{rdpsect}

In order to better trace the structure and the extension of the selected targets, we constructed 
radial density profiles (RDPs) around their newly determined centre coordinates and compared 
them with the mean stellar density of the surrounding field. Candidate clusters should present 
an stellar overdensity over the field. Moreover, a cluster limiting radius should be clearly defined 
as the radius where its local stellar density intersects that of the surrounding field.

Since the clusters structure is more easily discernable from the field when its brighter 
stellar population are considered (see Fig.~\ref{fig1}), we have conducted the RDP analysis by 
considering only stars brighter than the cluster limiting magnitude (see Sect.~\ref{lfsect}).
Each RDP was built by calculating the stellar density inside consecutive annular bins of
various sizes around the target centre coordinates, up to a radius of $\approx 100$ arcsec. 
The radius of the circle drawn in the blue density maps was adopted as what we called 
limiting radius and the mean stellar density of the surrounding field was calculated in an 
annulus immediately outside this limiting radius, with the same area like the internal region. 

\begin{figure}
\centering
\includegraphics[width=0.465\linewidth]{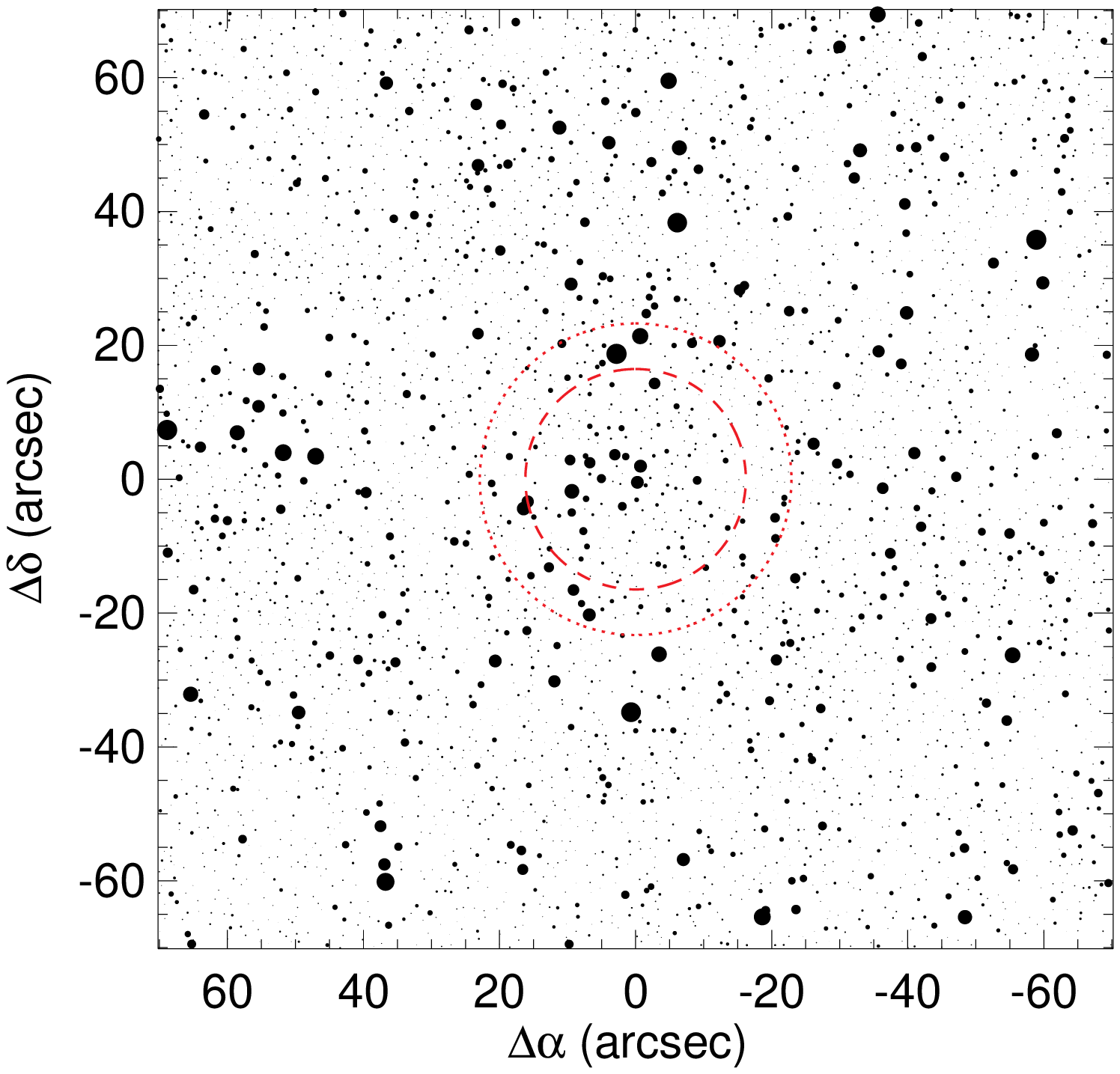} \hspace{0.04\linewidth}
\includegraphics[width=0.47\linewidth]{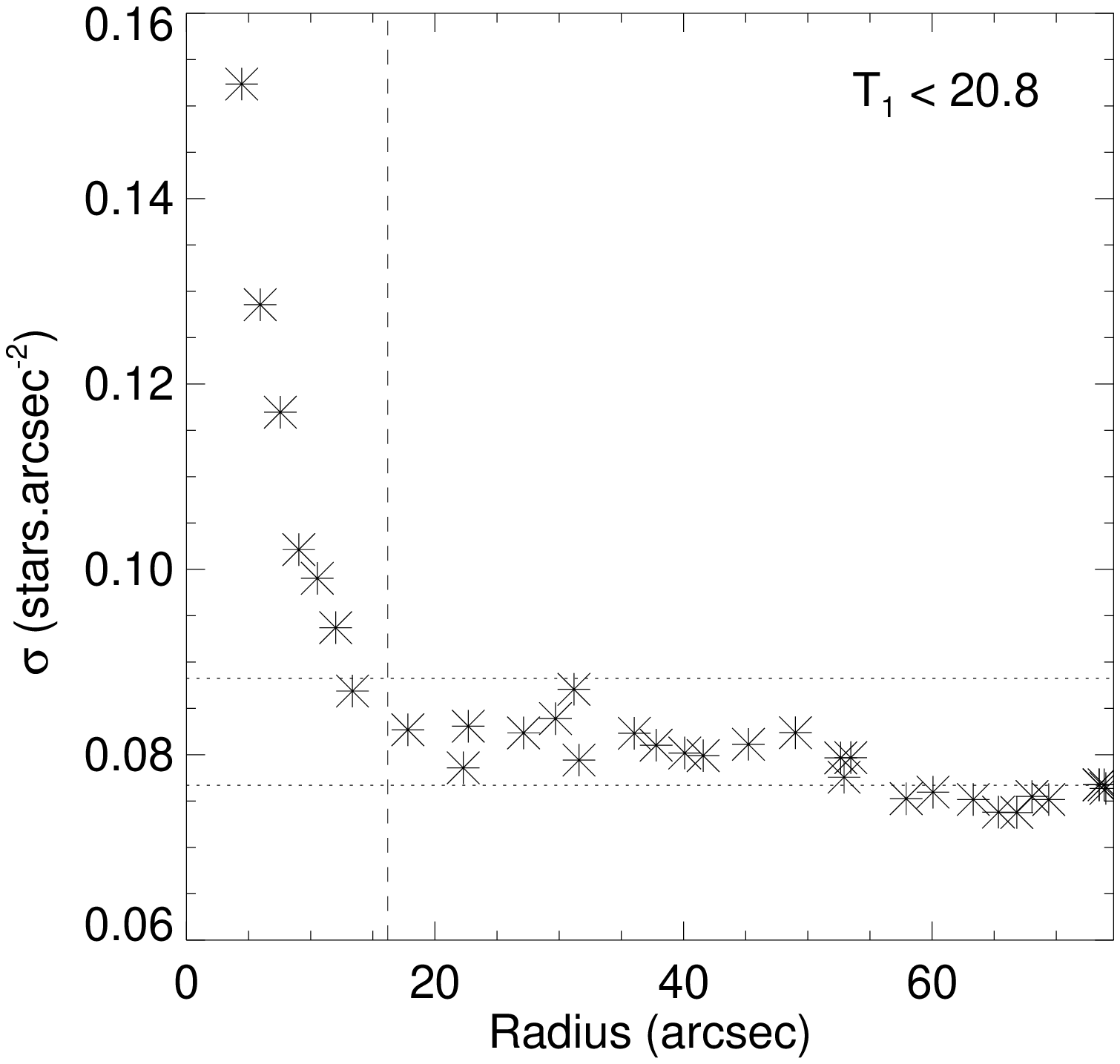}
\caption{ Schematic sky chart (left) and  magnitude limited RDP (right) of the cluster H86-76. 
The defined limiting radius and field sampling area are indicated in the chart by the dashed and 
dotted lines, respectively. The cluster limiting radius and the 1-$\sigma$ interval of the field star 
density are represented in the RDP by the dashed and dotted lines, respectively. }
\label{fig2}
\end{figure}

Because some targets are located at the borders of the images, or very close to another 
catalogued object, or on highly variable fields, the mean stellar densities of their surrounding 
field were calculated in an adjacent region with the same area located to the North, to the East, 
to the South or to the West directions of the targets, as appropriate. Fig~\ref{fig2} shows a 
schematic finding chart and the magnitude limited RDP of the cluster H86-76. The cluster 
limiting radius and the field sample adopted are indicated in the diagrams. 

Based on the analysis of their RDPs, 4 additional targets were removed from our candidate 
cluster list. These underpopulous targets could not be distinguished from the surrounding field 
density fluctuations. Table~\ref{tab1} lists the derived limiting radii (R$_\mathrm{l}$) and 
the mean stellar densities of the surrounding fields ($\sigma_\mathrm{bg}$) for the surviving
clusters.

In addition to the limiting radius, the central density and the core radius of the candidate 
clusters were also determined by fitting a 2-parameter \citet{King:1962} function to their 
magnitude limited radial profiles, according to the expression:

\begin{equation}
\sigma(r) = \sigma_{bg} + \frac{\sigma_0}{1 + (r/R_\mathrm{c})^2},
\end{equation}

\noindent
where $\sigma_{bg}$ represents the background density, $\sigma_0$ is the central density and
$R_\mathrm{c}$ is the core radius.

\begin{figure}
\centering
\includegraphics[width=0.95\linewidth]{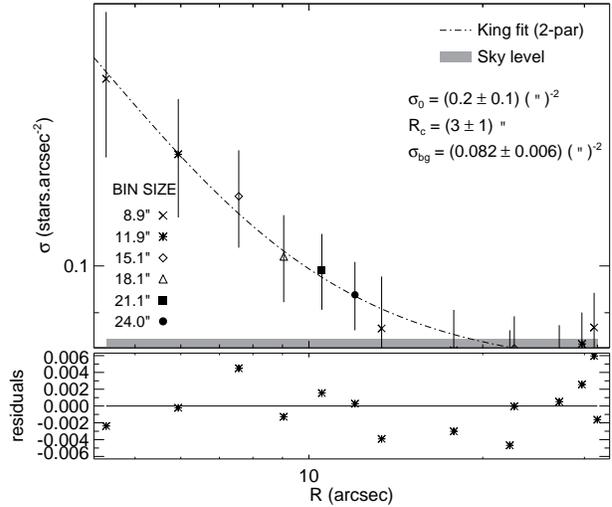}
\caption{King profile fitted to the RDP of the cluster H86-76 (top), considering only stars 
brighter than its magnitude limit ($\mathrm{T}_1 <  20.8$). The fitted function (dot-slashed line), 
the derived structural parameters and the level of the density 
fluctuations of the sky (grey bar) are shown. The Poisson uncertainties and the residuals of the fit,
in the sense (data - model) are also indicated for each annular bin (bottom).}
\label{fig3}
\end{figure}

The determined $\sigma_0$ and R$_\mathrm{c}$ of the targets are shown in Table~\ref{tab1}. 
Fig~\ref{fig3} shows the magnitude limited RDP of the cluster H86-76 and the fit for the 
determination of its structural parameters.

Moreover, a visual inspection in the derived density maps leads to conclude that 10 
clusters exhibit some sign of star field substructure. In order to avoid the bulk of the field 
population, only the "blue" maps presenting stars brighter than the derived magnitude limit were 
considered in this inspection. 
These 10 mostly young clusters resulted in a wide range of stellar background density, limiting 
radius and masses. Nevertheless,  one intermediate-age cluster
(BS75) and another of an interacting pair (NGC241) also present this feature.

\section{CMD analysis}

\subsection{Magnitude limit}
\label{lfsect}

The stellar density maps employed in the determination of the centres of our targets have shown 
the existence of a magnitude limit for which the cluster population has an enhanced density
contrast over the field. In order to better define this value, cumulative luminosity functions (CLFs) 
of our candidate clusters were built and compared to the CLFs of their surrounding fields.

At any given magnitude, the difference between the cluster region CLF and the surrounding 
field CLF should increase
when the cluster population is larger than that of the field, stall when the cluster population 
is comparable to that of the field or decrease when the field population is larger than that of the 
cluster. The magnitude limit was thus given by the bin where the difference of the CLFs stalls 
or reach a peak. Care was taken to exclude peaks at bright magnitudes corresponding to the
cluster turn-off.

\begin{figure}
\centering
\includegraphics[width=0.475\linewidth]{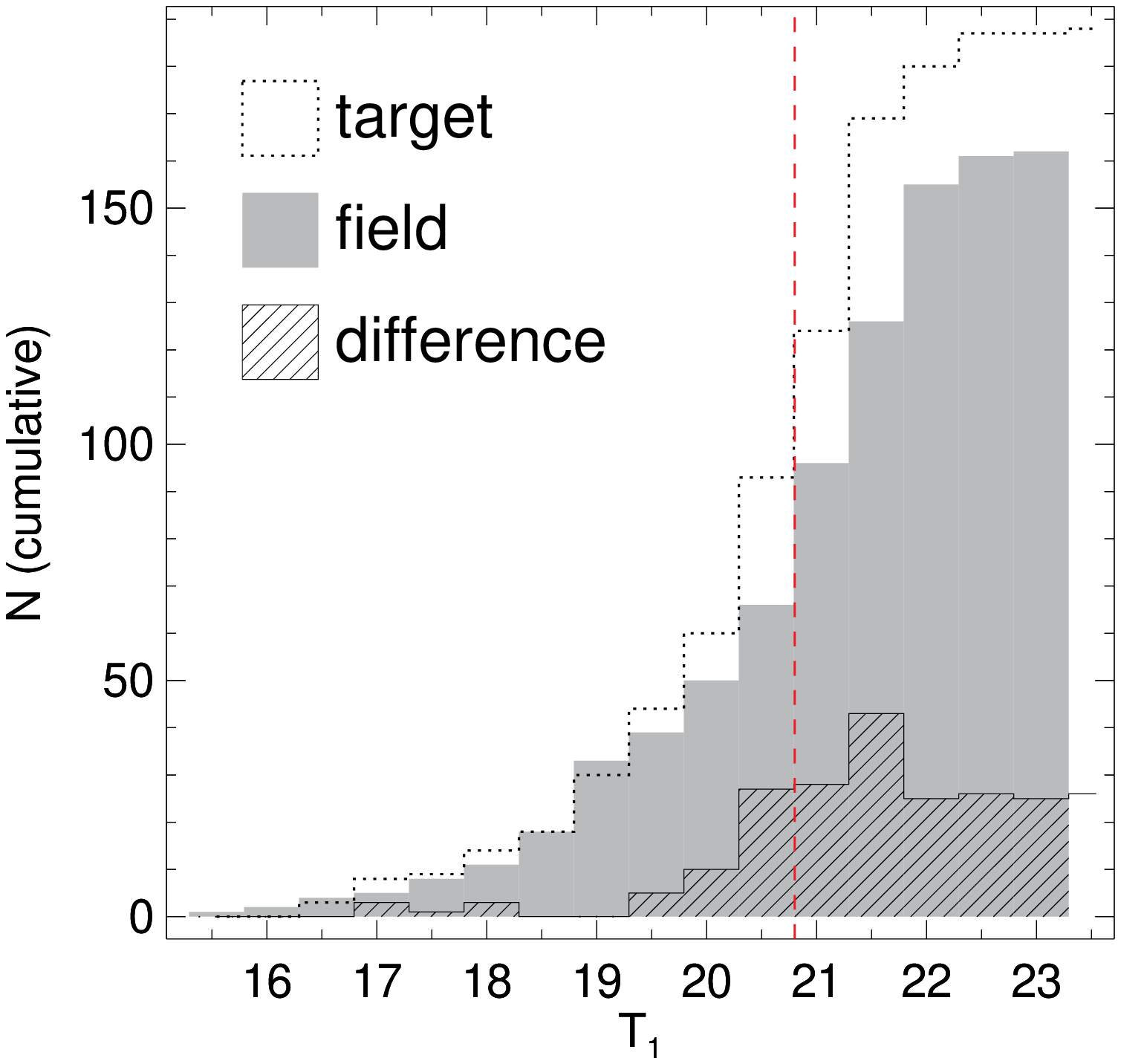} \hskip 0.02\linewidth
\includegraphics[width=0.47\linewidth]{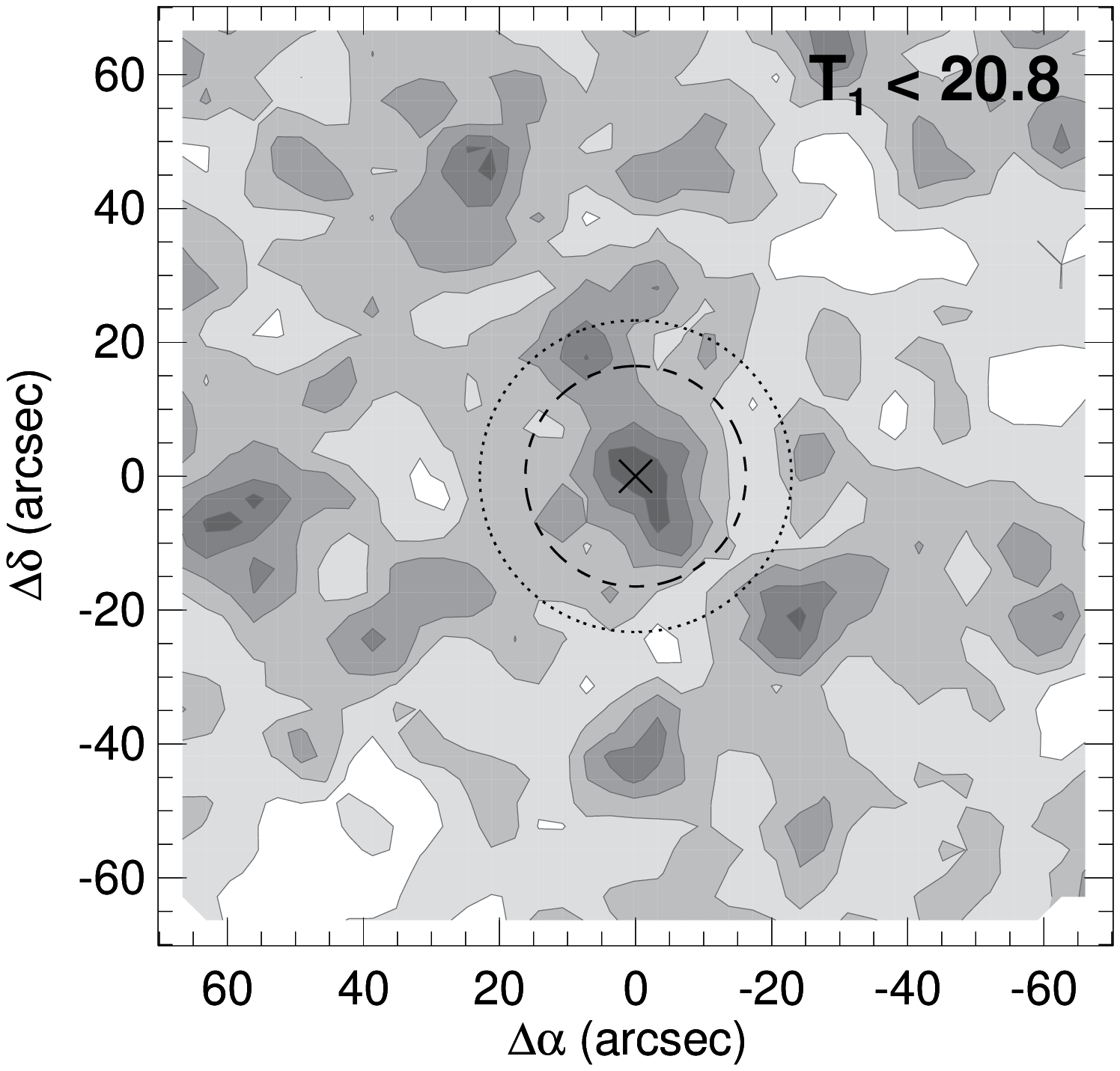}
\caption{Difference between the cumulative luminosity functions of H86-76 region 
and its surrounding field (left). The 
defined magnitude limit is indicated by the vertical dashed line. The stellar density map 
considering only stars brighter than this limit (right) allows for the identification of the cluster. Its 
limiting radius and field sampling area are indicated by the dashed and dotted lines, 
respectively.}
\label{fig4}
\end{figure}

Fig.~\ref{fig4} shows the difference between the CLF of the cluster region and the CLF of its 
surrounding field, built by counting stars within 0.5 magnitude bins starting from the brightest 
measured magnitude. The magnitude limit (M$_\mathrm{lim}$) found for each target is listed in 
Table~\ref{tab1}. A stellar density map constructed considering only stars brighter 
than the defined magnitude limit is also shown. The limiting radius and the field 
sampling area are indicated in this map.

\subsection{Isochrone fitting}
\label{decsect}

Because of the SMC field population is dominated by a mixture of stellar populations, the CMD 
analysis of its clusters can be severely biased by the presence of field stars. Even when young 
clusters 
are considered, the field contamination hampers the identification of the late evolutionary 
sequences as their sparse bright populations are often entangled with field giant stars.

In order to mitigate this effect, a decontamination procedure was used to statistically remove the
field population from the cluster CMD \citep{Maia:2010}. It works by (i) sampling the field 
photometric characteristics in a nearby region (see Sect.~\ref{rdpsect}); (ii) comparing with the 
cluster region containing members and field stars in the CMD; (iii) removing field stars from this 
region based on their local photometric similarity with the nearby field and on their distances 
from the cluster centre; (iv) assigning a photometric cluster membership value based on the 
local overdensity of stars in the CMD space. Previous validation tests of this method against 
proper motion selected members has shown that stars with membership probabilities larger 
than 0.3 comprises the best member sample.

Ages and colour excesses of the surviving 33 targets were then determined by means of 
isochrone fits to the $T_1 \times (C$-$T_1)$ decontaminated CMDs, giving higher priority to 
the most probable members (with higher assigned membership).  The Padova isochrones 
\citep{Marigo:2008} with Z=0.004 metallicity were used. Other metallicity values were also tested 
(Z=0.008 and Z=0.002) but discarded since they provided a poorer fit, particularly for stars at 
the turn-off and at the subgiant branch regions. 

Analysis of the decontaminated CMDs of our candidate clusters led us to further exclude 4 
additional targets as they do not show resemblance of any evolutionary sequence on the 
CMD. 
All the 39 rejected clusters candidates are gathered in Table~\ref{rejclu} (see Appendix~A), 
with their coordinates and
a remark about the criterion by which they were rejected.
The 29 targets that passed through this last criterion compose our final list of studied 
clusters. 
Their derived ages and colour excesses are shown in Table~\ref{tab1}. Uncertainties of these 
parameters were estimated by increasing/decreasing their values until a reasonable fit is no 
longer possible for the probable members. On average, errors from the isochrone fittings amount
to $\Delta \log t \approx 0.05$ and $\Delta$E(B-V)$\approx 0.05$.

\begin{figure}
\centering
\includegraphics[width=\linewidth]{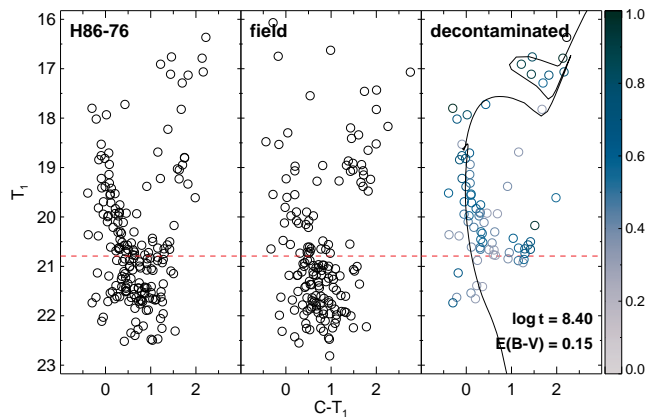}
\caption{Extracted (left), surrounding field (middle) and decontaminated (right) CMDs of H86-76. 
The best fitted isochrone and its derived parameters are shown in the right panel. 
The coloured-bar represents the assigned membership values. The limiting magnitude is 
represented by the dashed line.}
\label{fig5}
\end{figure}

Fig.~\ref{fig5} compares the CMDs of the target region and of the surrounding field with the one 
from the decontaminated sample for the cluster H86-76. 
The best isochrone fitted and the limiting magnitude derived in Sect.~\ref{lfsect} are also shown. 
For the majority of our targets the limiting magnitude derived showed good agreement with the 
faintest member stars left in the decontaminated CMDs.

\section{Mass distribution}
\label{mdistsect}

The distribution of mass in a stellar cluster can yield important information on its evolutionary 
state and on the external environment in which it is inserted. As none of the studied clusters
show any sign of their pre-natal dust or gas, their stellar components are the only source of their 
gravitational potential. Thus, the number of member stars and their concentration will determine, 
in addition to the galaxy potential, for how long clusters survive.

Since the majority of the studied clusters have ages between 100-500 Myr, mass loss due to 
stellar evaporation and tidal stripping play an important role in their structural evolution, which 
in turn may leave signs in their stellar mass distributions.
In principle, an unperturbed mass distribution should resemble the cluster IMF, whereas 
deviances from it can be interpreted as an effect of the tidal field and of the 
cluster internal dynamical evolution.

To derive the mass functions of the cluster sample, luminosity functions were 
built by counting stars inside 0.5 mag bins along the $T_1$ magnitude range. In order to account 
for the masses of 
member stars, two methods have been employed to discard field stars. The first method  
-hereafter called CMD method- consists in constructing the LF directly from the derived 
decontaminated CMD (Sect.~\ref{decsect}). In the second method -hereafter referred as
LF method- the LF is constructed using all measured stars inside the cluster limiting radius and 
then subtracted from a similar LF of the surrounding field region.

To ensure homogeneity, only stars brighter than the derived magnitude 
limit were considered in both methods. Moreover, some clusters had their magnitude limits 
shifted a bin towards brighter magnitudes in order to avoid scarcely populated regions or clear
field leftover in the decontaminated CMDs.

\begin{figure}
\centering
\includegraphics[width=0.85\linewidth]{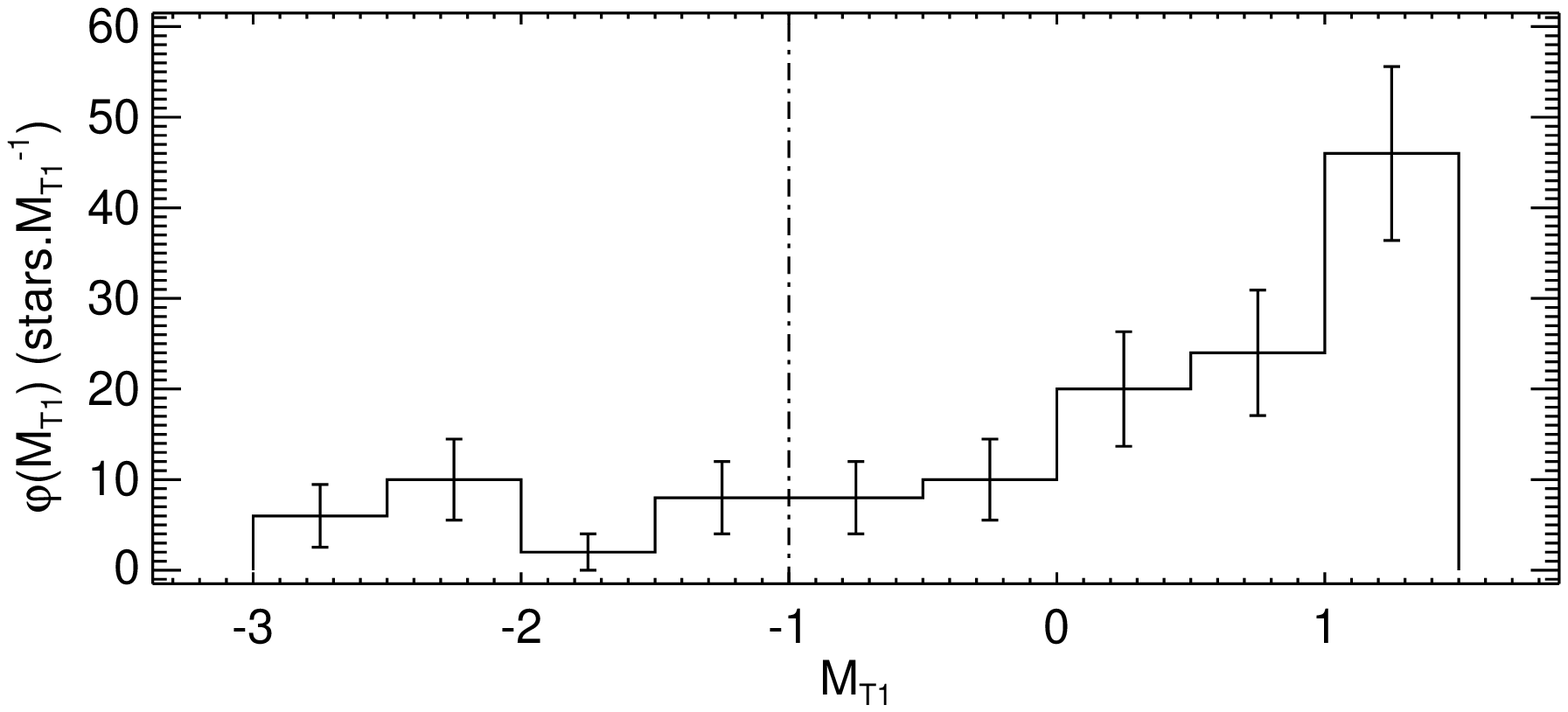} \\
\includegraphics[width=0.85\linewidth]{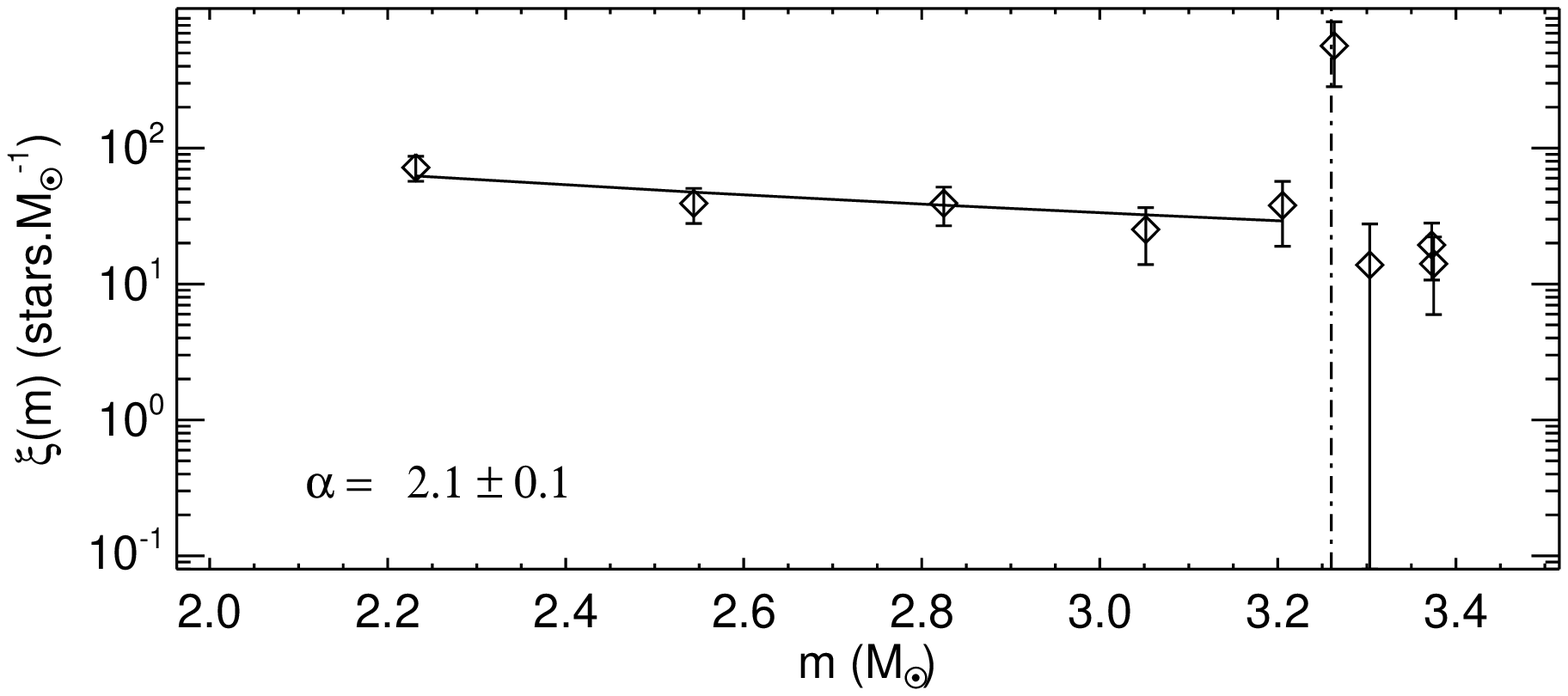}
\caption{Derived LF (top) and MF (bottom) of H86-76 using the CMD method for field 
removal. The turn-off magnitude and the corresponding mass are indicated by the dot-dashed 
vertical lines. 
The mass function fit over the main sequence stars is indicated by the solid line and 
by the power law exponent. The error bars correspond to the Poisson uncertainties.}
\label{fig6}
\end{figure}

\begin{figure}
\centering
\includegraphics[width=0.85\linewidth]{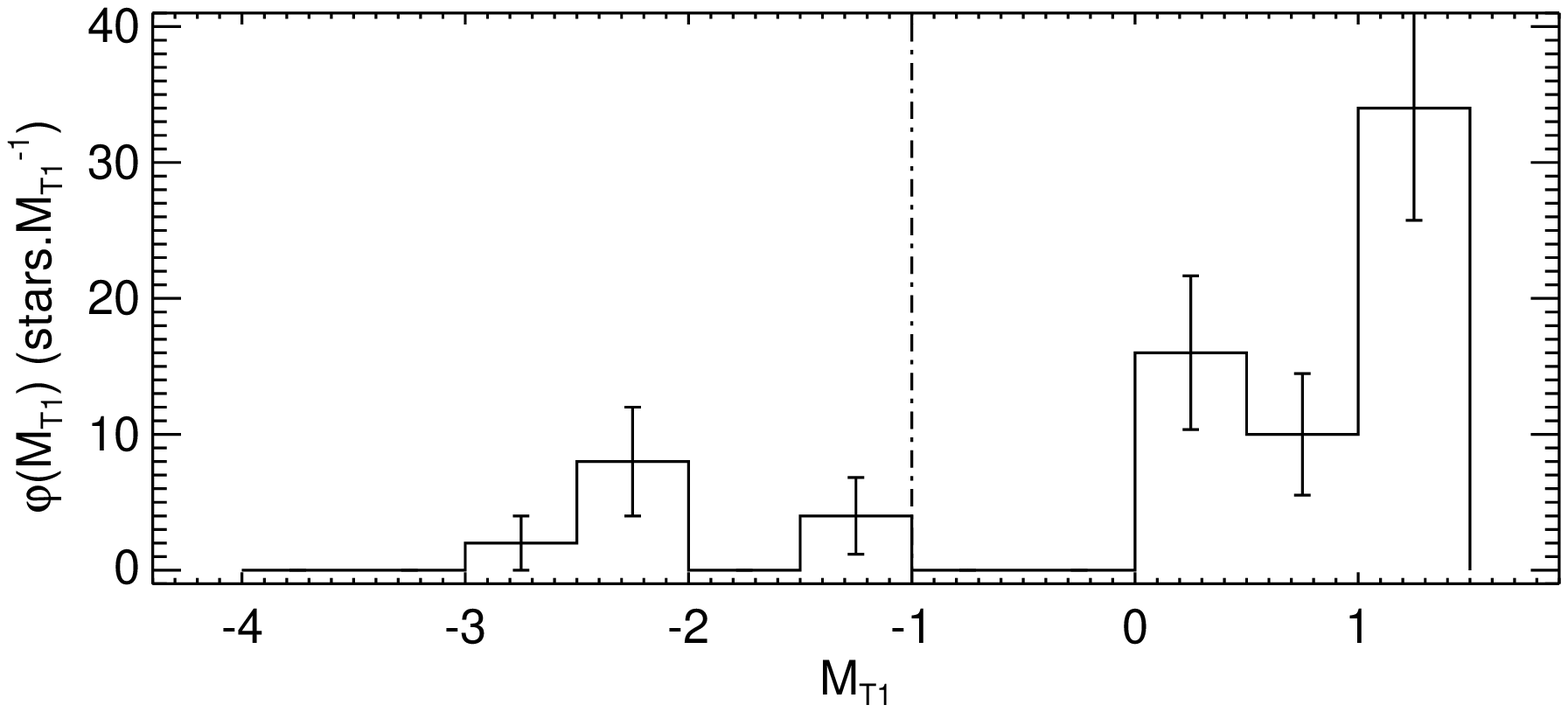} \\
\includegraphics[width=0.85\linewidth]{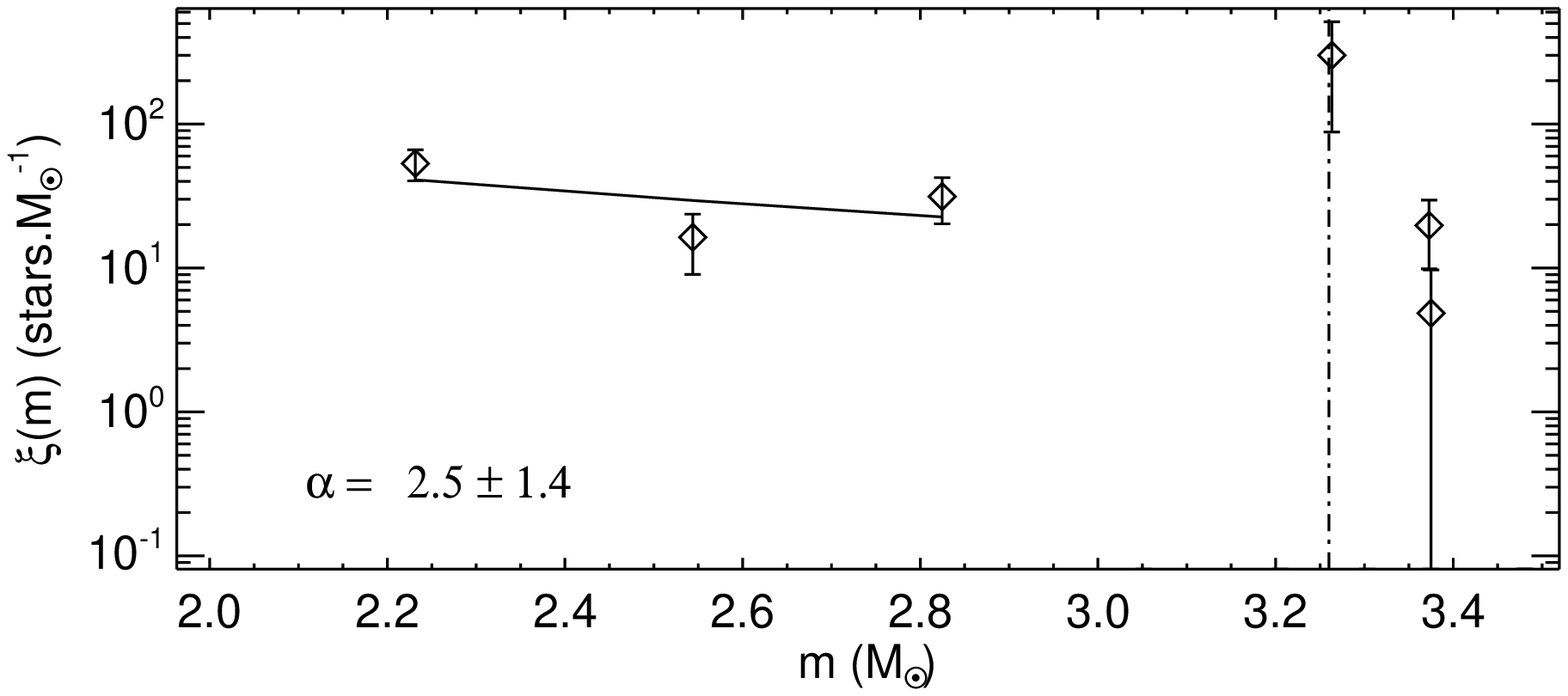}
\caption{Same as Fig~\ref{fig6}, except that the field removed LF was built from the difference 
between cluster and surrounding field LFs (LF method).}
\label{fig7}
\end{figure}

The mass distribution of each cluster was derived by using the mass-luminosity relationship
obtained from the isochrone corresponding to the cluster age. Fig.~\ref{fig6} 
shows the LF and the derived MF for H86-76 using its decontaminated sample (CMD method). 
Fig.~\ref{fig7} shows the LF and MF derived for H86-76, using the difference between the cluster 
and surrounding field LFs (LF method).

It can be seen that the LF built using the LF method presents some gaps, while the one built 
from the CMD method 
shows a smoother increase of the number of stars towards fainter magnitudes. This general
trend can be understood by noting that the difference between cluster and field LFs only 
takes into account the $T_1$ magnitudes of the stars, while the decontaminated sample (from 
which the CMD method comes from) also uses the $C$-$T_1$ colours and positions of the stars 
to better differentiate cluster and field populations. 

\subsection{Mass function slope}

To quantify the distribution of stellar masses in a given cluster two physical parameters are of 
particular interest: the total mass of the cluster and the MF 
slope. By comparing the observed MF slope with the \citet[][hereafter K01]{Kroupa:2001}'s IMF 
slope within the corresponding mass range, one can draw information about the
dynamical evolution of the cluster and diagnose processes such as mass segregation and 
mass loss.

The MF slope can be determined by fitting a power law function over the cluster mass 
distribution. Following the commonly used notation, we fitted an analytic mass function 
given by the power law: $\xi(m) = Am^{- \alpha}$, where $\xi(m)$ represents the mass 
distribution function, $\alpha$ is the power law exponent and $A$ is a normalisation constant. 
The power fit only considered masses smaller than the turn-off mass of the cluster. 
Moreover, it was only performed if 3 or more mass bins met that condition.

Although the mass distributions derived from the two methods are similar, the power 
law function fitted through LFs obtained from the CMD method presented, on average, lower 
uncertainties than those derived from the LF method. Nevertheless, 
whenever these two fits converged, we adopted the weighted mean of the derived 
exponents as the final slope of the MF and calculated its uncertainty by properly propagating 
the error derived from each fit.

Figs.~\ref{fig6} and \ref{fig7} show the resulting fits of the power law and the corresponding
exponent over the two mass distributions derived for H86-76. Table~\ref{tab1} lists the 
mean values and uncertainties of the exponent and the mass range of the power law 
fits for clusters for which these fits converged.

\subsection{Total mass}
\label{tmassect}

The total mass of a cluster was calculated by initially summing up its observable mass, i.e., the 
sum of the masses along the different bins of its mass distribution function, including those 
brighter than the cluster turn-off. 
Secondly, the values of the mass distribution function in the lower mass regime and the MF slope
given by K01 for such a mass interval were used to define the normalisation constant $A$, from 
which we extrapolated the power law function down to lower masses.
The mass contained in these low mass ranges was estimated by integrating the extrapolated
power law from the smallest observed mass bin to 0.1 solar mass. 
The total mass of a cluster was then estimated by adding the 
values obtained in these two steps. Its uncertainty was derived by 
propagating the errors at the individual mass bins in the observed mass distribution function and 
the intrinsic uncertainties of the K01 exponents in the extrapolated power law.

For H86-76 the total mass turned out to be $1405\pm585$ M$_\odot$ if the mass 
distribution function obtained from the CMD method is used, and $1002\pm452$ M$_\odot$ 
if that from the LF method is employed. Generally, our results indicate that the masses 
calculated through the CMD method were systematically higher than those calculated 
through the LF method, up to a factor of 2.
This result suggests that, in most cases, the simple subtraction of the cluster LF from that of the 
field tends to underestimate the actual cluster population.
The more elaborated field decontamination method not only retains a 
larger fraction of the cluster population, but also allow for an individual estimate of the 
member stars. Finally, since the uncertainties in the total masses are generally higher 
than the difference between the two values, we adopted the average of these masses 
weighted by their relative errors in the subsequent analysis. Its uncertainty 
was derived by propagating the errors of the two individual determinations. These results are 
shown in Table~\ref{tab1}.

Appendix~C compiles the general and magnitude limited density maps, the radial density profile, 
the cumulative luminosity function, the decontaminated CMD, and the field-subtracted LF and 
MF charts used in the analysis of each cluster. They are only available in the online version of 
the Journal.

\renewcommand{\tabcolsep}{0.11cm}
\begin{table*}
\raggedright
\caption{Determined parameters for the studied clusters}
\begin{tabular}{lccc r@{$\,\pm\,$}l r@{$\,\pm\,$}l r@{$\,\pm\,$}l ccc r@{$\,\pm\,$}l r@{$\,\pm\,$}l c r@{$\,\pm\,$}l}
\hline
Target & RA & Dec. & R$_\mathrm{l}$ &\multicolumn{2}{c}{$\sigma_\mathrm{bg}$} & \multicolumn{2}{c}{$\sigma_0$} & \multicolumn{2}{c}{R$_\mathrm{c}$} & M$_\mathrm{lim}$ & log\,t & E(B-V) & \multicolumn{2}{c}{total mass} &  \multicolumn{2}{c}{MF slope} & MF range & \multicolumn{2}{c}{R$_t$} \\ 
  & ($^\rmn{h}:^\rmn{m}:^\rmn{s}$) & ($\degr:\arcmin:\arcsec$) & ( \arcsec ) & \multicolumn{2}{c}{(arcsec$^{-2}$)} & \multicolumn{2}{c}{(arcsec$^{-2}$)} & \multicolumn{2}{c}{( \arcsec )} &  &  &  & \multicolumn{2}{c}{(10$^3$ M$_\odot$)} & \multicolumn{2}{c}{} & (M$_\odot$) & \multicolumn{2}{c}{( \arcsec )} \\ \hline
  NGC241 & 00:43:33 & -73:26:20 & 23 & 0.186&0.027 & 0.10&0.02 &  9.5&1.4 & 19.9 & 8.35 & 0.00 & 2.2&0.7 & 2.2&0.8 & 2.60$-$3.26 & 38.6&4.2 \\
  NGC242 & 00:43:38 & -73:26:26 & 20 & 0.196&0.028 & 0.08&0.01 &  6.2&0.7 & 20.0 & 7.80 & 0.05 & 1.1&0.4 & 0.8&0.2 & 3.01$-$6.05 & 31.1&3.3 \\
  H86-76 & 00:46:01 & -73:23:44 & 16 & 0.196&0.016 & 0.23&0.13 &  3.1&1.2 & 20.8 & 8.40 & 0.15 & 1.2&0.4 & 2.1&0.7 & 2.23$-$3.23 & 28.7&2.9 \\
  H86-85 & 00:46:55 & -73:25:24 & 20 & 0.175&0.019 & 0.16&0.11 &  2.7&1.1 & 19.8 & 7.90 & 0.15 & 1.4&0.6 & 2.3&0.2 & 3.51$-$5.50 & 29.4&3.9 \\
  H86-87 & 00:47:06 & -73:22:23 & 24 & 0.192&0.012 & 0.04&0.01 & 10.4&1.4 & 19.7 & 8.10 & 0.10 & 3.1&1.7 & \multicolumn{2}{c}{$-$} & $-$ & 37.6&6.7 \\
  H86-90 & 00:47:25 & -73:27:29 & 16 & 0.176&0.027 & 0.32&0.51 &  2.0&1.8 & 20.4 & 8.40 & 0.00 & 0.8&0.3 & 1.9&0.6 & 2.23$-$3.23 & 24.4&3.4 \\
  H86-97 & 00:48:12 & -73:26:47 & 23 & 0.155&0.018 & 0.16&0.03 &  5.6&0.7 & 19.5 & 8.10 & 0.05 & 3.3&1.3 & 2.4&0.7 & 3.27$-$4.41 & 37.7&4.8 \\
     B48 & 00:48:37 & -73:24:56 & 31 & 0.171&0.010 & 0.03&0.00 & 12.9&1.9 & 18.4 & 7.90 & 0.00 & 3.4&1.6 & \multicolumn{2}{c}{$-$} & $-$ & 37.0&5.9 \\
     L39 & 00:49:18 & -73:22:18 & 16 & 0.200&0.017 & 0.21&0.07 &  4.2&1.3 & 20.0 & 8.05 & 0.05 & 1.5&0.5 & 3.1&0.2 & 2.85$-$4.61 & 26.9&3.0 \\
SOGLE196$^a$ & 00:49:27 & -73:23:53 & 16 & 0.123&0.107 & 0.09&0.01 &  7.6&1.0 & 19.9 & 8.35 & 0.00 & 1.0&0.3 & 1.4&0.6 & 2.60$-$3.39 & 23.6&2.6 \\
     B55 & 00:50:21 & -73:23:14 & 24 & 0.181&0.025 & 0.67&0.73 &  2.2&1.3 & 19.9 & 8.40 & 0.00 & 1.3&0.6 & 1.8&1.0 & 2.54$-$3.23 & 24.6&3.9 \\
    BS75 & 00:54:31 & -74:11:07 & 27 & 0.105&0.011 & 0.11&0.03 & 10.1&2.3 & 22.4 & 9.25 & 0.00 & 1.2&0.4 & 2.3&0.6 & 1.17$-$1.43 & 32.4&3.9 \\
    BS80 & 00:56:14 & -74:09:22 & 27 & 0.092&0.011 & 0.15&0.02 &  7.0&0.6 & 22.4 & 9.45 & 0.00 & 1.5&0.4 & \multicolumn{2}{c}{$-$} & $-$ & 33.4&3.1 \\
 H86-174 & 00:57:18 & -72:55:58 & 16 & 0.128&0.013 & 0.20&0.16 &  2.7&1.3 & 20.4 & 8.65 & 0.00 & 0.6&0.2 & \multicolumn{2}{c}{$-$} & $-$ &  9.9&1.0 \\
    HW32 & 00:57:20 & -71:10:13 & 24 & 0.043&0.009 & 0.05&0.01 &  7.4&1.4 & 21.9 & 7.90 & 0.00 & 0.3&0.1 & 1.9&0.1 & 1.44$-$5.24 & 15.4&1.2 \\
 H86-188 & 01:00:14 & -72:27:30 & 32 & 0.054&0.005 & 0.02&0.01 & 13.5&1.3 & 19.9 & 8.10 & 0.00 & 1.0&0.4 & 1.1&0.4 & 2.82$-$4.21 &  5.5&0.7 \\
 H86-190 & 01:00:33 & -72:15:30 & 16 & 0.050&0.018 & 0.02&0.01 &  7.1&3.2 & 20.4 & 7.70 & 0.00 & 0.4&0.1 & 1.3&0.2 & 2.51$-$6.93 &  4.5&0.5 \\
     K43 & 01:00:49 & -73:20:56 & 23 & 0.145&0.014 & 0.08&0.01 & 10.2&2.0 & 20.2 & 8.10 & 0.10 & 2.1&0.7 & 1.5&0.2 & 2.82$-$4.21 & 19.2&2.1 \\
    B103 & 01:00:56 & -73:09:06 & 22 & 0.111&0.013 & 0.08&0.02 &  5.9&0.8 & 19.7 & 8.40 & 0.10 & 1.3&0.5 & \multicolumn{2}{c}{$-$} & $-$ & 12.8&1.6 \\
     B99 & 01:01:24 & -73:14:25 & 24 & 0.108&0.016 & 0.06&0.01 &  6.3&1.0 & 20.2 & 8.10 & 0.10 & 1.0&0.3 & 2.5&0.5 & 2.82$-$4.41 & 12.9&1.5 \\
    B111 & 01:01:58 & -71:01:13 & 22 & 0.038&0.006 & 0.21&0.13 &  3.5&1.2 & 22.4 & 9.15 & 0.00 & 0.5&0.1 & 2.1&0.9 & 1.19$-$1.64 & 19.1&1.7 \\
     K47 & 01:03:11 & -72:16:25 & 22 & 0.050&0.013 & 0.03&0.01 & 10.6&1.2 & 19.9 & 7.90 & 0.00 & 0.6&0.3 & 0.9&0.3 & 2.95$-$5.24 &  3.3&0.5 \\
    B124 & 01:05:02 & -73:02:34 & 16 & 0.134&0.015 & 0.28&0.17 &  2.5&0.9 & 20.4 & 8.00 & 0.00 & 0.4&0.1 & 2.3&0.2 & 2.42$-$4.82 &  6.8&0.6 \\
    HW52 & 01:06:57 & -73:14:06 & 24 & 0.128&0.015 & 0.13&0.02 &  8.2&1.0 & 21.5 & 8.10 & 0.05 & 0.8&0.2 & 1.8&0.2 & 1.68$-$4.41 & 11.8&1.0 \\
     K55 & 01:07:31 & -73:07:11 & 32 & 0.127&0.014 & 0.25&0.05 &  8.9&1.3 & 21.9 & 8.45 & 0.00 & 1.9&0.4 & 2.0&0.1 & 1.41$-$2.85 & 13.7&1.0 \\
     K57 & 01:08:14 & -73:15:25 & 32 & 0.116&0.013 & 0.19&0.02 &  7.1&0.5 & 20.9 & 8.65 & 0.00 & 1.9&0.5 & 2.2&0.5 & 1.82$-$2.48 & 16.7&1.5 \\
    B134 & 01:09:01 & -73:12:24 & 24 & 0.106&0.011 & 0.05&0.01 & 12.3&1.9 & 20.9 & 8.15 & 0.00 & 0.6&0.2 & 1.7&0.2 & 1.99$-$4.21 & 11.0&1.1 \\
     K61$^a$ & 01:09:02 & -73:05:11 & 23 & 0.131&0.032 & 0.07&0.01 & 11.1&1.6 & 20.9 & 8.30 & 0.00 & 1.1&0.3 & 2.6&0.1 & 1.95$-$3.39 & 11.7&1.1 \\
     K63 & 01:10:47 & -72:47:31 & 32 & 0.099&0.009 & 0.10&0.01 & 13.2&0.8 & 21.4 & 8.25 & 0.00 & 1.0&0.2 & 1.3&0.2 & 1.66$-$3.51 &  9.1&0.7 \\

\hline
\end{tabular}

\noindent
Note: $^a$ structural parameters were derived by using semi-annular bins 
to avoid a nearby CCD gap. Likewise, the LF and MF of these targets were corrected to account 
for the cluster area lost in the gap.
\label{tab1}
\end{table*}

\subsubsection{SSP models}

The cluster masses were also determined by using their integrated magnitudes and ages and by 
employing single-burst stellar population (SSP) models as built from Padova isochrones. The 
SSPs contain stars in the mass range $0.08<m$(M$_\odot$)$<120$ distributed according to a 
Kroupa IMF with the total mass normalised to one. SSPs were generated for ages $6.6 < 
\log{t(yr)} < 10.1$ and for metallicities $Z=0.019$, 0.008 and 0.004.

\begin{figure}
\centering
\includegraphics[width=0.85\linewidth]{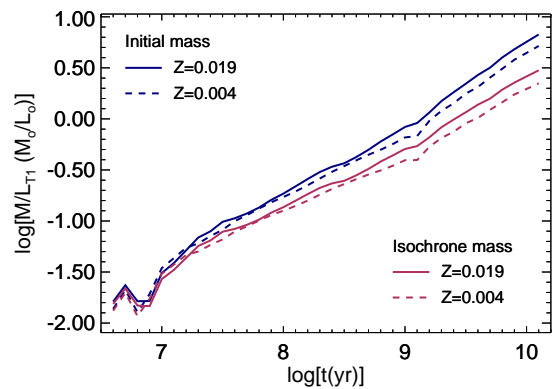}
\caption{The $T_1$ magnitude evolution of the mass-luminosity ratio according to SSP models
of different metallicities and different ways of computing the total mass.}
\label{ageml}
\end{figure}

We started by first computing the evolution of the SSP mass-luminosity ratio
($\mathcal{M}/\mathcal{L}$), which does not depend on the IMF normalisation constant.
Operationally, for a cluster of a given age, its $\mathcal{M}/\mathcal{L}$ is derived from the
models; then, its mass is determined using the integrated absolute magnitude.
Fig.~\ref{ageml} shows the $\mathcal{M}/\mathcal{L}$ evolution in the $T_1$ magnitude for 
SSPs with metallicities  $Z=0.004$ and 0.019. 
The label 'Initial mass' refers to models where the total SSP mass remains 
constant (equal to one) along the cluster evolution, while the label 'Isochrone mass' refers to the 
total mass computed using the actual isochrone stellar masses, which naturally changes with 
age. For the $T_1$ mag, it can be seen that the difference in $\mathcal{M}/\mathcal{L}$ ratio for 
ages larger than 100 Myr due to the different total mass prescriptions adopted are twice as
large as that produced by a metallicity variation of $\Delta Z =0.015$.

The models for which the total mass was computed from the isochrones should reproduce 
better the evolution of real clusters because mass loss effects due to the stellar evolution are 
accounted for. 
Stellar remnants are also excluded from the total mass computed from the isochrones. 
However, even if a cluster retains a substantial content of stellar remnants, its mass fraction is 
small (see Appendix~B), which makes their contribution to the total mass (and light) also small. 

\begin{figure}
\centering
\includegraphics[width=0.85\linewidth]{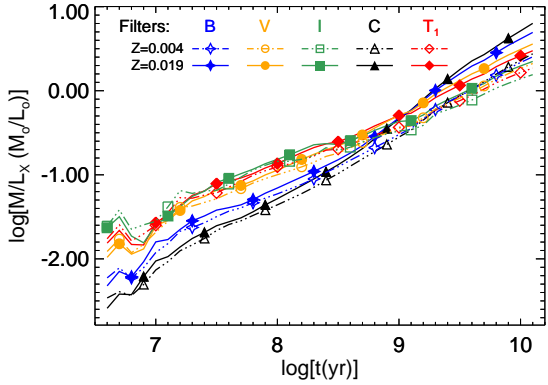}
\caption{Multicolour evolution of the mass-luminosity ratio according 
to SSP models of different metallicities and including mass loss effects.}
\label{logmlfil}
\end{figure}

The $\mathcal{M}/\mathcal{L}$ ratio evolution in various filters ($BVICT_1$) for models including 
mass loss is shown in Fig.~\ref{logmlfil} for $Z=0.019$ and 0.004. The $T_1$ filter is very 
similar to the Johnson $R$ filter. It is worth noticing that the $\mathcal{M}/\mathcal{L}$ ratio has a 
narrow range ($\sim 0.4$\,dex) around $\log{t(yr)}\sim 9$ for the various filters and 
metallicities. After $\log{t(yr)}=9$, the SSPs' $\mathcal{M}/\mathcal{L}$ ratio spread, reaching 
$\sim 0.6$\,dex at the age of $\log{t(yr)}=10$ for the extreme wavelength filters $C$ and $I$. The 
largest $\mathcal{M}/\mathcal{L}$  ratio difference occurs for the youngest SSPs, reaching about 
1.0\,dex at $\log{t(yr)}=6.7$. For ages older than 1\,Gyr, the $\mathcal{M}/\mathcal{L}$ ratio 
in the $I$ band is less sensitive to metallicity and its evolution is 
smoother than that for the shorter wavelength filters. The contrary occurs for the $C$ filter. 

The cluster $\mathcal{M}/\mathcal{L}$ ratio is determined by interpolating its derived age in the 
relations shown in Fig.~\ref{logmlfil} for the filters $T_1$ and $C$, generated 
from the SSP models with $Z=0.004$ (SMC global metallicity) and taking into account mass loss 
effects. Cluster masses were then obtained by means of their integrated $M_{T_1}$ and $M_C$
magnitudes computed from two methods (Sect.~\ref{mdistsect}):
(i) by summing up the flux of the member stars as coming from the decontaminated sample 
(CMD method), and (ii) by integrating the cluster luminosity function after subtracting the 
surrounding field luminosity function (LF method).

\begin{figure}
\centering
\includegraphics[width=0.85\linewidth]{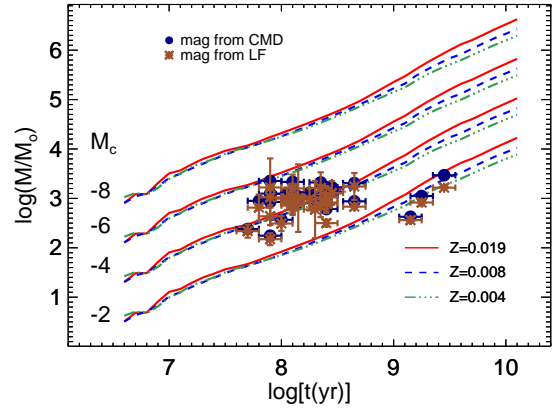}
\caption{Distribution of masses as a function of age for the clusters in our sample. 
Masses were derived from the mass-luminosity ratio and the integrated $C$ magnitude 
according to the CMD (blue circles) and LF (brown asterisks) methods, respectively.
Models of constant absolute $C$ magnitude for metallicities $Z=0.004$, $Z=0.008$, $Z=0.019$
are superimposed with dot-dashed, dashed and continuous lines, respectively.}
\label{agem_c}
\end{figure}

\begin{figure}
\centering
\includegraphics[width=0.85\linewidth]{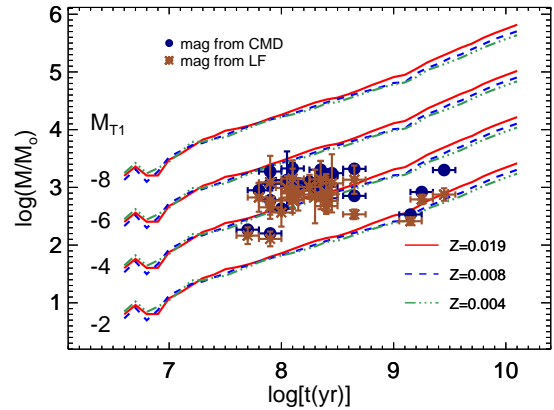}
\caption{As in Fig.~\ref{agem_c} but for integrated $T_1$ magnitude.}
\label{agem_t1}
\end{figure}

The resulting cluster masses as a function of their ages are presented in Figs.~\ref{agem_c} and 
\ref{agem_t1}. Mass uncertainties were propagated from the integrated magnitudes and ages.
Because a SSP fades with time as an effect of the stellar 
evolution, the SSP mass at a fixed luminosity depends on the age. At any age, the SSP mass 
increases with its luminosity, reflecting the population size.  According to these models, our 
cluster sample consists of systems with masses between 300 and 3000 M$_\odot$.

\begin{figure}
\centering
\includegraphics[width=0.85\linewidth]{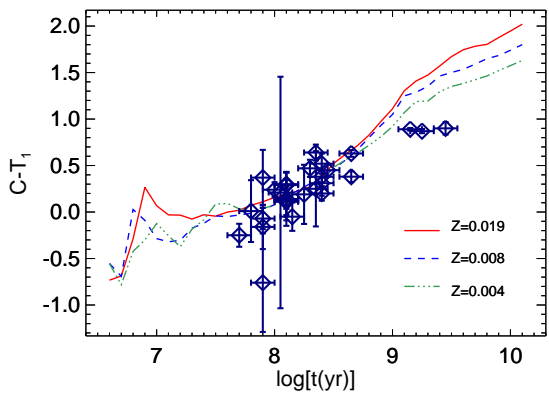}
\caption{The colour evolution of the cluster sample is compared to that resulting from SSP 
models for different metallicities.}
\label{agec}
\end{figure}

Figs.~\ref{agem_c} and \ref{agem_t1} can be also employed to estimate our photometric mass
limits. For instance, from Fig.~\ref{agem_t1} it is possible to infer the photometric depth needed 
to reach a 1000 M$_\odot$ cluster in the SMC (true distance modulus ($m-M$)$_\circ$=18.9). 
Thus, for a 10 Myr old cluster, 
its integrated $T_1$ magnitude should be $T_1=M_{T_{1}} + A_{T_{1}} +18.9 \sim 12.0$, 
neglecting the extinction. For a 1~Gyr old cluster, the integrated mag limit results $T_1 \sim 
16.0$. Such a difference is a consequence of the clusters fading as they become older. 
Similarly, the cluster integrated $C-T_1$ colours, calculated from the CMD method, also
match the respective evolution of the SSP models, as shown in Fig.~\ref{agec}.
Note that stochastic variations in the cluster light 
produced by bright stars may lead to significant colour fluctuations, especially for the youngest 
and less populous clusters.  

\begin{figure}
\centering
\includegraphics[width=0.85\linewidth]{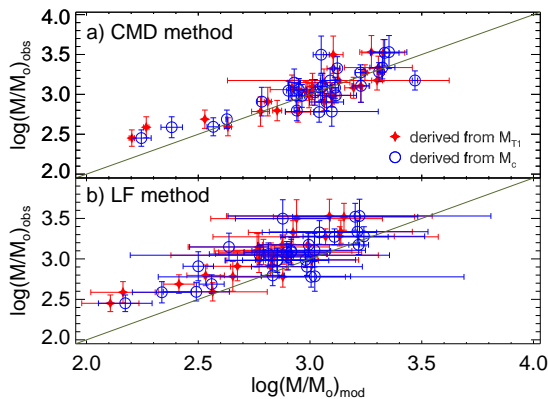}
\caption{Comparison of cluster masses derived from star counting ('obs') and SSP 
modelling ('mod'). The continuous line is the one-to-one relation.}
\label{mmass}
\end{figure}

A comparison between the masses obtained from SSP models using the integrated $C$ and 
$T_1$ magnitudes and those estimated using star counts (Sect.~\ref{tmassect}) shows a
reasonable agreement (Fig.~\ref{mmass}). It is clear that the cluster masses derived from the 
CMD method (panel a) provides a better match than those obtained from the LF method (panel 
b), where a larger 
spread and a systematic deviation for lower masses is seen. In addition, the resulting masses 
from SSP models do not seem to depend on whether the integrated $C$ or $T_1$ mag is used 
as input. Although both mass computations make use of the same isochrone set, the 
significantly different approaches leading to compatible values strongly support their reliability. 
For this reason we derived analytic relations in order to 
estimate the cluster mass from the knowledge of its age and integrated magnitude. Their 
implementation is based on a least square fit of a straight line passing over the flat interval of 
the $\mathcal{M}/\mathcal{L}$ relationship for SSPs older than $\log{t(yr)}=7.3$ (20 Myr).
The fits were performed for all the filters presented in Fig.~\ref{logmlfil}, according to the 
eq.~\ref{ml1}: 
\begin{equation}
\label{ml1}
\log\left[{\frac{\mathcal{M}}{\mathcal{L}}}\left(\frac{{\rm M}_\odot}{{\rm L}_\odot}\right)\right]=a+b
\log{[t(yr)]} \ \ \ (\log{t}>7.3)
\end{equation}
\noindent
The resulting correlation coefficients were superior to 0.99 in all cases. The 
fitted coefficients at different metallicities are summarised in Table~\ref{linfit}.

\begin{table}
\centering
\caption{Linear fit coefficients for the $\mathcal{M}/\mathcal{L}$ evolution}
\label{linfit}
\begin{tabular}{l c r@{$\,\pm\,$}l r@{$\,\pm\,$}l r@{$\,\pm\,$}l}
\hline
Band & coef & \multicolumn{6}{c}{$Z$} \\
      &  &  \multicolumn{2}{c}{0.019} & \multicolumn{2}{c}{0.008} & \multicolumn{2}{c}{0.004} \\
\hline
$B$   & $a$ & -7.9 & 0.2     & -7.6  & 0.1    &-7.1  & 0.1    \\
      & $b$ & 0.84 & 0.02    & 0.80  & 0.01   &0.74  & 0.01   \\
$V$   & $a$ & -6.4 & 0.1     & -6.14 & 0.08   &-5.87 & 0.07   \\
      & $b$ & 0.68 & 0.01    & 0.644 & 0.009  &0.608 & 0.008  \\
$I$   & $a$ & -5.2 & 0.1     & -4.9  & 0.1    &-4.8  & 0.1    \\
      & $b$ & 0.54 & 0.01    & 0.50  & 0.01   &0.49  & 0.01   \\
$C$   & $a$ & -8.9 & 0.1     & -8.4  & 0.1    &-7.9  & 0.1    \\
      & $b$ & 0.95 & 0.02    & 0.89  & 0.01   &0.83  & 0.01   \\
$T_1$ & $a$ &-5.76 & 0.08    & -5.49 & 0.08   &-5.31 & 0.07   \\
      & $b$ &0.612 & 0.009   & 0.573 & 0.009  &0.547 & 0.008  \\

\hline
\end{tabular}
\end{table}

The mass can be obtained from the integrated magnitude by rewriting  eq.~\ref{ml1} as:

\begin{equation}
\label{efm}
\log\mathcal{M}=a+b\ \log{t} - 0.4 (M_n - M_{n,\odot})
\end{equation}

\noindent
where $n=B, V, I, C, T_1$ and $M_{B,\odot}=5.49$, $M_{V,\odot}=4.83$, $M_{I,\odot}=4.13$, 
$M_{C,\odot}=5.68$, $M_{T_1,\odot}=4.47$ are the Sun absolute magnitudes and $M_n$ the 
integrated absolute magnitude in the corresponding filter. Its uncertainty results in:

\begin{equation}
\label{erefm}
\sigma_{\log\mathcal{M}}=\sqrt{\sigma_a^2+(\log^2{t})\sigma_b^2 + b^2 \sigma_{\log{t}}^2 + 
0.4^2 \sigma_M^2}
\end{equation}

\begin{figure}
\centering
\includegraphics[width=0.85\linewidth]{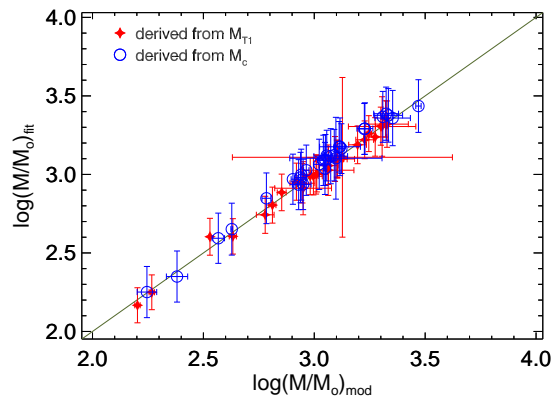}
\caption{Comparison of cluster masses derived from interpolations (abscissa) or fits to the 
$\mathcal{M}/\mathcal{L}$ relations (ordinate). The continuous line is the one-to-one relation.}
\label{massfit}
\end{figure}

Fig.~\ref{massfit} compares the masses derived by using eq. \ref{efm} and 
those obtained from integrated magnitudes and the interpolated 
$\mathcal{M}/\mathcal{L}$ ratios at the cluster ages. As can be seen, there is a tight correlation 
between masses coming from both procedures. 
We recall however, that the applicability range of these relations
is limited to clusters after the embedded initial phases, i.e., older than 20 Myr.
The above analysis shows that the cluster masses estimated in Sect.~\ref{tmassect} are 
compatible with those derived from their integrated properties and from the analytic relations 
provided.
  
\section{Discussion}

\begin{figure}
\centering
\includegraphics[width=0.47\linewidth]{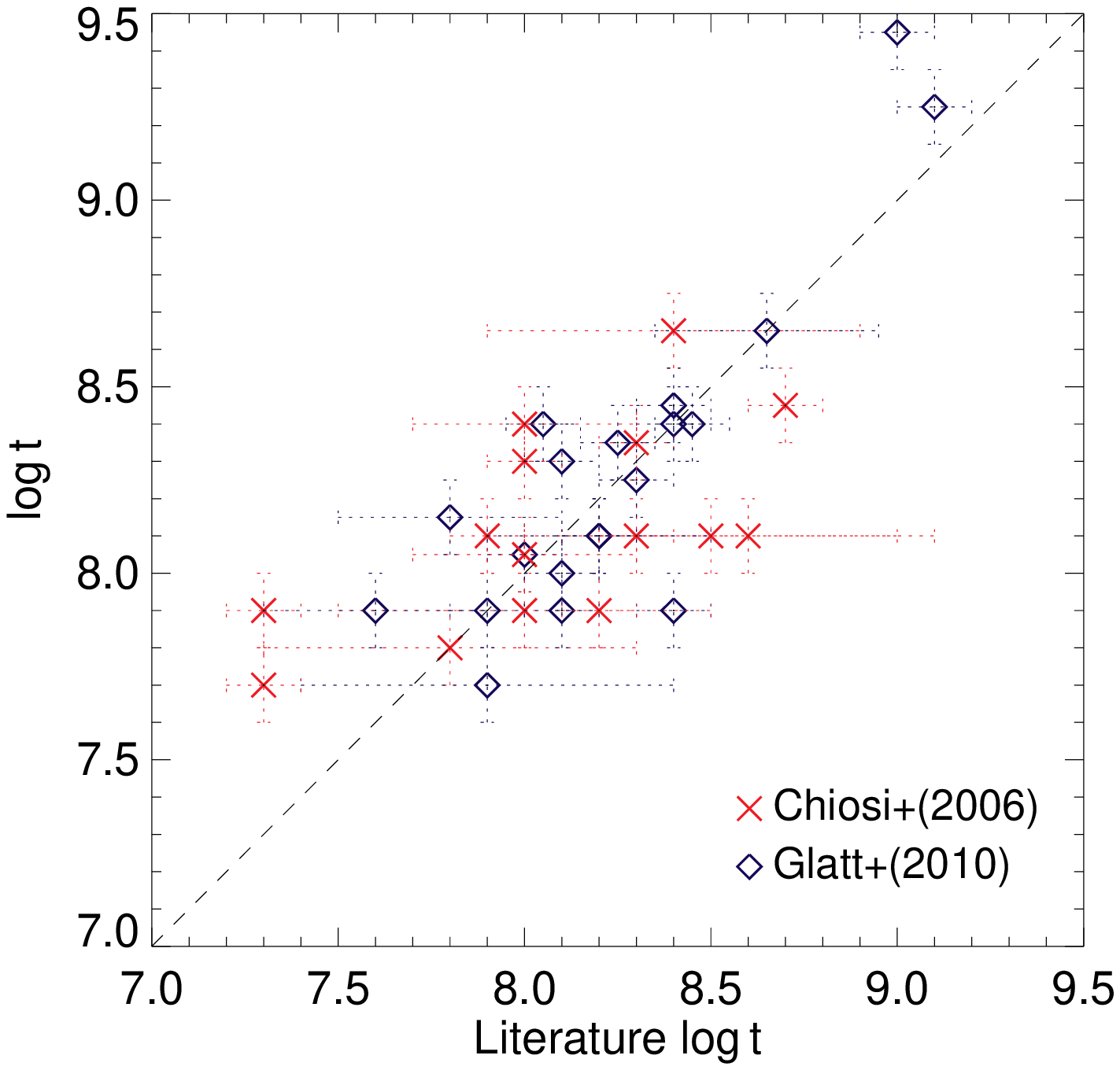}
\includegraphics[width=0.47\linewidth]{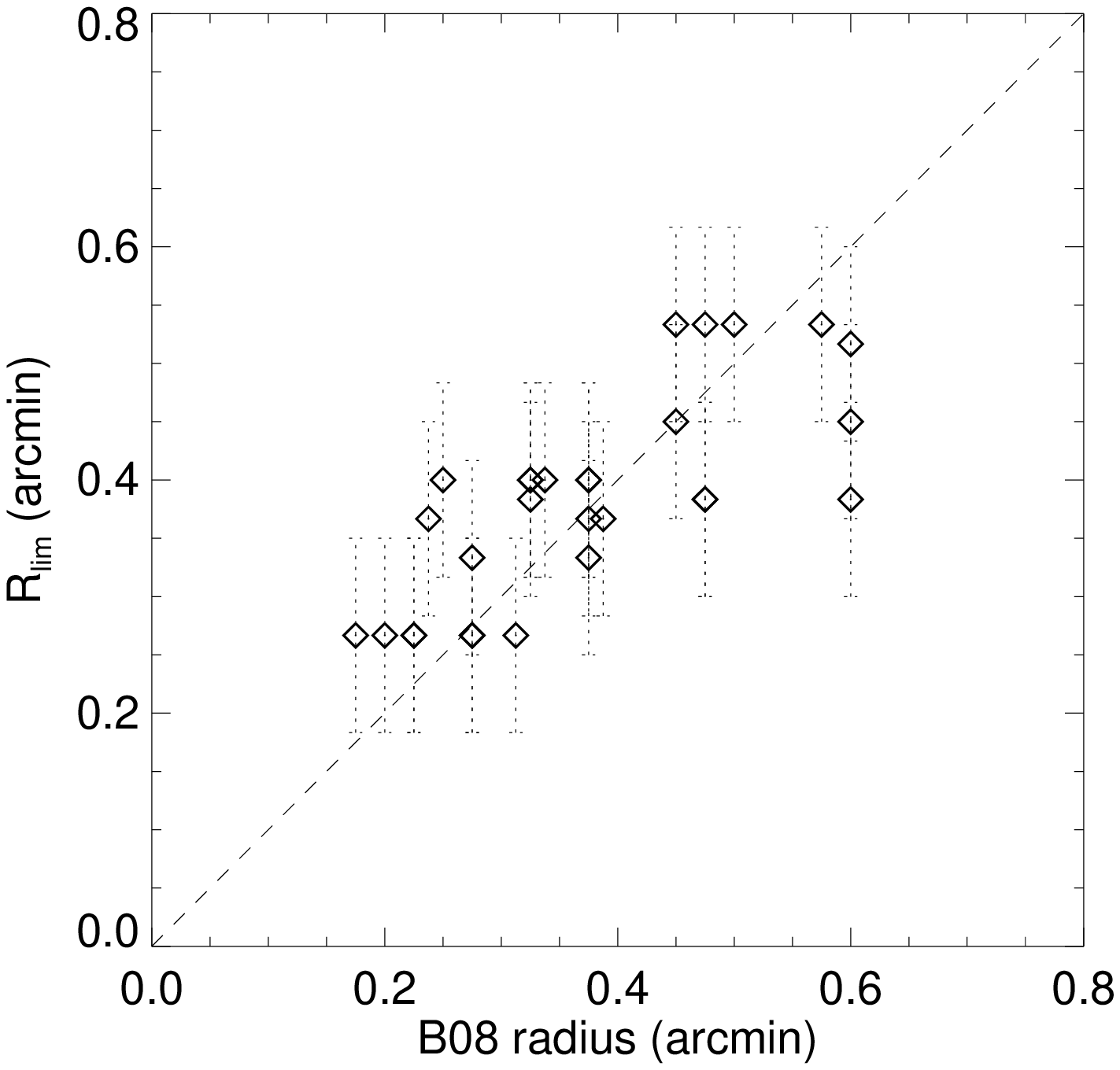}
\caption{Correlation between derived and published ages (left) 
and radii (right). The dashed line corresponds to the one-to-one relation. }
\label{parcomp}
\end{figure}

Fig.~\ref{parcomp} (left panel) compares our age estimates with the values obtained by 
\citet{Chiosi:2006} and \citet{Glatt:2010} for 15 and 18 clusters in common, respectively. 
Although we found a good 
agreement in the age range between $7.5 < \log t < 9.0$, it seems that we have overestimated 
the age of the young clusters K47 and H86-190. This could likely be caused by the 
saturation limit of our images which prevented us from identifying a turn-off brighter than 
T$_1 \sim 16$. On the other hand, previous age estimates for the two oldest
clusters (BS75 and BS80) may have their published ages biased to younger values, since 
these clusters present a faint turn-off and the authors did not account for field star 
decontamination on their CMD analysis.

The comparison between the derived limiting radii (R$_\mathrm{l}$) and the values 
from the B08 catalogue also shows a general good agreement (see Fig.~\ref{parcomp}, right 
panel). However, we found that 
some of the smaller clusters (e.g. B111, H86-190) appear to be substantially larger than 
previously known, while the radii of some of the largest clusters (e.g. K43) were truncated by 
CCD gaps in our images, leading to deceivingly smaller limiting radii. 

Concerning the MF slope variation, clusters H86-188, H86-190, K47 and K63 
clearly present flatter MF slopes than prescribed by K01 IMF ($\alpha=2.3\pm0.7$). 
Similarly, NGC242, which does have the 
lowest MF slope value within the studied cluster sample ($\alpha = 0.8$), forms a known 
interacting pair with NGC241 ($\alpha = 2.2$). Its flat MF slope could be the consequence 
of tidal stripping by the larger, more massive cluster, NGC241. 
On the other hand, L39, located in a crowded field and placed close to a CCD 
gap, presents the steepest MF slope in the sample. In this case, the field subtraction 
method resulted more prone to include field leftover stars in their decontaminated sample.

\begin{figure}
\centering
\includegraphics[width=0.75\linewidth]{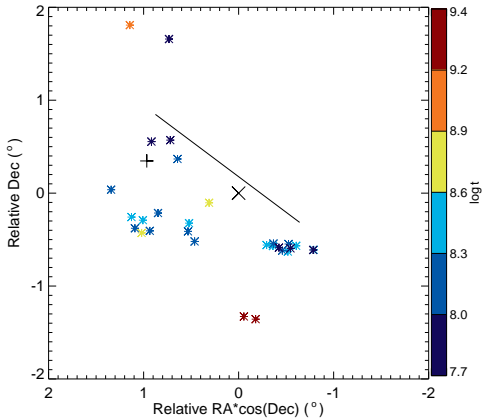}
\caption{Cluster positions relative to the optical centre (cross). The SMC bar (line) and the 
rotational centre (plus) are also shown. The right-hand colourbar is scaled with the cluster 
age.}
\label{chart}
\end{figure}

Although our cluster sample is not homogeneously distributed, neither spatially nor 
chronologically, the analysis of the derived cluster parameters can still provide important 
information regarding the galactic environment and its impact on the evolution of the cluster 
population. Fig.~\ref{chart} shows the positions of the studied clusters with respect to the SMC 
optical centre \citep[RA$(J2000)=00^\rmn{h} 52^\rmn{m} 45^\rmn{s}$, Dec.$(J2000)=-72\degr 
49\arcmin 43\arcsec$;][]{de-Vaucouleurs:1976}. It can be seen that the rotational centre, 
represented with a plus sign, is displaced from both the SMC bar \citep{Westerlund:1997} and 
the optical centre. In addition, the youngest clusters are preferentially found near the bar, while 
the oldest ones are located more than $\sim$ 1$\degr$ away.

\begin{figure}
\centering
\includegraphics[width=0.483\linewidth]{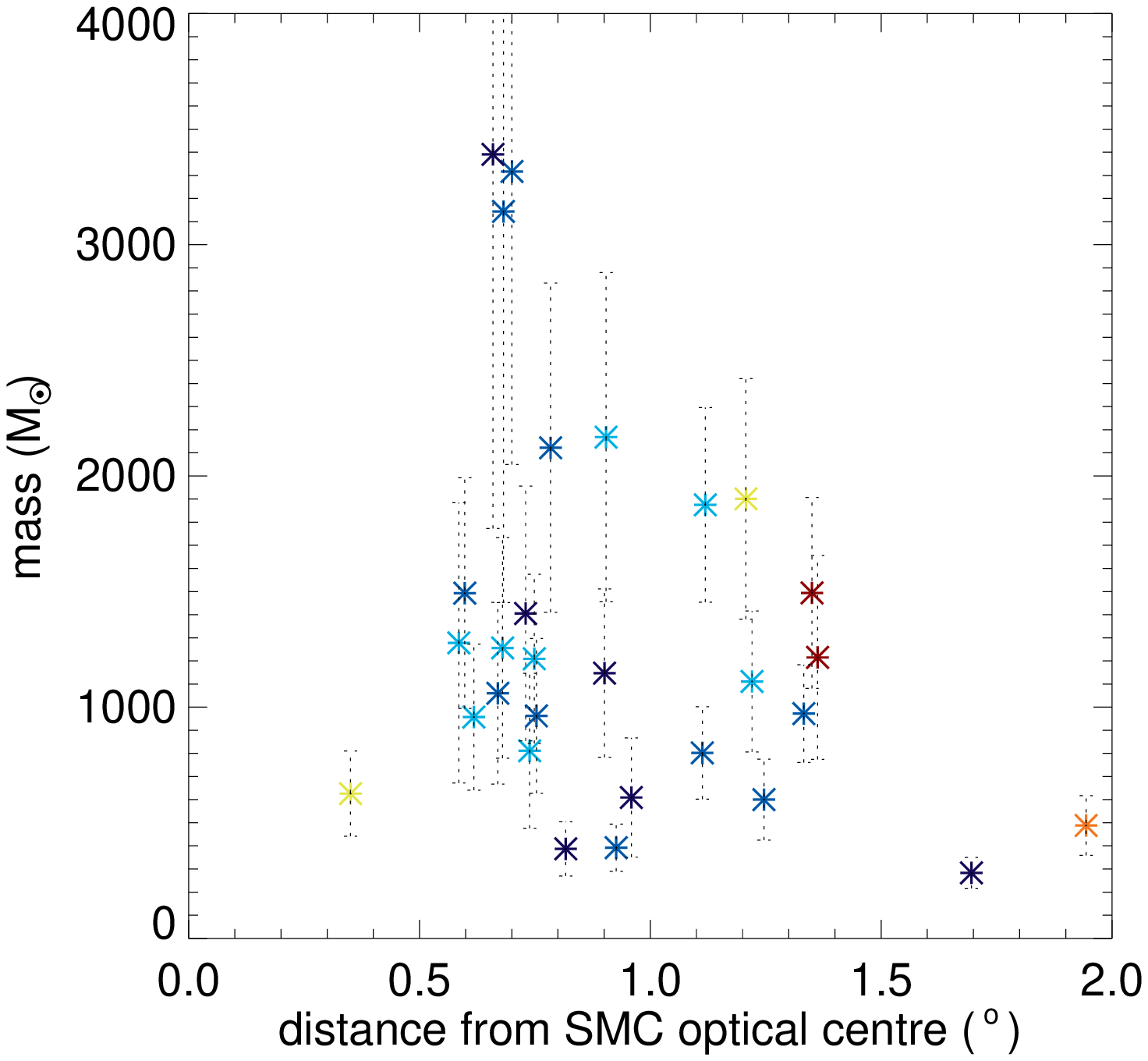}
\includegraphics[width=0.48\linewidth]{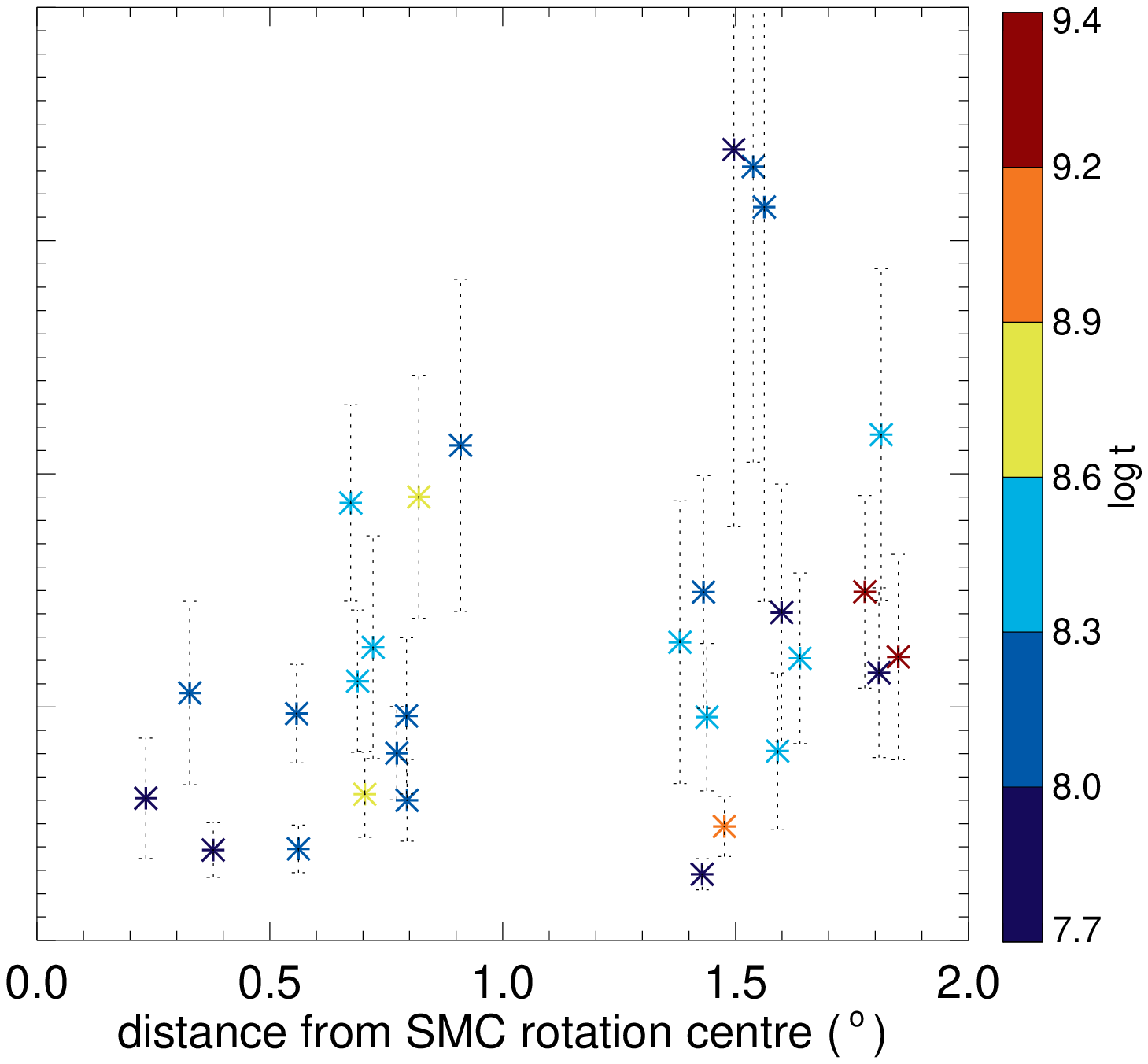}
\caption{Cluster spatial distribution respect to the SMC optical (left) and rotational (right) 
centres as a function of their masses. The right-hand colourbar is as in Fig.~\ref{chart}.}
\label{massrad}
\end{figure}

Fig.~\ref{massrad} depicts the spatial distribution of the clusters with respect to the SMC 
optical and rotational centres as a function of their total masses. As shown in Fig.~\ref{chart}, 
there is a segregation of the oldest clusters to the outer regions of the SMC. 
In addition, there seems to exist a trend of the clusters maximum mass with the distance from the 
SMC rotational centre.

\begin{figure}
\centering \hskip 0.25cm
\includegraphics[width=0.470\linewidth]{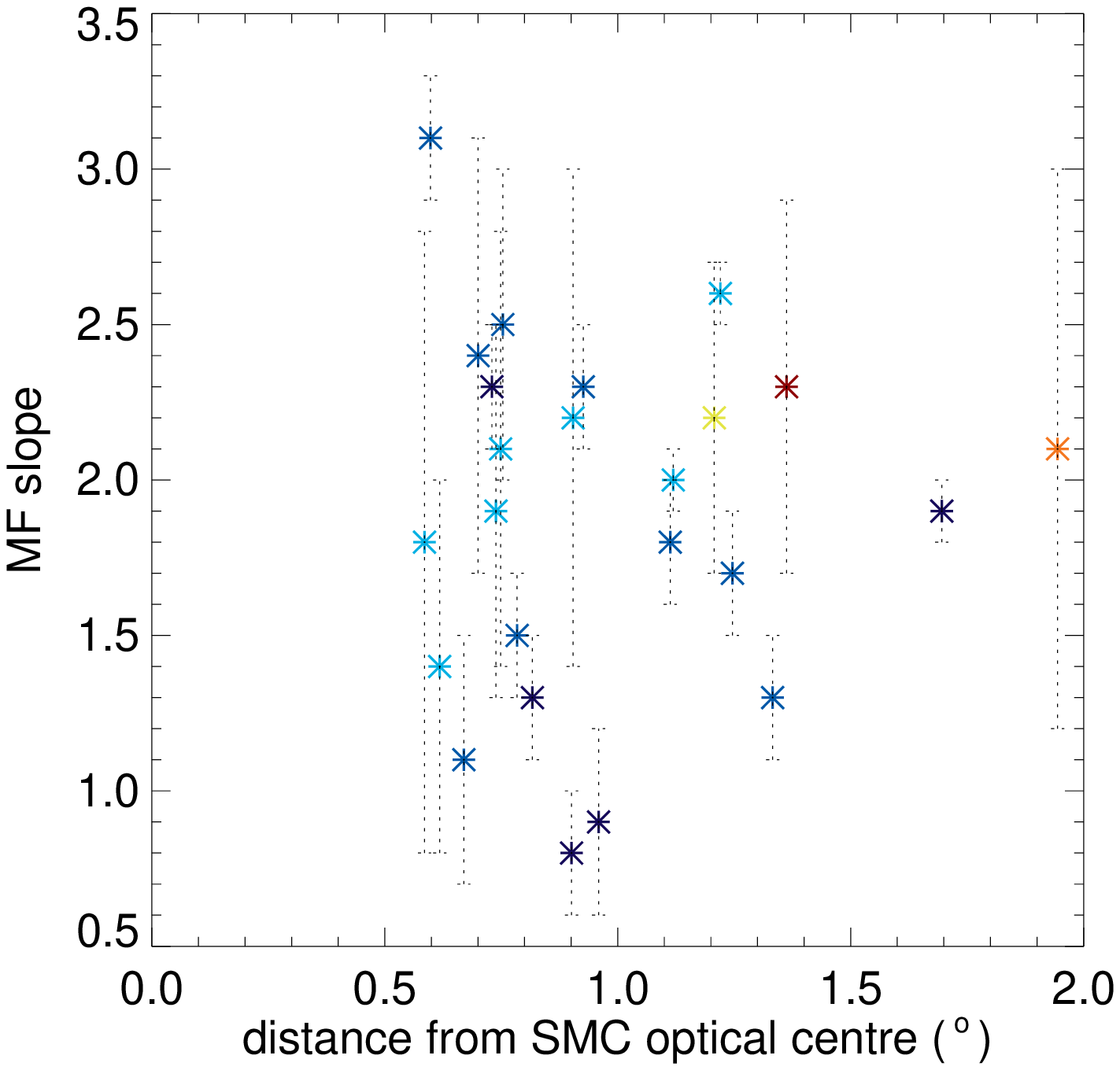}
\includegraphics[width=0.480\linewidth]{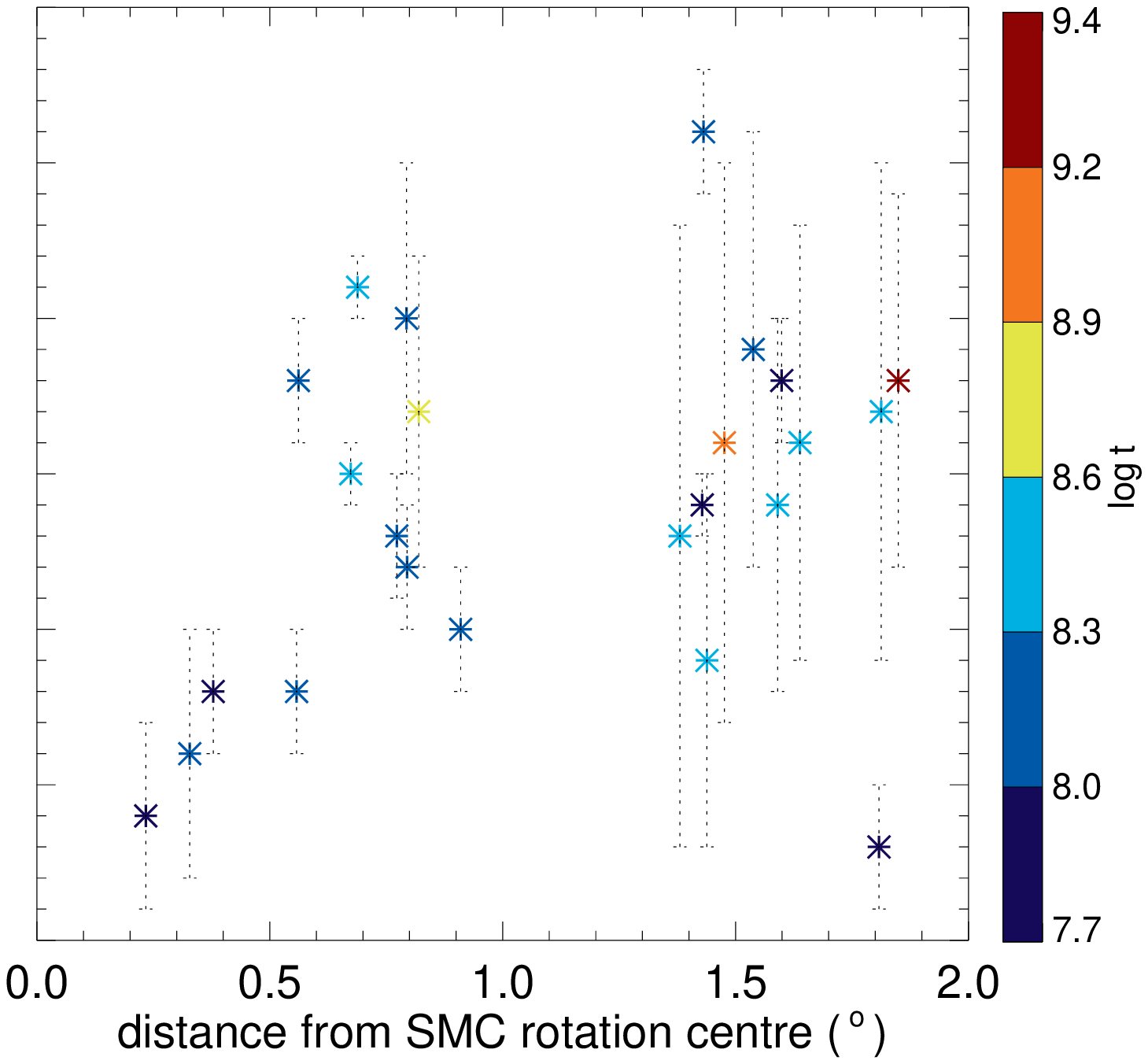}
\caption{Cluster spatial distributions respect to the SMC optical (left) and rotational (right) 
centres as a function of their MF slopes. The right-hand colourbar is as in Fig.~\ref{chart}.}
\label{mfsrad}
\end{figure}

We also analysed the cluster spatial distributions with respect to both optical and rotational SMC
centres in terms of their MF slopes. The result is shown in Fig.~\ref{mfsrad}.
As it can be seen, clusters with any MF slope are found inside $\sim$ 1$\degr$ from the optical 
centre, while those located in outer regions present slopes more similar to the canonical value 
expected for an undisturbed population (e.g. $\alpha \sim 2.3$). 
Concerning the rotational centre, it seems that the clusters located inside $\sim$0.6$\degr$ 
present, in average, flatter MF slopes than those outside this radius.

The cluster distances with respect to the rotational centre and their derived masses were also 
used to calculate the tidal radii through eq.~\ref{eqtr} (see Table~\ref{tab1}). As already 
mentioned, these values are 
probably underestimated, as the distances used are 2D projections of the real distances. 

\section{Concluding Remarks}

An initial sample of 68 candidate clusters was considered for investigation. Analysis of their 
structures showed that 31 objects do not present a concentration of stars over the various 
magnitude limited density maps consistent with with the existence of a genuine star cluster. 
Furthermore, 4 additional 
objects were removed from the initial sample due to their RDPs could not be 
distinguished from the background fluctuations, and others 4 for showing CMD star distributions 
that do not correspond to cluster sequences. However, we do not rule out the possibility that 
some of these 39 rejected objects might still be bound systems. The 100 per cent completeness 
level of our photometry is reached at $T_1$ $\sim$ 21.5 mag. \citep{Piatti:2012a}, so that 
sparse clusters older than $\sim$3 Gyr could be easily overwhelmed by the 
field. In addition, CCD saturation caused by very bright stars contaminate the photometry of 
fainter stars, leaving "holes" in the field spatial distribution of some targets. Although this effect 
compromised the 
RDP of some discarded objects, some of these targets presenting a well defined CMD were
still included in the list of surviving clusters (e.g. NGC241, NGC242).
Finally, since the early disruption of star clusters is a very common occurrence 
\citep{Lada:2003}, it should be expected that many young objects no longer have the 
concentrated stellar structures found in the more populous clusters, but rather a much 
sparser stellar content. Such targets would certainly fail our centre finding and RDP selection 
methods, as their structural characteristics are akin to those of associations.
Therefore, because the employed methods are not optimised to the investigation of these more
challenging targets, we postpone their analysis to a forthcoming paper without placing any 
classification on their physical nature. These rejected candidate clusters are gathered in the 
Table~\ref{rejclu} (Appendix~A).

The remaining 29 objects compose our list of studied clusters. For these clusters we
derived central coordinates, central stellar density, core, limiting and tidal radii, field stellar 
density, age, interstellar reddening and total mass. We also derived the MF slope for 24 
clusters in our sample and found 5 clusters presenting slopes flatter than K01 IMF.
Although ages and structural 
parameters for some of the studied clusters are available in the literature, the mass and 
the MF slope estimates are derived for the first time for most of them.

The cluster integrated colours and SSP mass-luminosity ratios show 
that the derived masses are internally consistent. Based on these results, we provide 
equations for computing the cluster mass using its integrated magnitude and 
age. These equations were derived for $B$, $V$, $I$, $C$ and $T_1$ magnitudes and 
should be useful for stellar population studies using integrated colours.

The cluster spatial distribution shows that most of the young clusters seem projected towards 
the SMC bar, as our cluster sample is likely to suffer from inhomogeneity and 
selection effects. Their maximum age, maximum mass and MF slope seem to 
increase with their distance from the rotational centre. 

\section*{Acknowledgements}
We would like to thank the anonymous referee for the valuable suggestions that helped us to 
improve this work. We also acknowledge the PI of the archival data used, Doug Geisler.
F.~F.~S.~Maia thanks the INCT-A and the CNPq for funding. This work was partially supported
by the Argentinian institutions CONICET and Agencia Nacional de Promoci\'on Cient\'{\i}fica y 
Tecnol\'ogica (ANPCyT).

{\footnotesize
\bibliographystyle{mn2e}
\bibliography{mps13} }

\appendix

\section{Rejected candidate clusters}
Rejected clusters are gathered in Table~\ref{rejclu} along with their coordinates and the 
criterion of rejection.

\renewcommand{\tabcolsep}{0.25cm}
\begin{table}
\caption{List of rejected candidate clusters}
\label{rejclu}
\begin{tabular}{l c c l} \hline
Target & RA (J2000) & Dec. (J2000) & Reject criterion \\
 & ($^\rmn{h}:^\rmn{m}:^\rmn{s}$) & ($\degr:\arcmin:\arcsec$) & \\ \hline
     B31 & 00:43:38 &-72:57:32 & colour-magnitude diagram \\ 
    BS27 & 00:44:55 &-73:10:31 & radial density profile \\
  H86-74 & 00:45:14 &-73:13:09 & centre finding \\ 
    BS30 & 00:45:30 &-73:29:06 & centre finding \\
 SOGLE30 & 00:45:33 &-73:06:27 & centre finding \\
SOGLE183 & 00:46:10 &-73:03:57 & radial density profile \\
 SOGLE37 & 00:46:41 &-73:00:00 & centre finding \\
  H86-89 & 00:47:06 &-73:15:24 & centre finding \\
  H86-93 & 00:47:24 &-73:12:20 & centre finding \\
SOGLE192 & 00:48:26 &-73:00:25 & colour-magnitude diagram \\
SOGLE193 & 00:48:37 &-73:10:50 & centre finding \\
 SOGLE50 & 00:48:59 &-73:09:04 & centre finding \\
    BS42 & 00:49:16 &-73:14:57 & centre finding \\
 SOGLE53 & 00:49:17 &-73:12:36 & colour-magnitude diagram \\
     B52 & 00:49:40 &-73:03:12 & centre finding \\
     B53 & 00:50:03 &-73:23:03 & centre finding \\
SOGLE199 & 00:50:15 &-73:03:15 & radial density profile \\
     B54 & 00:50:28 &-73:12:13 & colour-magnitude diagram \\
    BS46 & 00:50:39 &-72:58:44 & centre finding \\
 H86-116 & 00:50:40 &-72:57:55 & centre finding \\
 SOGLE65 & 00:50:54 &-73:03:26 & radial density profile \\
     B56 & 00:50:55 &-73:12:11 & centre finding \\
  NGC290 & 00:51:14 &-73:09:41 & centre finding \\
     B82 & 00:55:36 &-71:58:57 & centre finding \\
  NGC330 & 00:56:19 &-72:27:50 & centre finding \\
 H86-170 & 00:56:20 &-72:21:10 & centre finding \\
 H86-175 & 00:57:50 &-72:26:24 & centre finding \\
 H86-179 & 00:57:57 &-72:26:34 & centre finding \\
     B92 & 00:58:14 &-72:00:14 & centre finding \\
   BS269 & 00:58:19 &-72:13:10 & centre finding \\
 H86-181 & 00:58:19 &-72:17:57 & centre finding \\
 H86-183 & 00:58:33 &-72:16:44 & centre finding \\
 H86-186 & 00:59:57 &-72:22:24 & centre finding \\
    B100 & 01:00:23 &-72:05:05 & centre finding \\
 H86-193 & 01:01:18 &-72:13:42 & centre finding \\
    B105 & 01:01:37 &-72:24:25 & centre finding \\
SOGLE233 & 01:02:40 &-72:23:50 & centre finding \\
    B114 & 01:02:53 &-72:24:53 & centre finding \\
    B135 & 01:09:19 &-73:11:15 & centre finding \\
 \hline
\end{tabular}
\end{table}

\section{Contribution of stellar remnants and gas to the total mass}

What are the mass contribution of stellar remnants, i.e., white dwarfs (WDs),
neutron stars (NSs) and black holes (BHs) to the total mass of a star
cluster of a certain age?
Since the mass locked in
a remnant is a fraction of the initial star mass at the Main Sequence (MS), it
should be also questioned how much gas is lost or locked into the system for
the subsequent star formation.
Padova isochrones include both the initial mass of a star at the MS
and the actual mass at an age $t$. They are different as a consequence of mass
loss during the stellar evolution. \citet[][hereafter K09]{Kruijssen:2009} studied the 
evolution of the mass
function in star clusters providing analytic expressions that link the star's
initial mass with the correspondent remnant mass.

To quantify the mass locked in stellar remnants and gas yielded by mass loss
as predicted by Padova isochrones, we considered the relationships between
the initial mass and the remnant mass given
by K09 and references therein. Stars were distributed according to
the Kroupa's IMF in the mass range $0.08 < m_{\circ}($M$_\odot) < 120$
and the total mass of the SSP was normalised
to 1. At a given age, stars whose initial mass ($m_{\circ}$) are higher
than the initial mass of the most massive star  still represented
in the isochrone ($m^{iso}_{\circ}$) evolved, losing mass,
to a state characterised by stellar remnants, namely WDs
(if  $m_{\circ} < 8$ M$_\odot$), NSs
(if  8 M$_\odot \le m_{\circ} < 30$ M$_\odot$) and BHs (if
$m_{\circ} \ge 30 $M$_\odot$). The actual mass of the stellar remnant is then
calculated using $m_\mathrm{WD}=0.109 m_{\circ} + 0.394$, 
$m_\mathrm{NS} = 0.03636\,(m_{\circ} - 8.) + 1.02$ and
$m_\mathrm{BH} = 0.06\,(m_{\circ} - 30.) + 8.3$, 
with the SSP age defining which type of remnant is produced and the IMF giving
how many remnants are formed.
Notice that the younger the SSP the smaller the number of remnants and their
total mass. For solar metallicity isochrones, the ages at which the different
remnants have their initial mass boundaries are $t^{\ >}\!\!\!\!_\sim$ 40 Myr
($m^{iso}_{\circ} < 8$ M$_\odot$) to form WDs,
$6.5^{\ >}\!\!\!\!_\sim \ {t}^{\ >}\!\!\!\!_\sim$ 40 Myr
($8$ M$_\odot \le m^{iso}_{\circ} < 30$ M$_\odot$) to form NSs, and
${t}^{\ <}\!\!\!\!_\sim$ 6.5 Myr
($m^{iso}_{\circ} \ge 30$ M$_\odot$) to form BHs, respectively.
Because the actual mass in remnants is lower than their initial masses due
to evolutionary mass loss, it is also possible to quantify the amount of
gas released that should increase as the SSP gets older. Operationally,
the difference between the total initial mass of the SSP (1 M$_\odot$)
and the total actual mass of the stars in the isochrone plus the total
actual mass in remnants gives the mass of gas yielded by the SSP.

\begin{figure}
\centering
\includegraphics[width=0.85\linewidth]{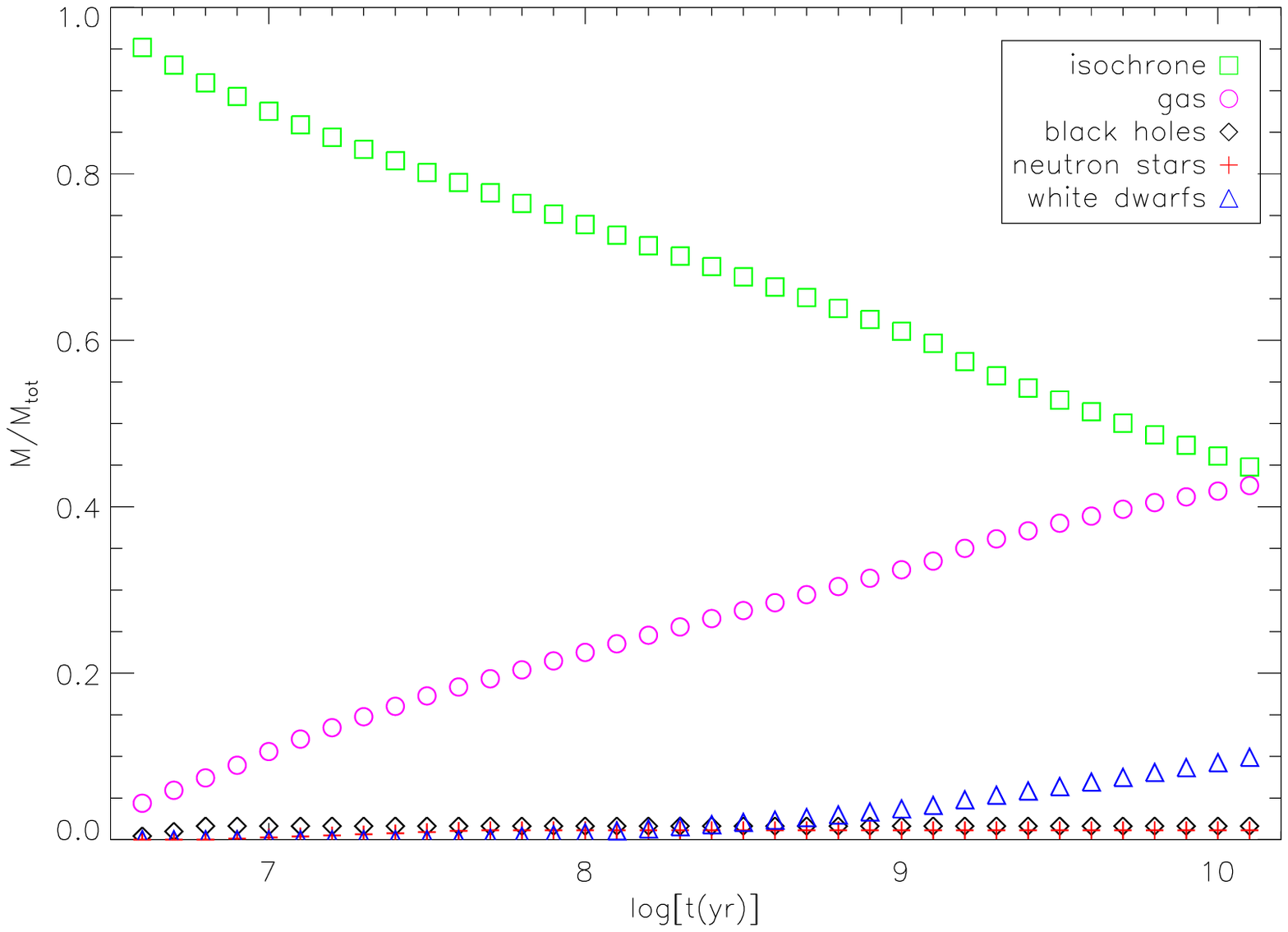}
\caption{Mass contribution of stellar remnants and gas for SSPs
of different ages.}
\label{mass_rem}
\end{figure}

Fig. \ref{mass_rem} shows the mass contribution from the different
components of solar metallicity SSPs as a function of their ages. 
The mass contribution of BHs and NSs is small
regardless the SSP age, reaching at most 1.6\% and 1.1\%, respectively.
The WDs component mass builds up after $\approx$ 40 Myr and reaches
about 10\% at the age of 12.6 Gyr.
The gas released during the stellar evolution provides a sizeable amount of mass
and dominates over the remnant contribution to the total mass as the SSP ages.
It reaches about 20\% and 40\% of the total mass at the age of 60 Myr
and 5 Gyr, respectively. In real clusters, this mass should have been
transformed into stars in a secondary burst of star formation or may have
been expelled from the cluster in its initial phases by stellar winds and
supernova explosions.
The remnant mass contribution mainly reflects the IMF combined with
stellar evolution, in which WDs outnumber BHs and NSs for SSPs older than
$\approx$ 500 Myr.
The mass contained in an isochrone of any age is above all other
contributions, but its importance decreases as the age increases.

The same analysis was done for isochrones of metallicity $Z=0.004$. 
The relative mass contribution as a function of age is qualitative 
similar to that for solar metallicity. 
The mass contribution of BHs and NSs are 
nearly identical to that for solar metallicity models, while 
the mass in WDs is slightly higher, reaching about 12\% at 
the age of 12.6 Gyr. 
The gas mass reaches about 20\% and 40\% of the total mass at 
the age of 60 Myr and 4 Gyr, respectively.

It is worth noting that the above relative mass values are overestimates
because the ignored star cluster dynamical evolution affects its original
mass, especially depleting low mass stars.

\section{Selected clusters charts}

All figures used in the analysis of the cluster sample are compiled in this appendix for 
reference. They include: magnitude limited density maps for the determination of the cluster 
centre, the schematic sky chart showing nearby objects and field region selected, the radial 
density profile for the estimation of structural parameters, the cumulative luminosity function for 
the estimation of the magnitude limit, the decontaminated CMD for the isochrone fitting and 
the field-subtracted LFs and MFs for the estimation of the mass distribution. They are only 
available in the online version of the Journal.

%
%



\begin{figure*} 
\centering
\begin{sideways}
\begin{minipage}{230mm}
\includegraphics[width=16cm,angle=90]{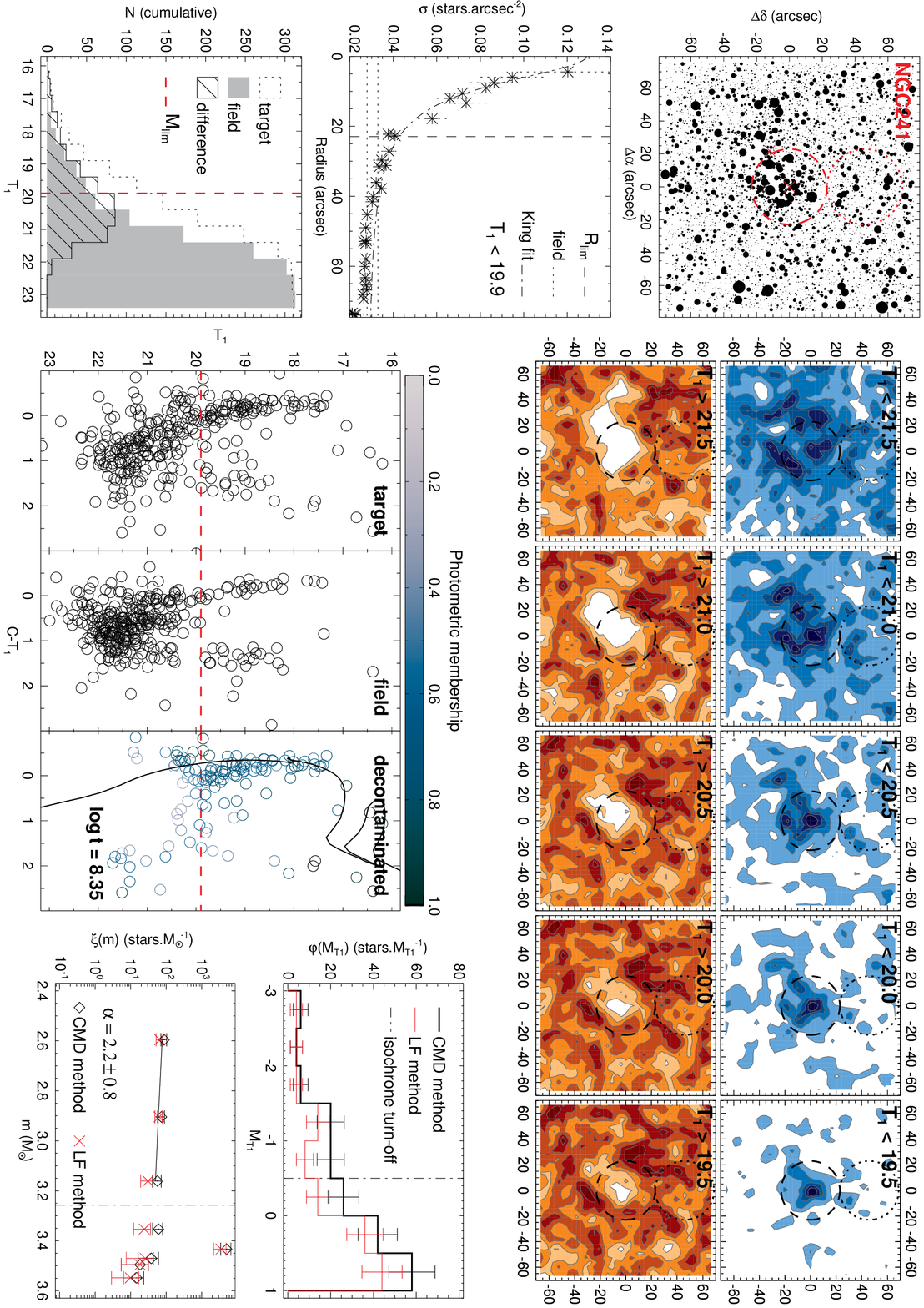}
\caption{NGC241 analysis charts. Top-left: schematic sky chart showing the cluster 
(dashed line) and field (dotted line) regions adopted, along with catalogued objects 
(cross). Top-right: magnitude limited density maps. Middle-left: magnitude limited RDP 
showing the limiting radius (dashed line), the field stellar density (dotted lines) and the 
2-parameter King fit (dot dashed line). Bottom-left: Cumulative LFs showing the adopted 
magnitude limit. Bottom-middle: CMDs comparing the cluster region, the surrounding field 
and the decontaminated samples; the magnitude limit derived is also shown (dashed line). 
Bottom-right: field-subtracted LFs and MFs according to the two methods employed; the 
turn-off magnitude and mass are also shown.}
\label{onlinefig} 
\end{minipage}
\end{sideways}
\end{figure*}

\clearpage

\begin{figure*} 
\centering
\begin{sideways}
\begin{minipage}{230mm}
\includegraphics[width=16cm,angle=90]{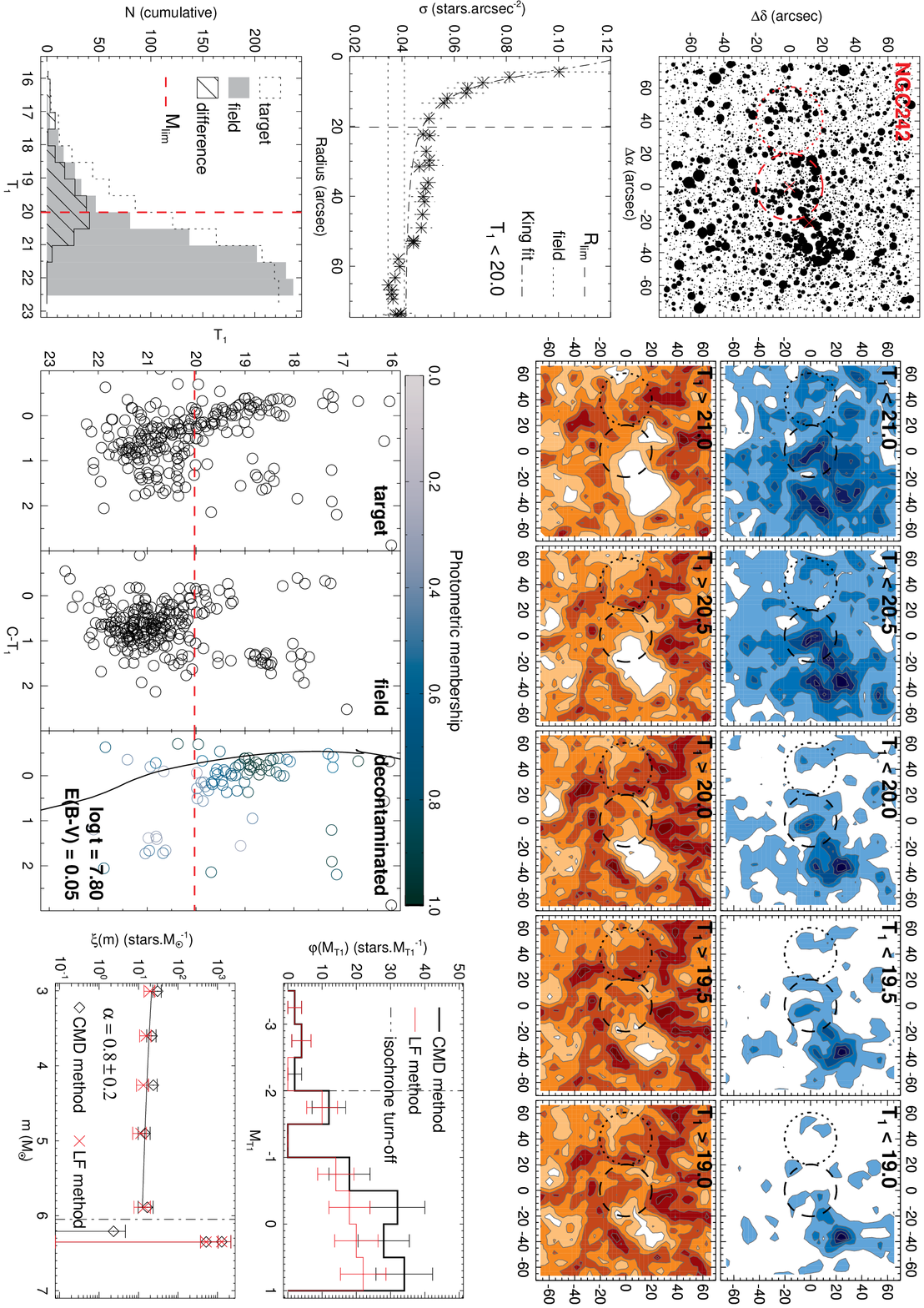}
\caption{NGC242 analysis charts. Panels are the same as in Fig.~\ref{onlinefig}}
\end{minipage}
\end{sideways}
\end{figure*}

\clearpage

\begin{figure*} 
\centering
\begin{sideways}
\begin{minipage}{230mm}
\includegraphics[width=16cm,angle=90]{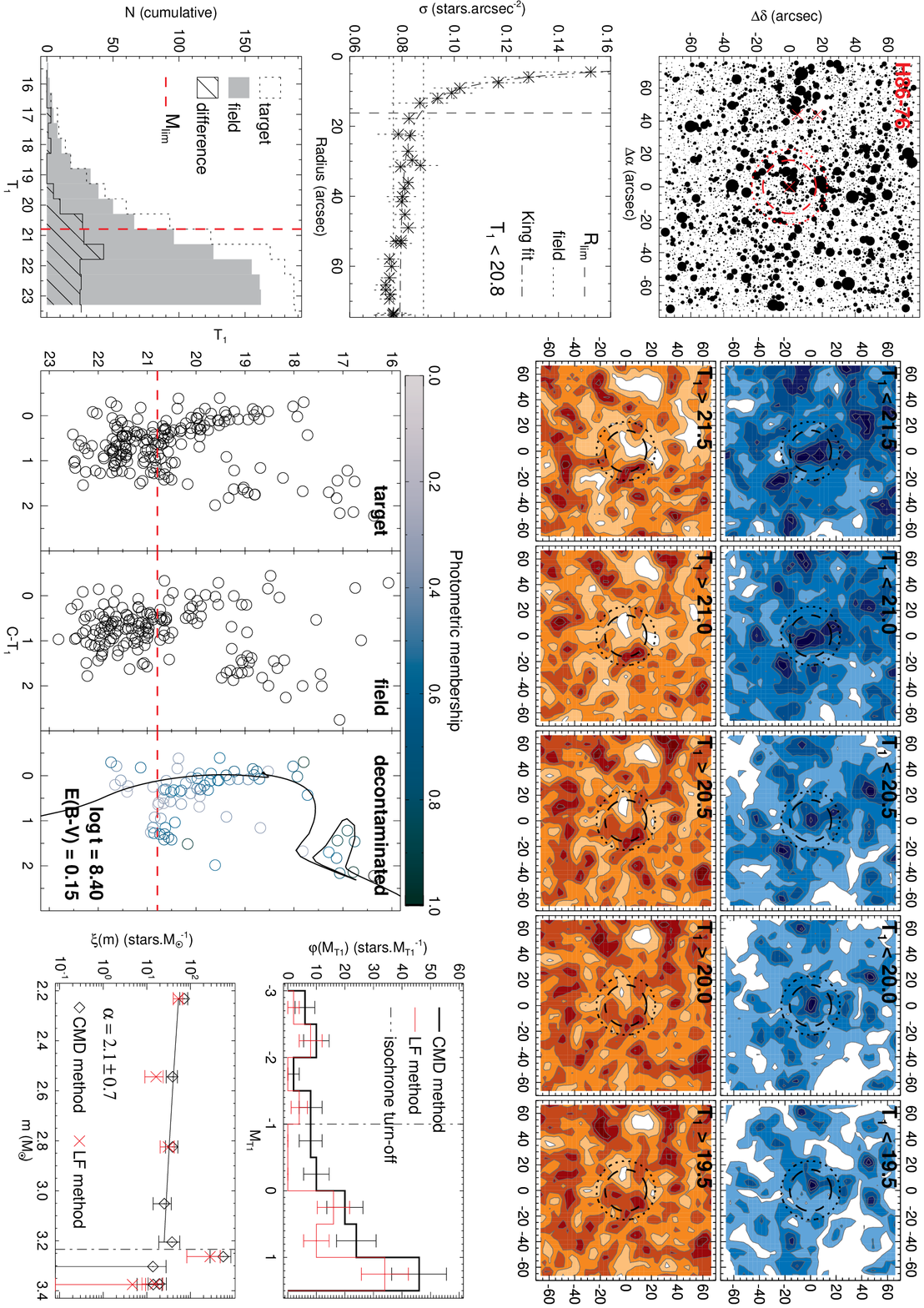}
\caption{H86-76 analysis charts. Panels are the same as in Fig.~\ref{onlinefig}}
\end{minipage}
\end{sideways}
\end{figure*}

\clearpage

\begin{figure*} 
\centering
\begin{sideways}
\begin{minipage}{230mm}
\includegraphics[width=16cm,angle=90]{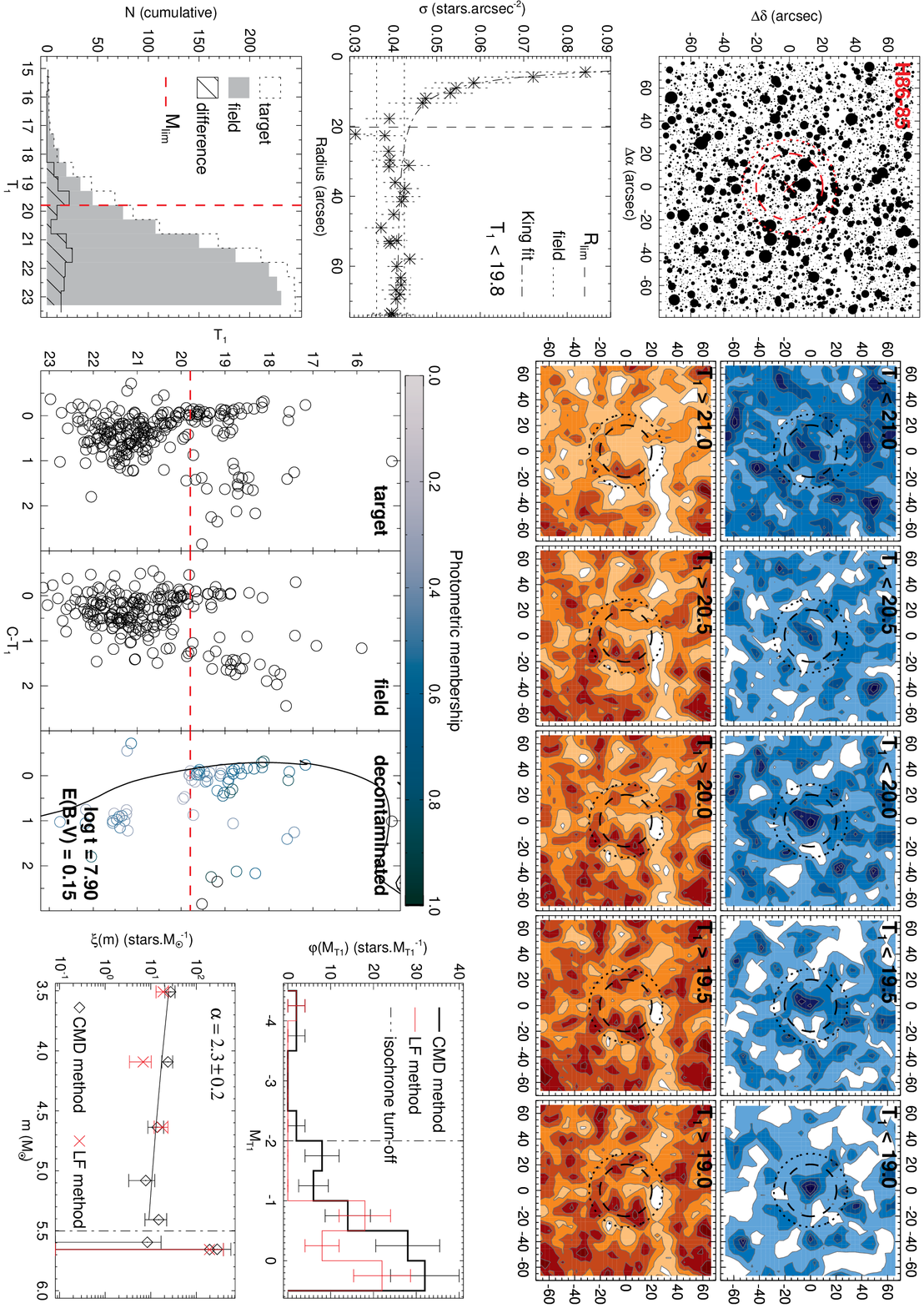}
\caption{H86-85 analysis charts. Panels are the same as in Fig.~\ref{onlinefig}}
\end{minipage}
\end{sideways}
\end{figure*}

\clearpage

\begin{figure*} 
\centering
\begin{sideways}
\begin{minipage}{230mm}
\includegraphics[width=16cm,angle=90]{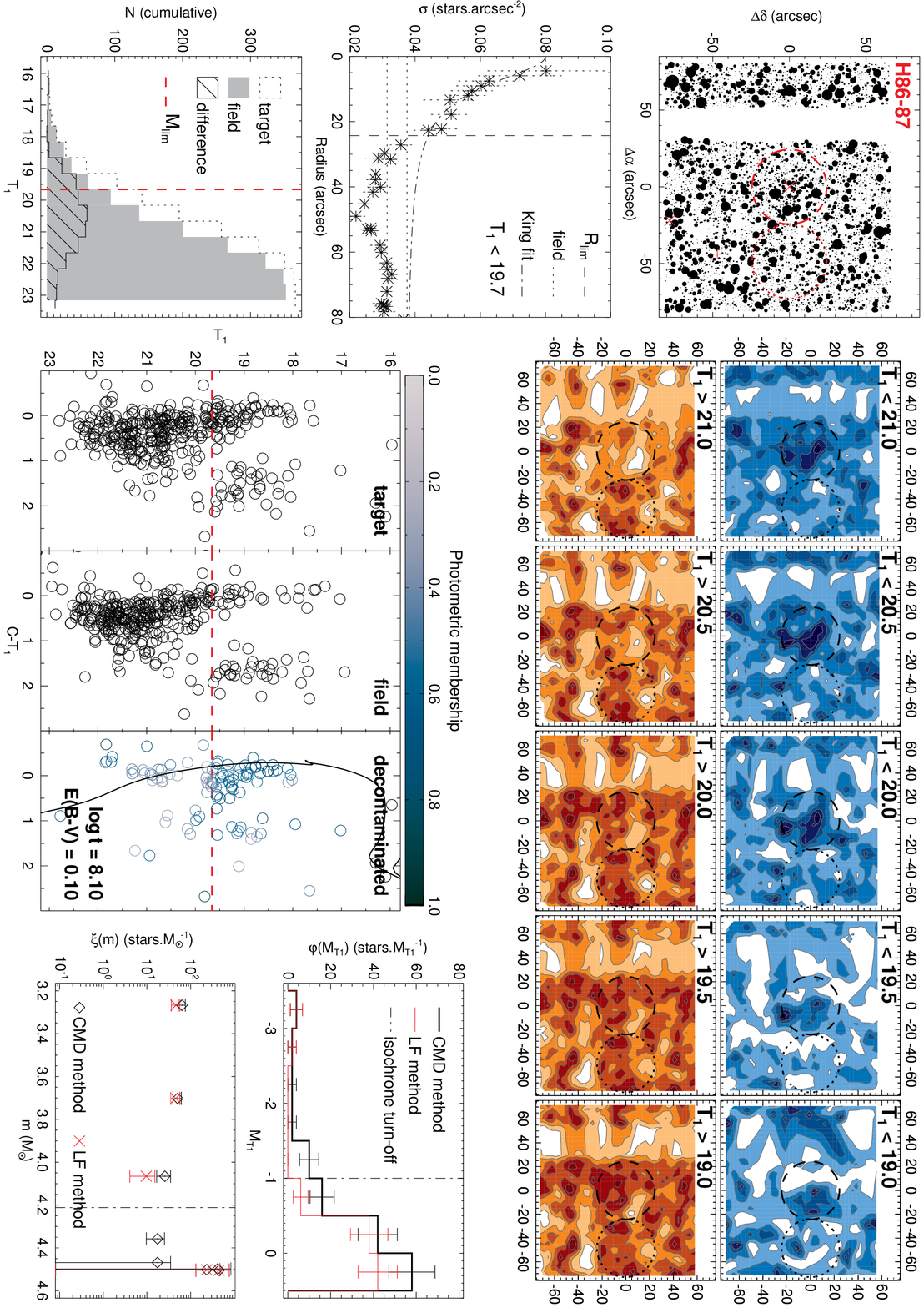}
\caption{H86-87 analysis charts. Panels are the same as in Fig.~\ref{onlinefig}}
\end{minipage}
\end{sideways}
\end{figure*}

\clearpage

\begin{figure*} 
\centering
\begin{sideways}
\begin{minipage}{230mm}
\includegraphics[width=16cm,angle=90]{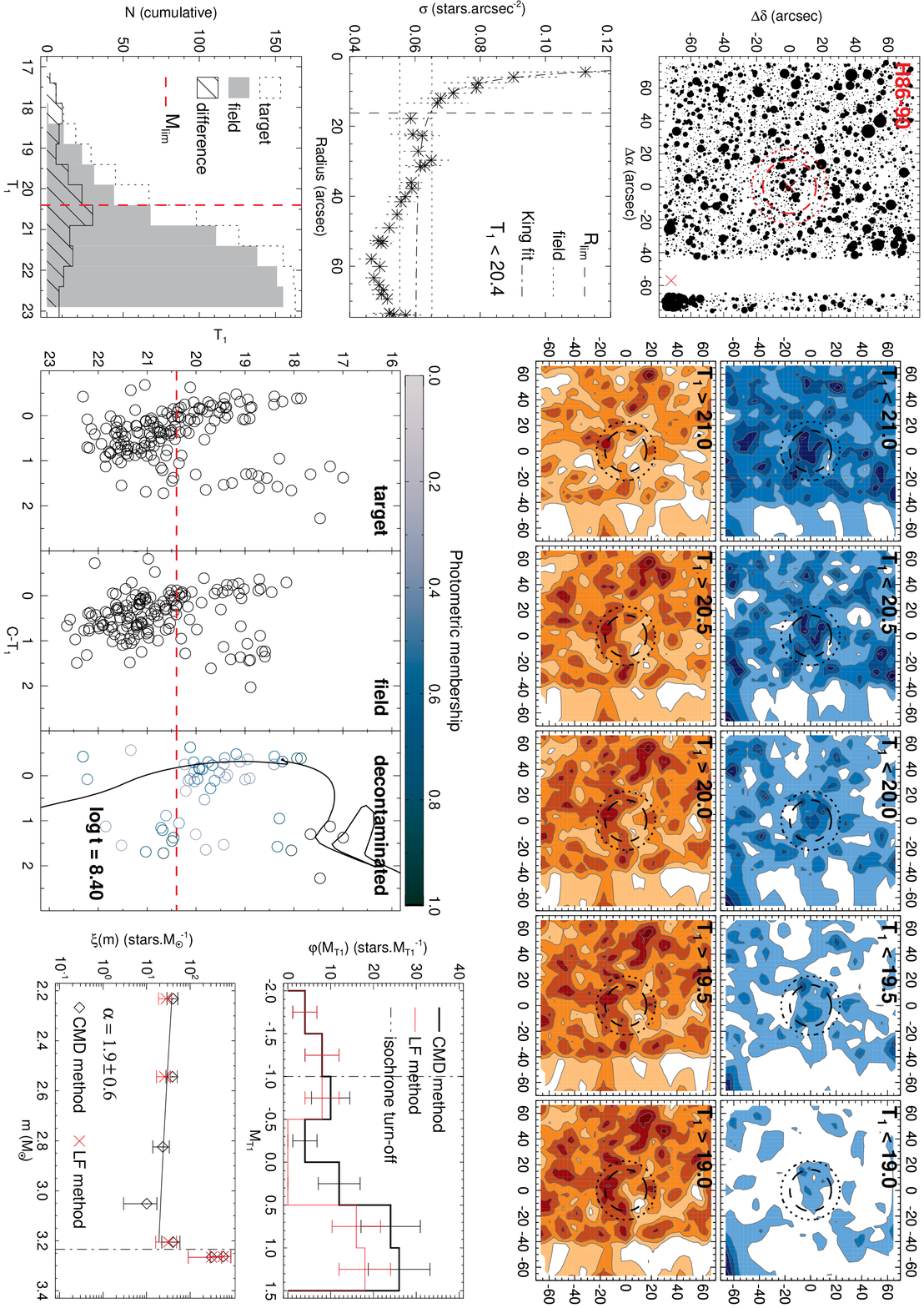}
\caption{H86-90 analysis charts. Panels are the same as in Fig.~\ref{onlinefig}}
\end{minipage}
\end{sideways}
\end{figure*}

\clearpage

\begin{figure*} 
\centering
\begin{sideways}
\begin{minipage}{230mm}
\includegraphics[width=16cm,angle=90]{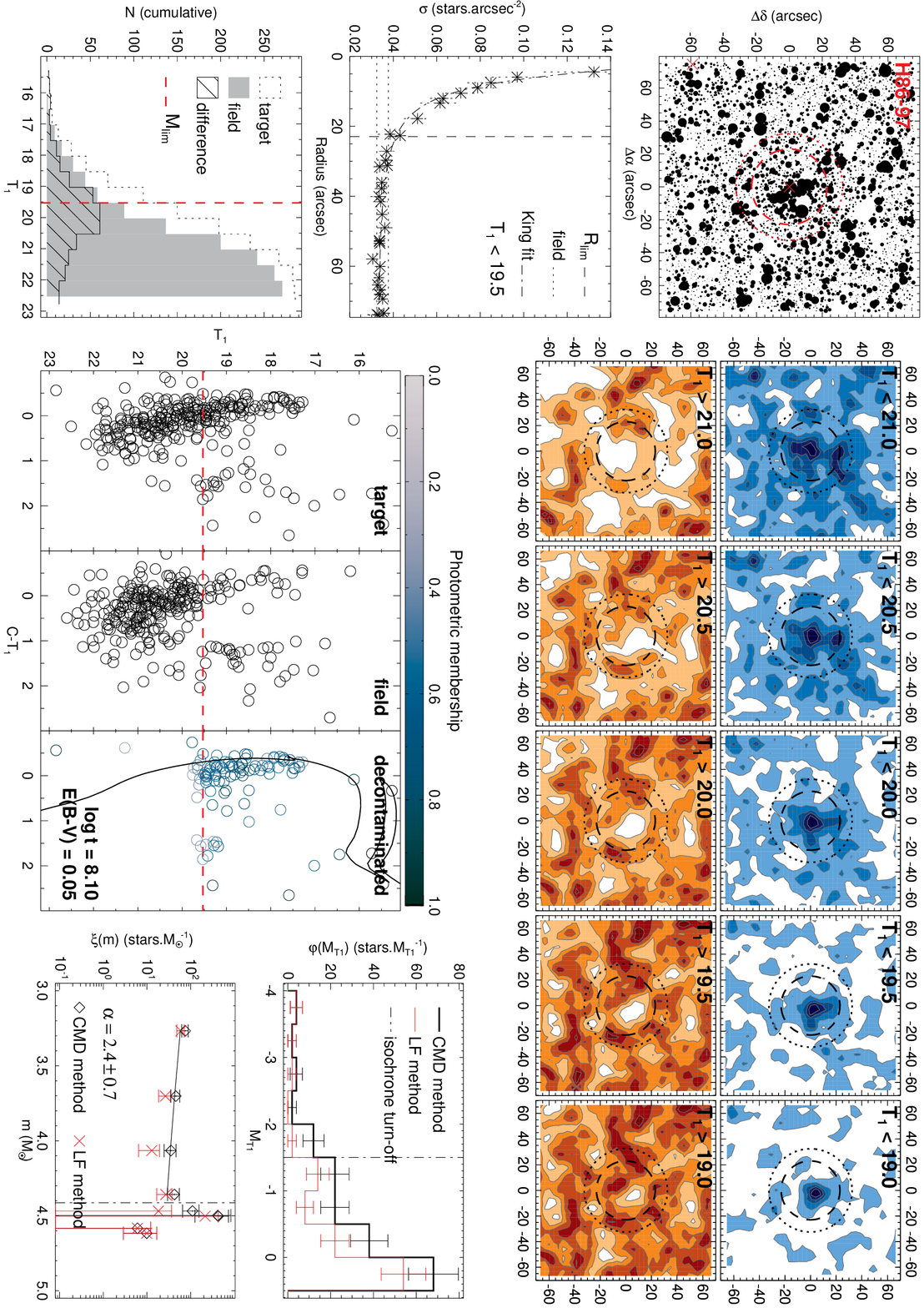}
\caption{H86-97 analysis charts. Panels are the same as in Fig.~\ref{onlinefig}}
\end{minipage}
\end{sideways}
\end{figure*}

\clearpage

\begin{figure*} 
\centering
\begin{sideways}
\begin{minipage}{230mm}
\includegraphics[width=16cm,angle=90]{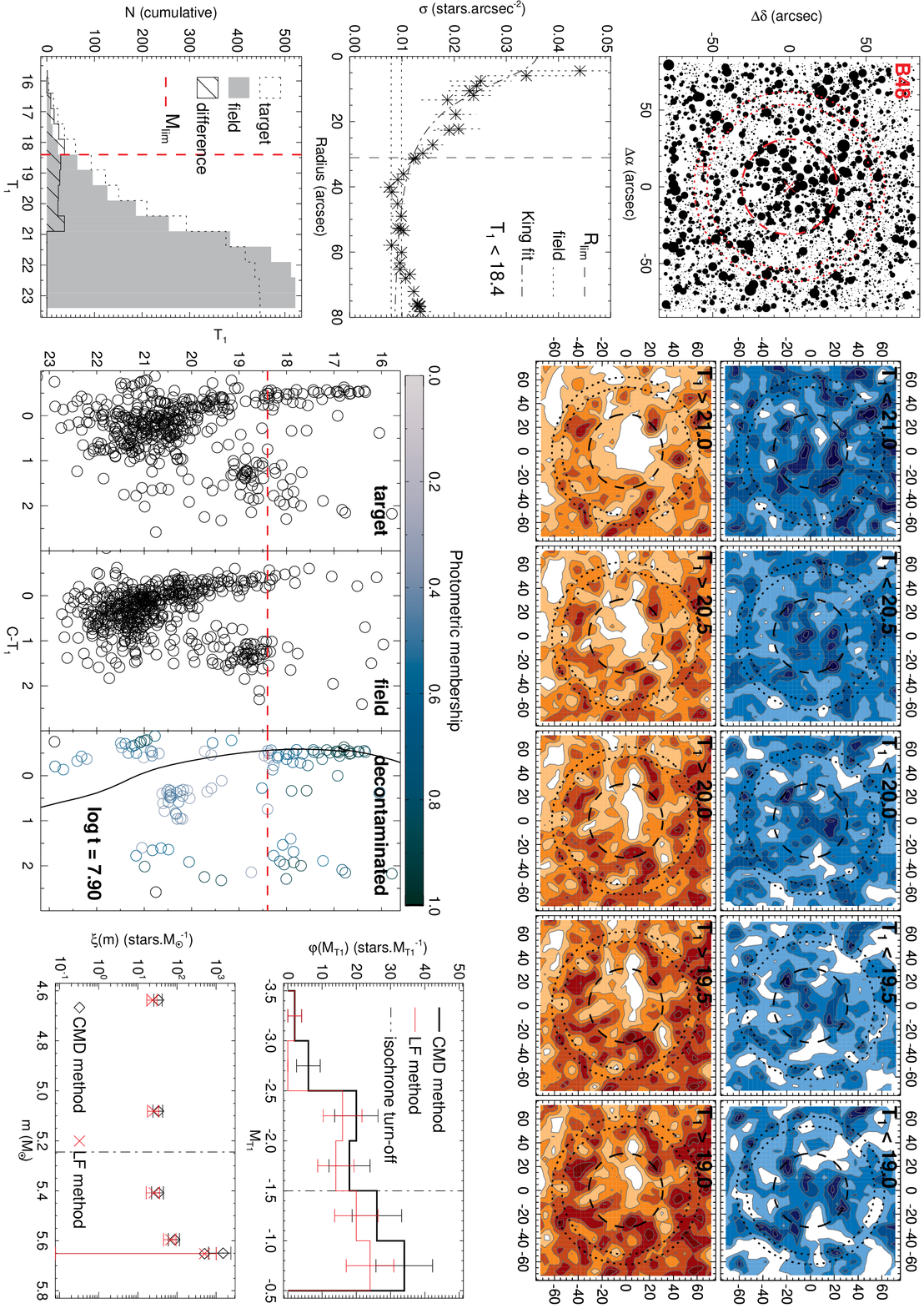}
\caption{B48 analysis charts. Panels are the same as in Fig.~\ref{onlinefig}}
\end{minipage}
\end{sideways}
\end{figure*}

\clearpage

\begin{figure*} 
\centering
\begin{sideways}
\begin{minipage}{230mm}
\includegraphics[width=16cm,angle=90]{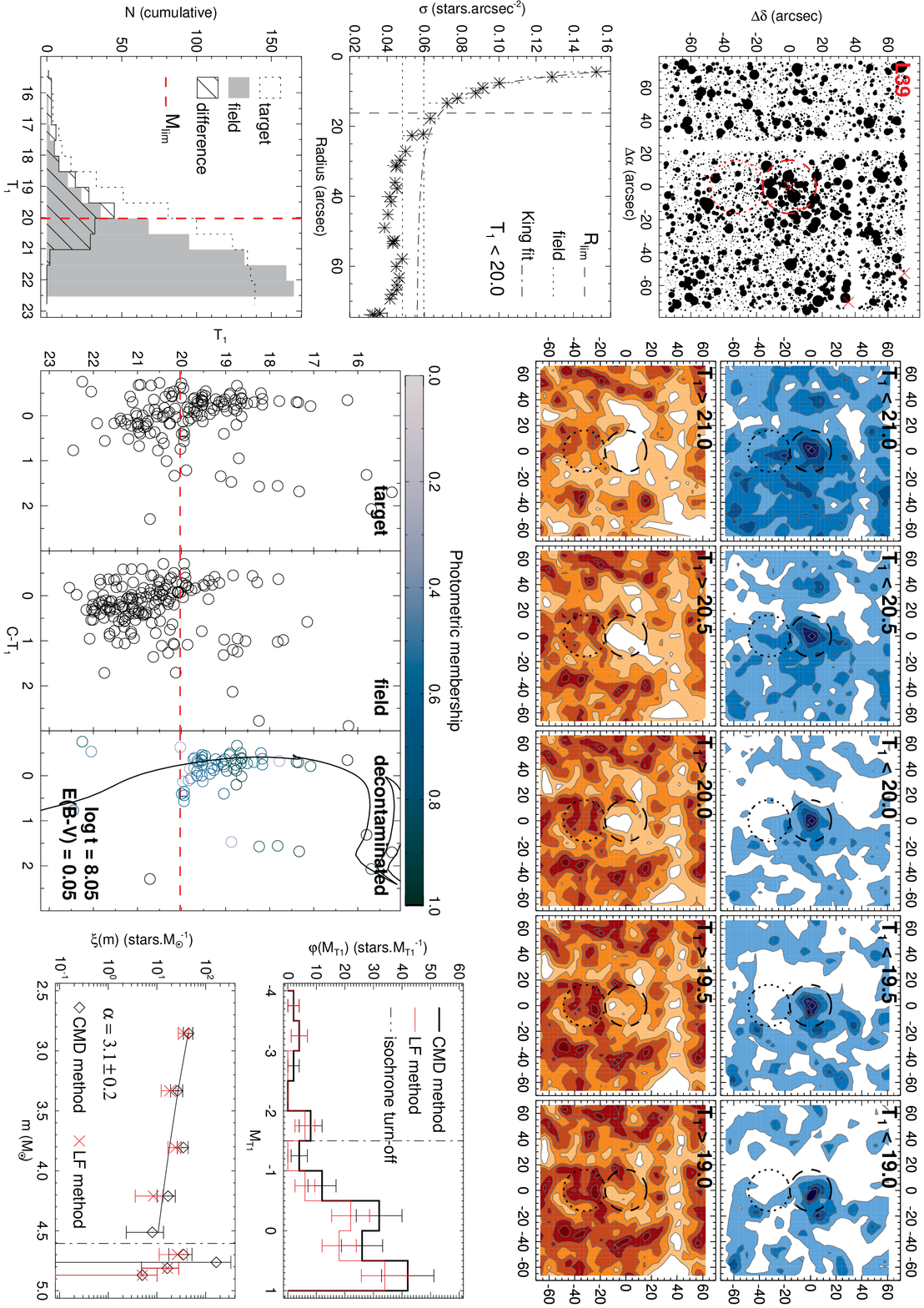}
\caption{L39 analysis charts. Panels are the same as in Fig.~\ref{onlinefig}}
\end{minipage}
\end{sideways}
\end{figure*}

\clearpage

\begin{figure*} 
\centering
\begin{sideways}
\begin{minipage}{230mm}
\includegraphics[width=16cm,angle=90]{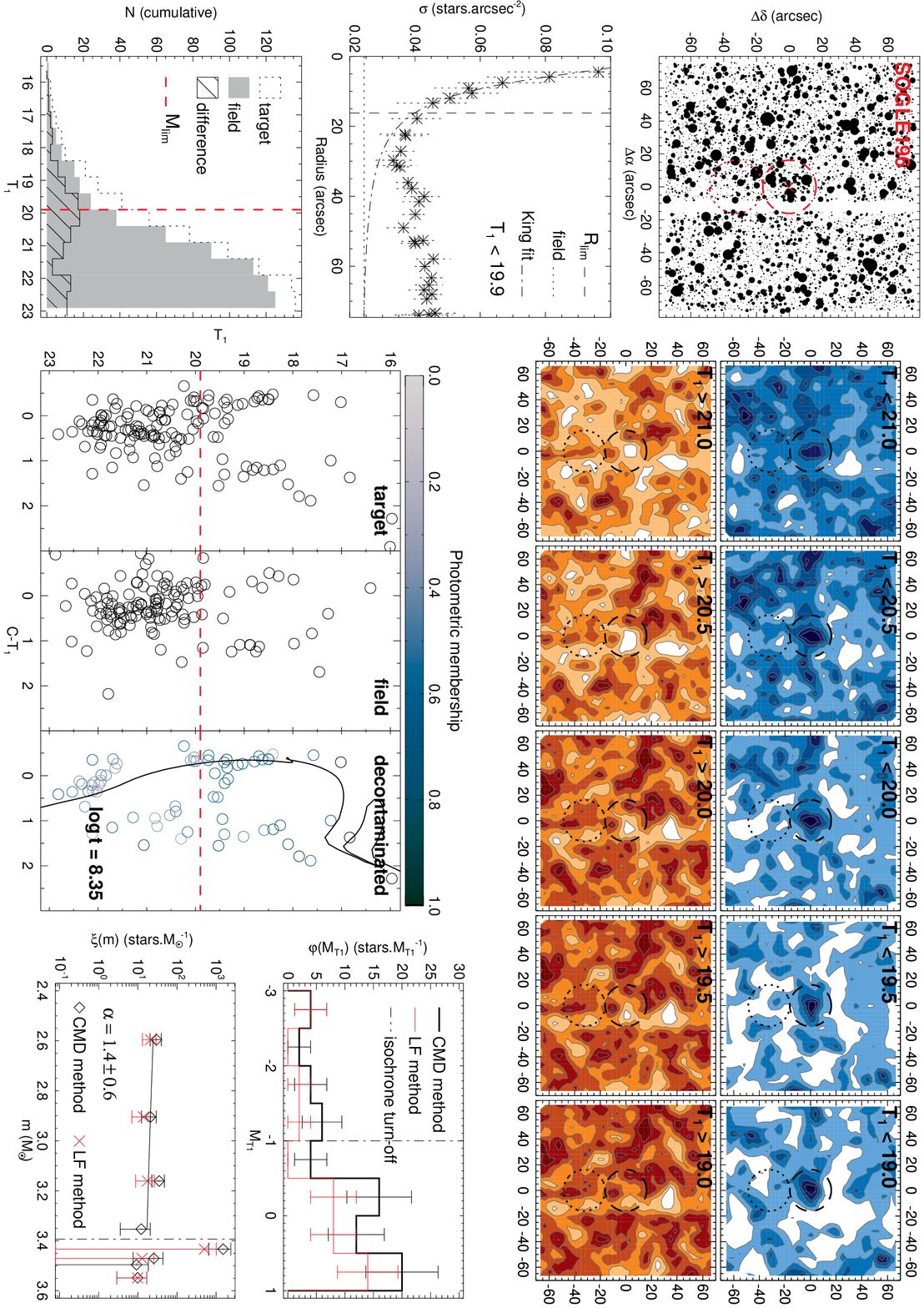}
\caption{SOGLE196 analysis charts. Panels are the same as in Fig.~\ref{onlinefig}}
\end{minipage}
\end{sideways}
\end{figure*}

\clearpage

\begin{figure*} 
\centering
\begin{sideways}
\begin{minipage}{230mm}
\includegraphics[width=16cm,angle=90]{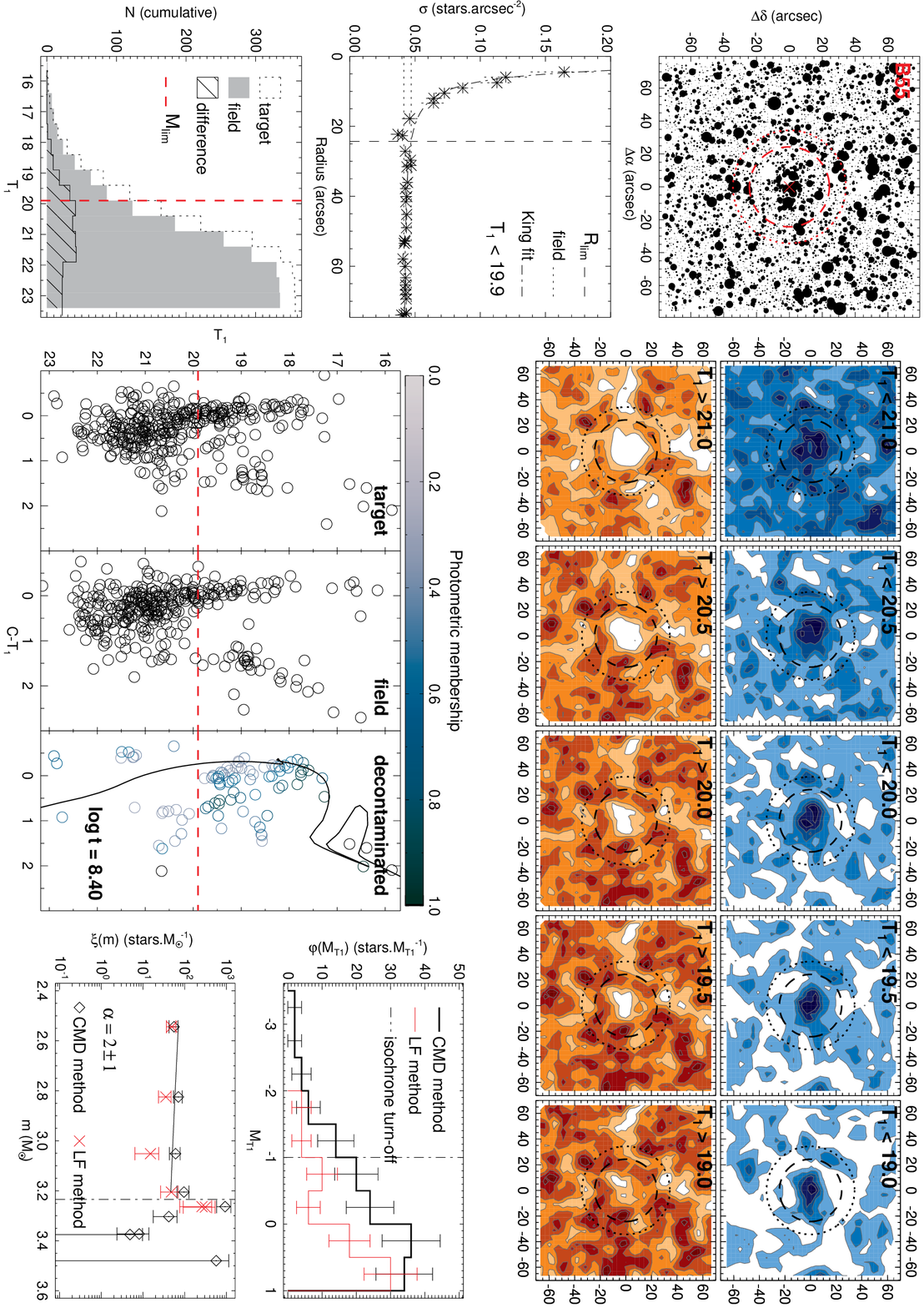}
\caption{B55 analysis charts. Panels are the same as in Fig.~\ref{onlinefig}}
\end{minipage}
\end{sideways}
\end{figure*}

\clearpage

\begin{figure*} 
\centering
\begin{sideways}
\begin{minipage}{230mm}
\includegraphics[width=16cm,angle=90]{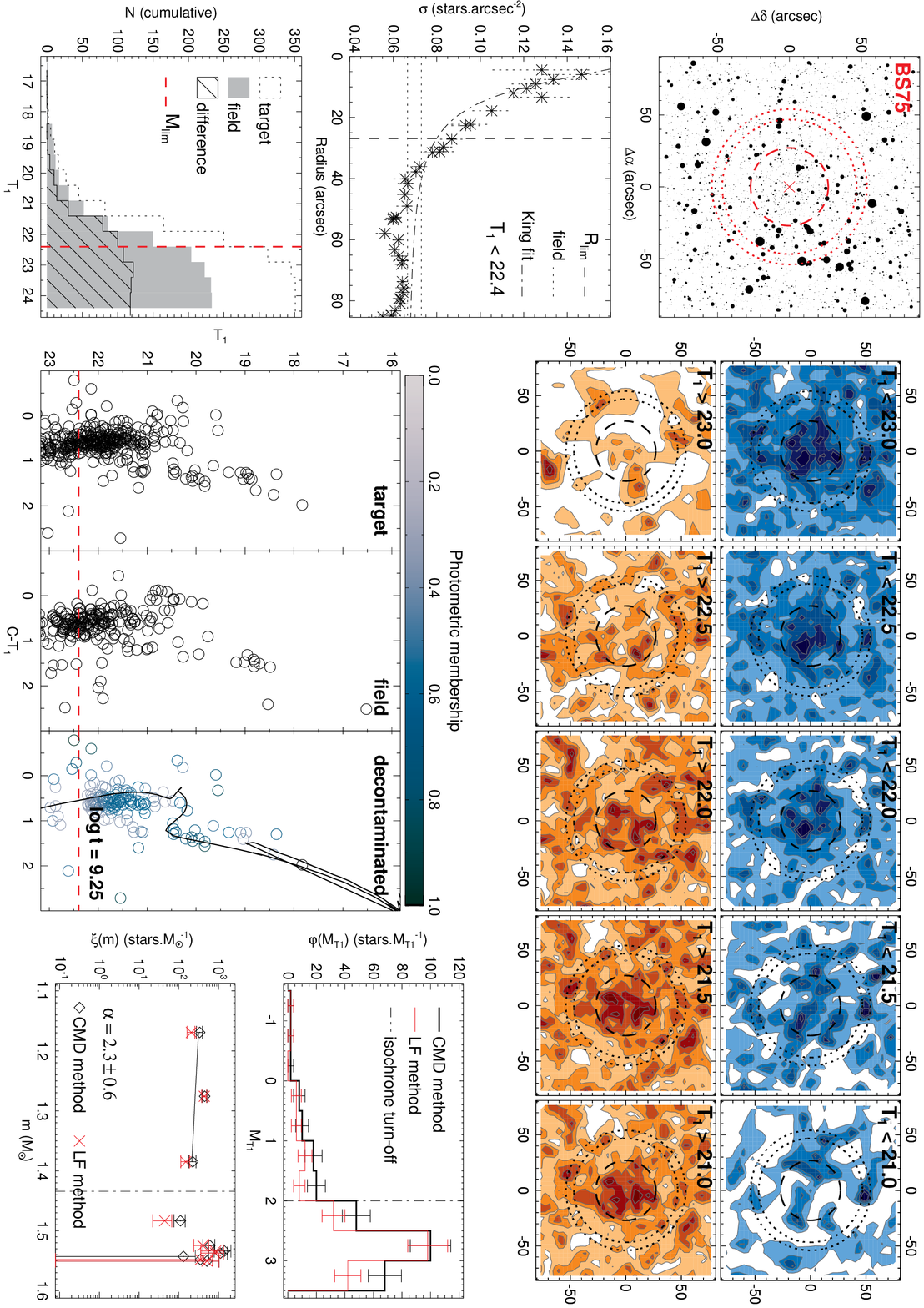}
\caption{BS75 analysis charts. Panels are the same as in Fig.~\ref{onlinefig}}
\end{minipage}
\end{sideways}
\end{figure*}

\clearpage

\begin{figure*} 
\centering
\begin{sideways}
\begin{minipage}{230mm}
\includegraphics[width=16cm,angle=90]{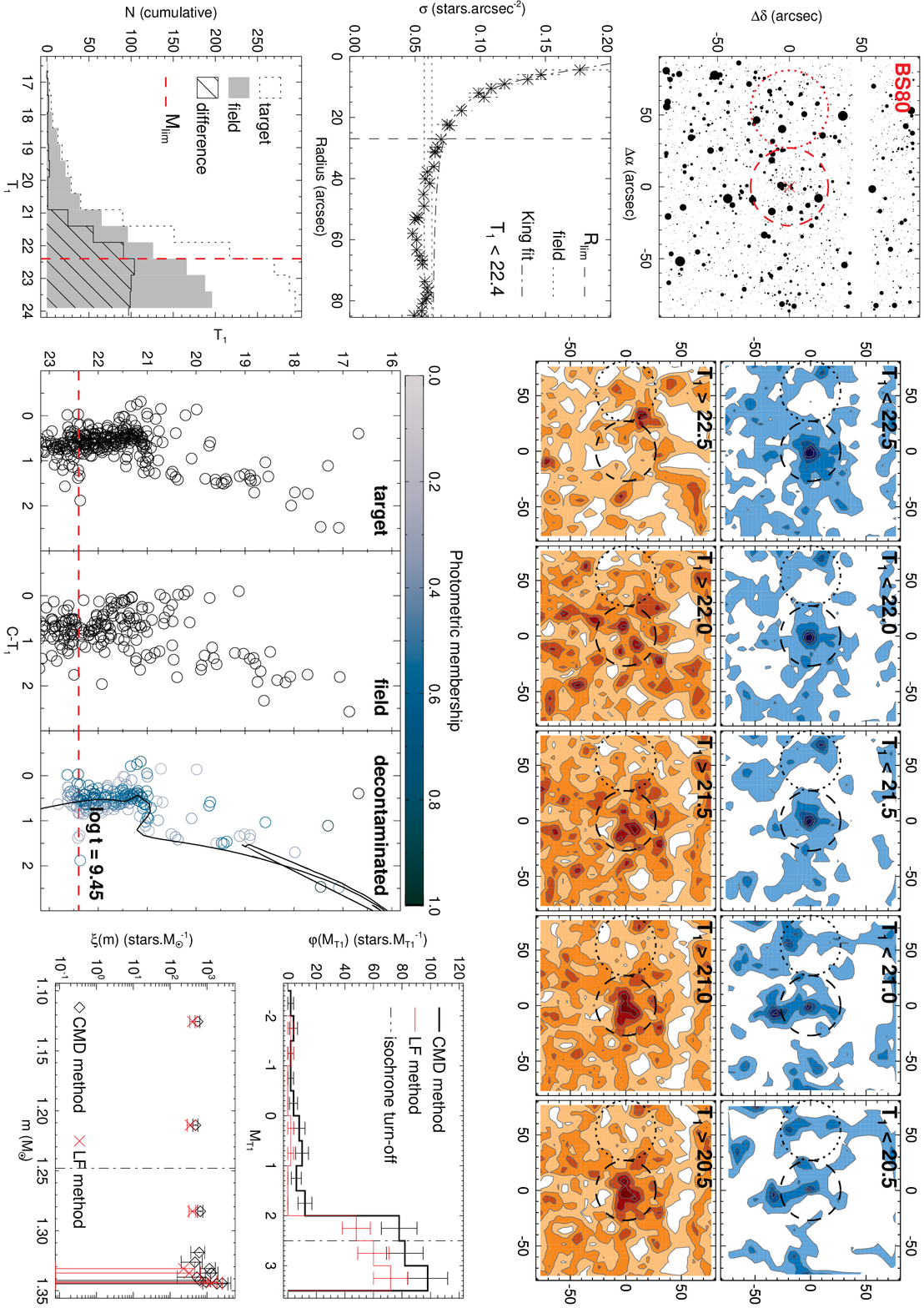}
\caption{BS80 analysis charts. Panels are the same as in Fig.~\ref{onlinefig}}
\end{minipage}
\end{sideways}
\end{figure*}

\clearpage

\begin{figure*} 
\centering
\begin{sideways}
\begin{minipage}{230mm}
\includegraphics[width=16cm,angle=90]{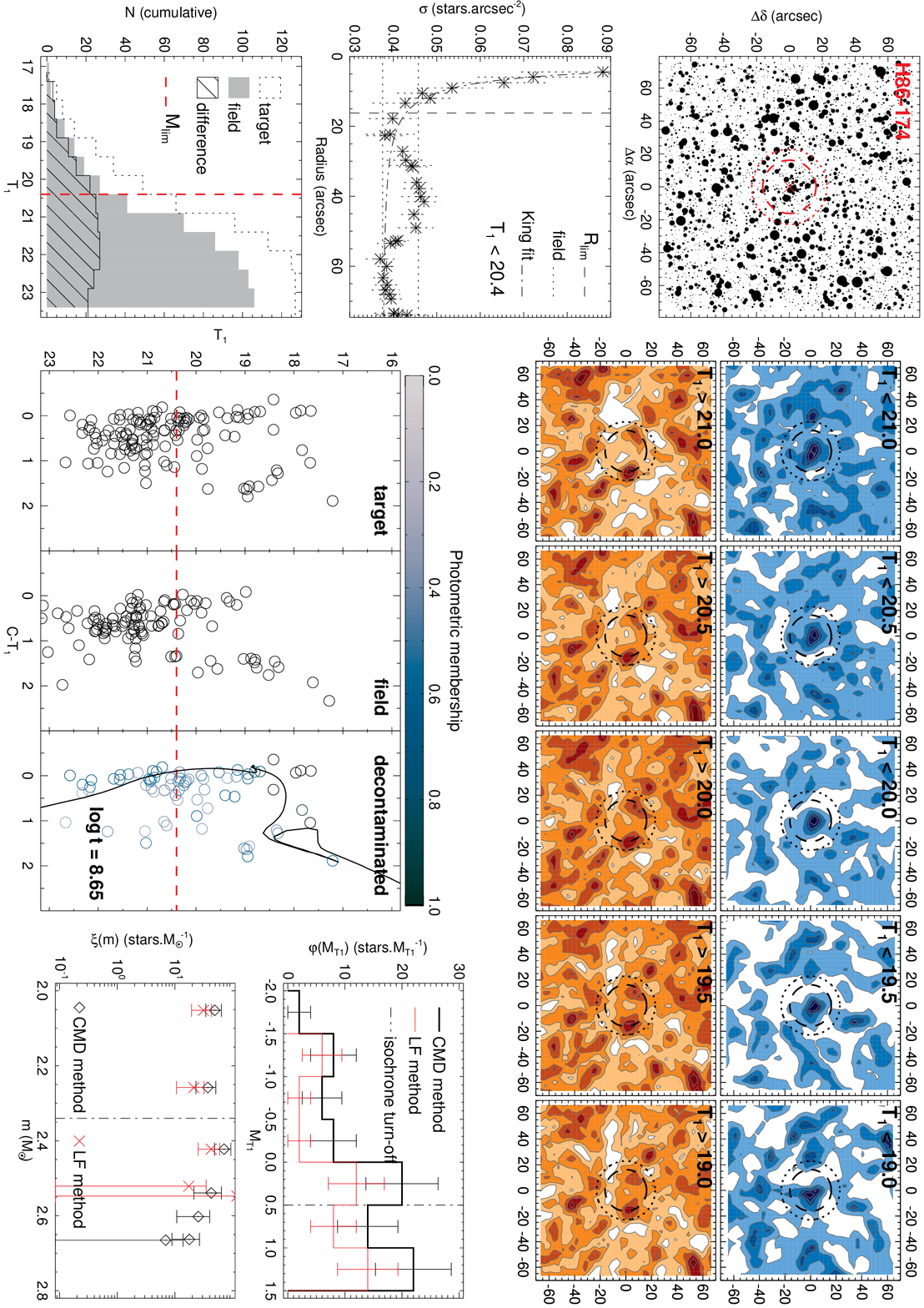}
\caption{H86-174 analysis charts. Panels are the same as in Fig.~\ref{onlinefig}}
\end{minipage}
\end{sideways}
\end{figure*}

\clearpage

\begin{figure*} 
\centering
\begin{sideways}
\begin{minipage}{230mm}
\includegraphics[width=16cm,angle=90]{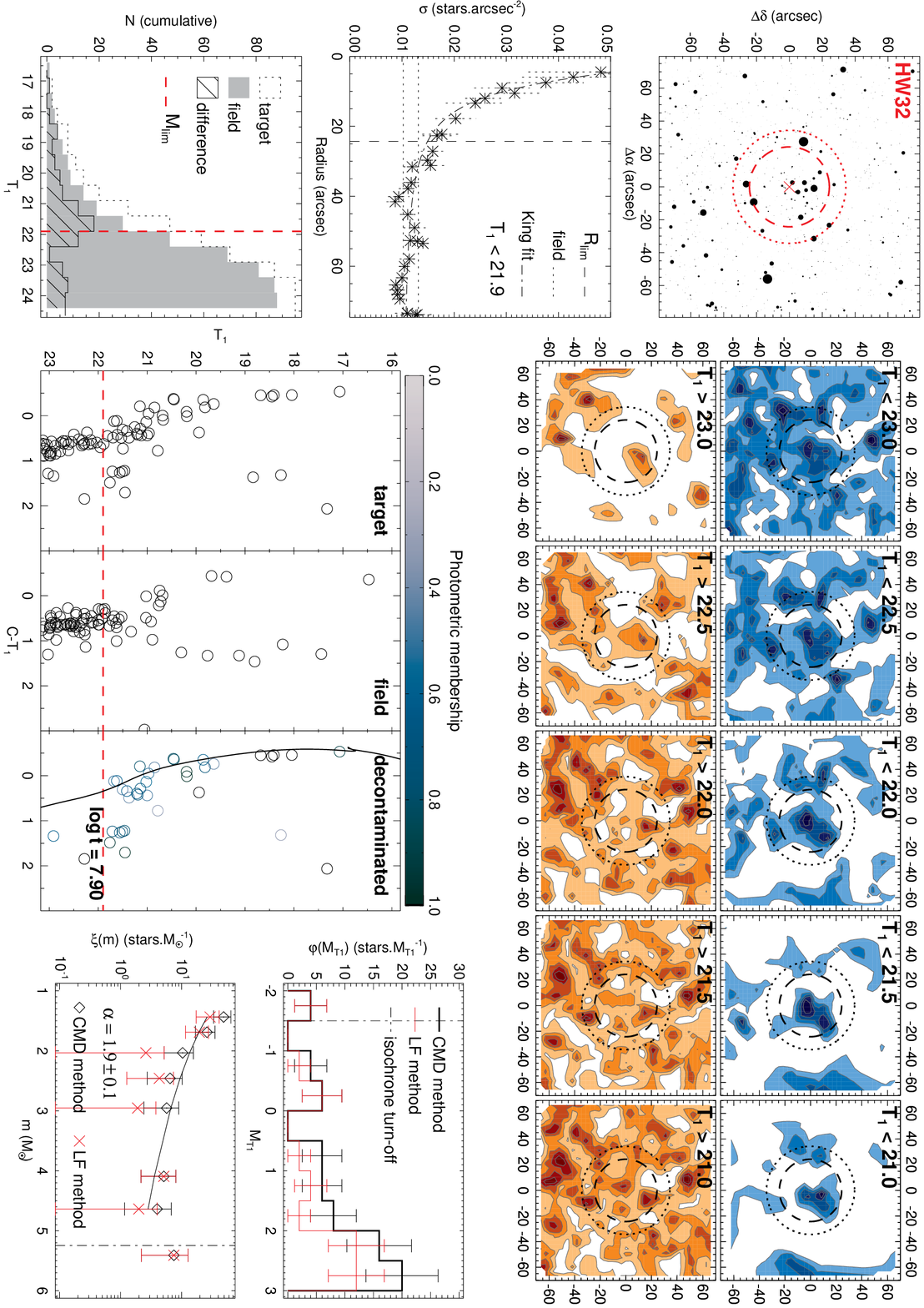}
\caption{HW32 analysis charts. Panels are the same as in Fig.~\ref{onlinefig}}
\end{minipage}
\end{sideways}
\end{figure*}

\clearpage

\begin{figure*} 
\centering
\begin{sideways}
\begin{minipage}{230mm}
\includegraphics[width=16cm,angle=90]{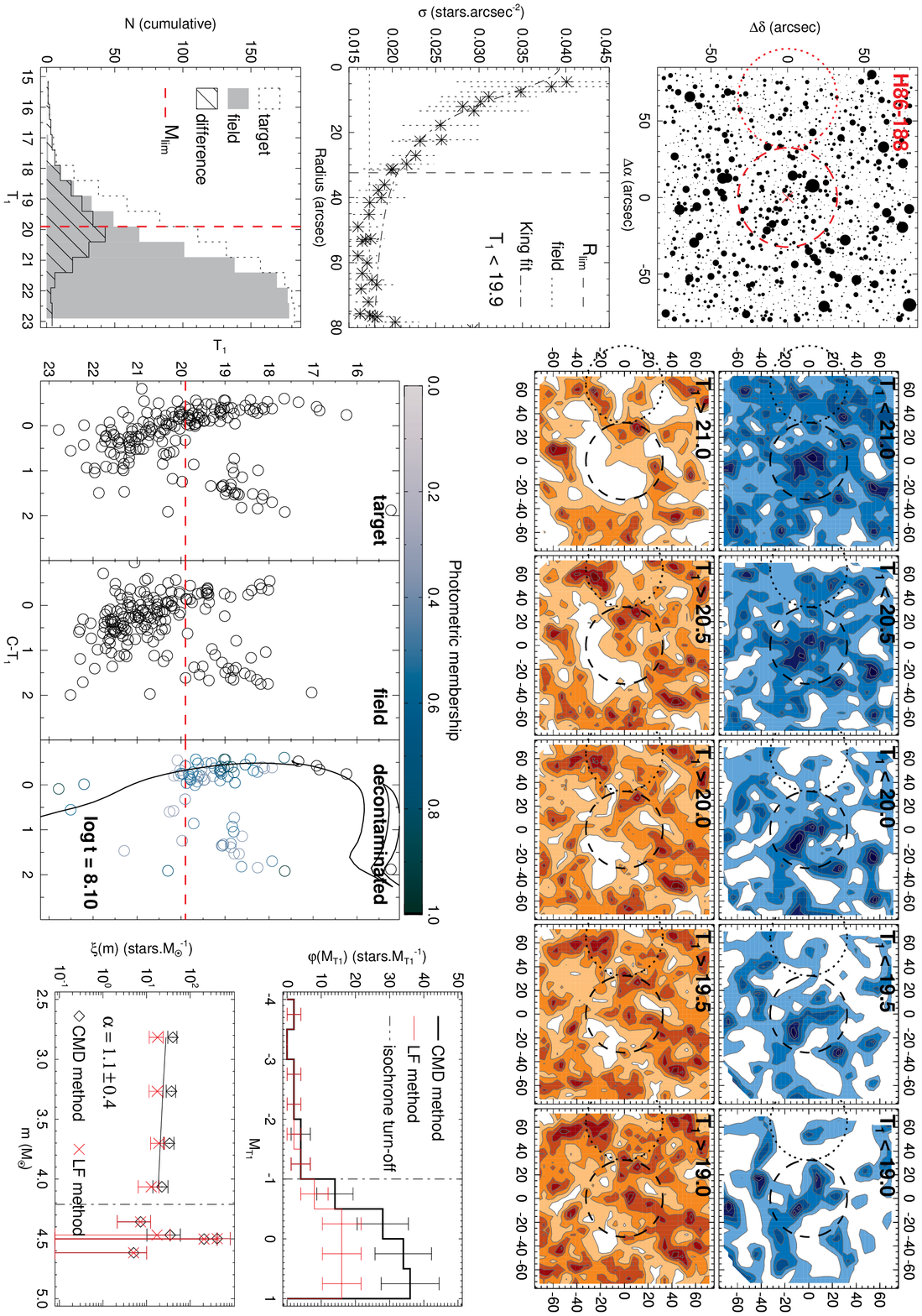}
\caption{H86-188 analysis charts. Panels are the same as in Fig.~\ref{onlinefig}}
\end{minipage}
\end{sideways}
\end{figure*}

\clearpage

\begin{figure*} 
\centering
\begin{sideways}
\begin{minipage}{230mm}
\includegraphics[width=16cm,angle=90]{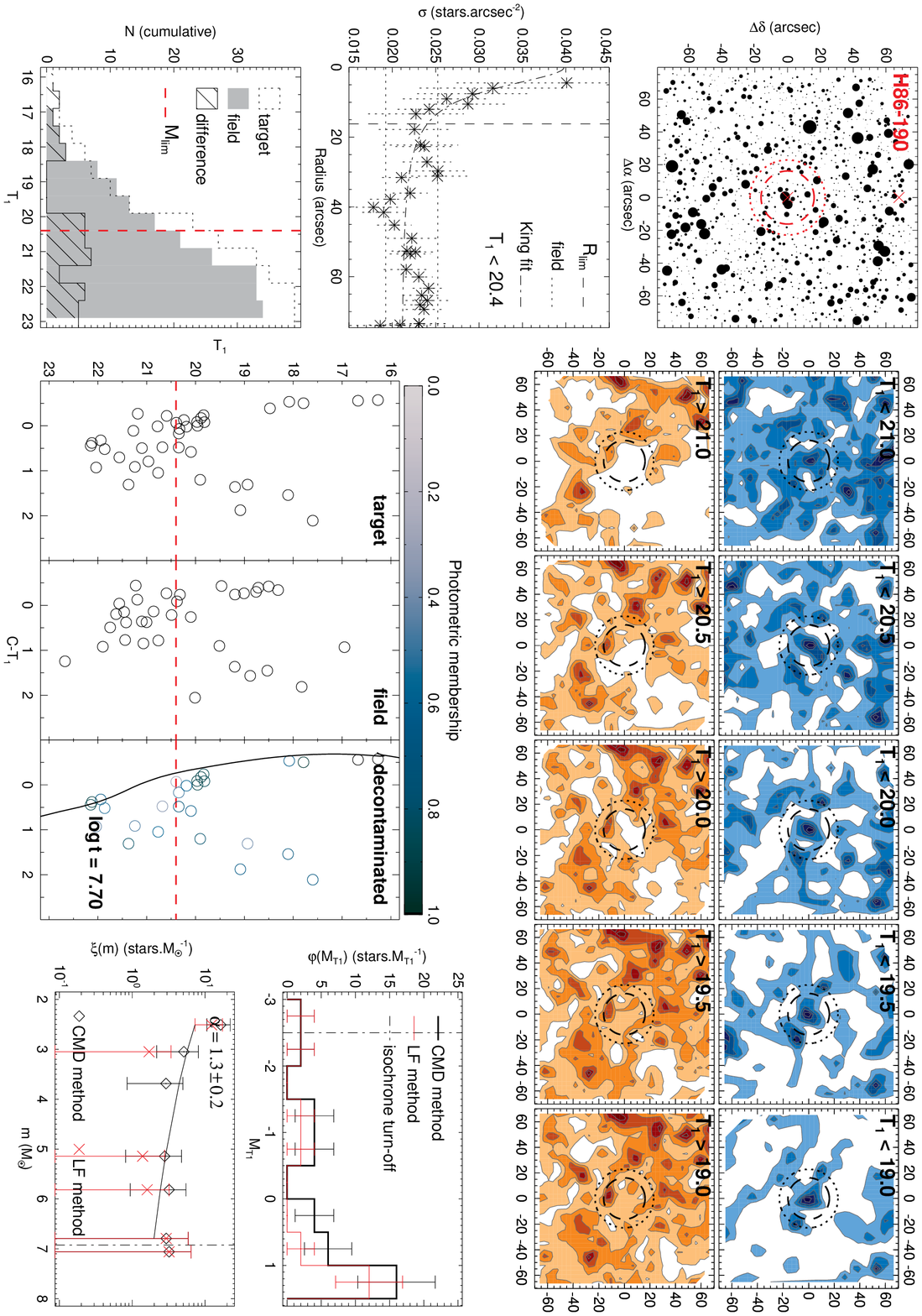}
\caption{H86-190 analysis charts. Panels are the same as in Fig.~\ref{onlinefig}}
\end{minipage}
\end{sideways}
\end{figure*}

\clearpage

\begin{figure*} 
\centering
\begin{sideways}
\begin{minipage}{230mm}
\includegraphics[width=16cm,angle=90]{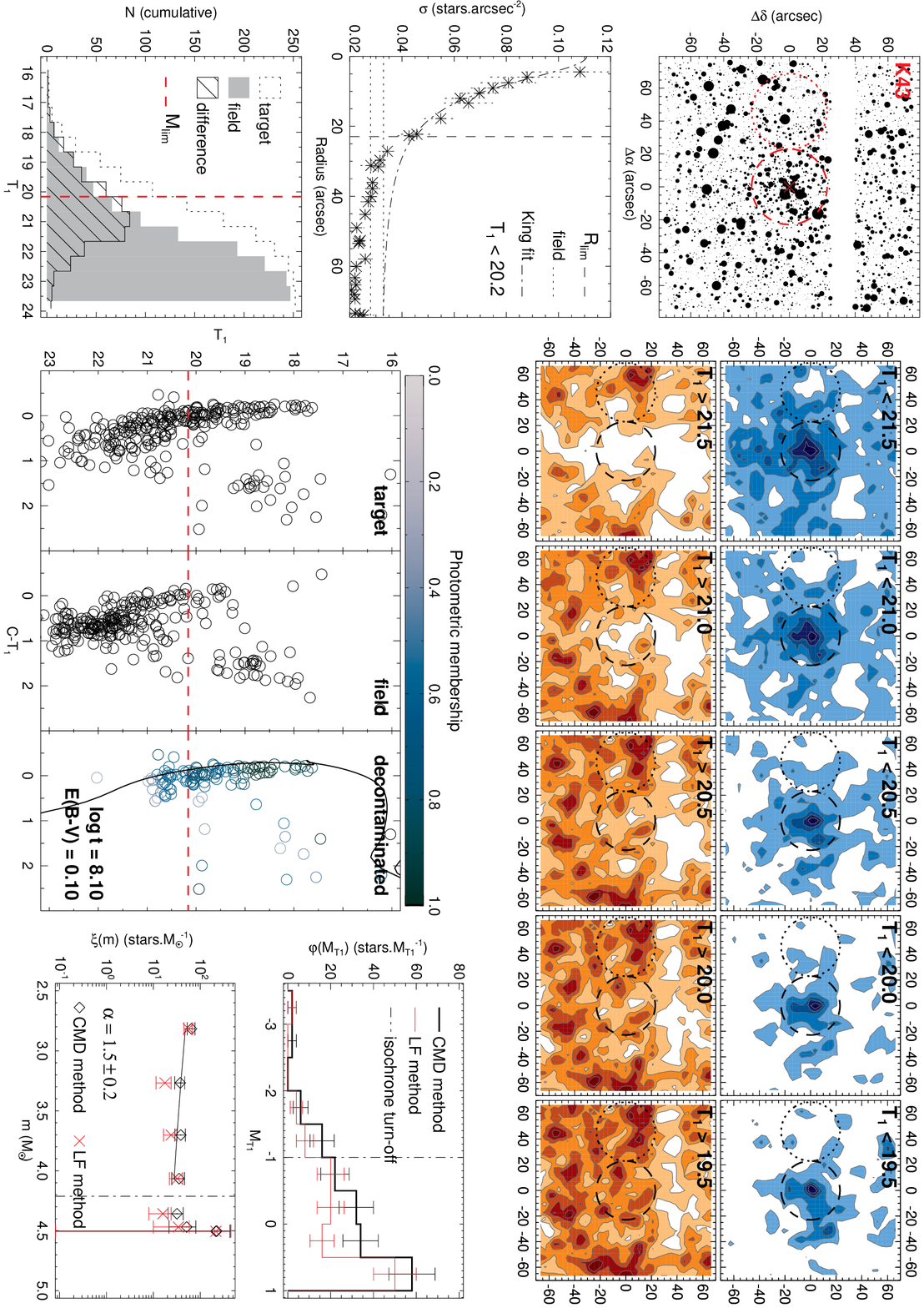}
\caption{K43 analysis charts. Panels are the same as in Fig.~\ref{onlinefig}}
\end{minipage}
\end{sideways}
\end{figure*}

\clearpage

\begin{figure*} 
\centering
\begin{sideways}
\begin{minipage}{230mm}
\includegraphics[width=16cm,angle=90]{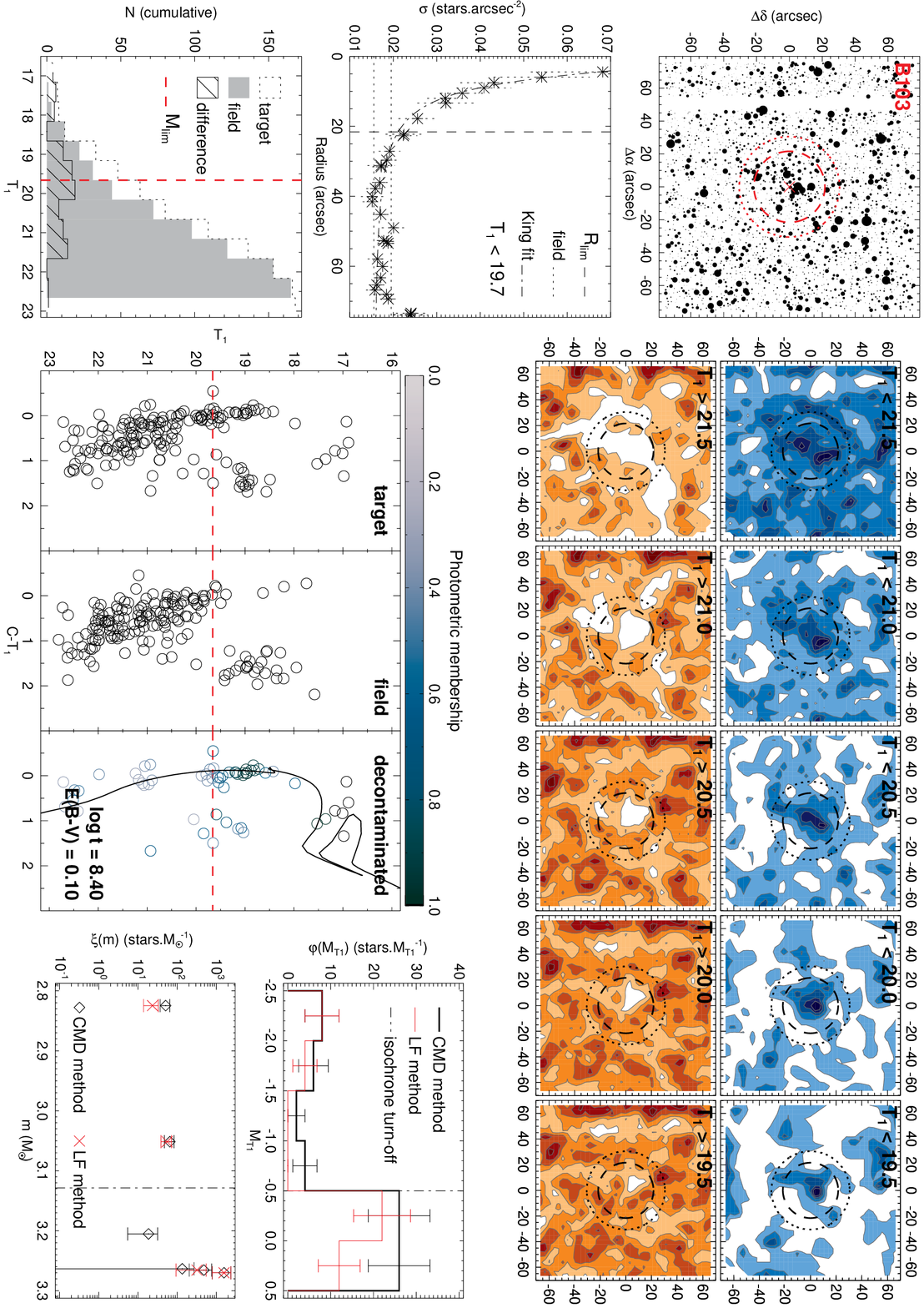}
\caption{B103 analysis charts. Panels are the same as in Fig.~\ref{onlinefig}}
\end{minipage}
\end{sideways}
\end{figure*}

\clearpage

\begin{figure*} 
\centering
\begin{sideways}
\begin{minipage}{230mm}
\includegraphics[width=16cm,angle=90]{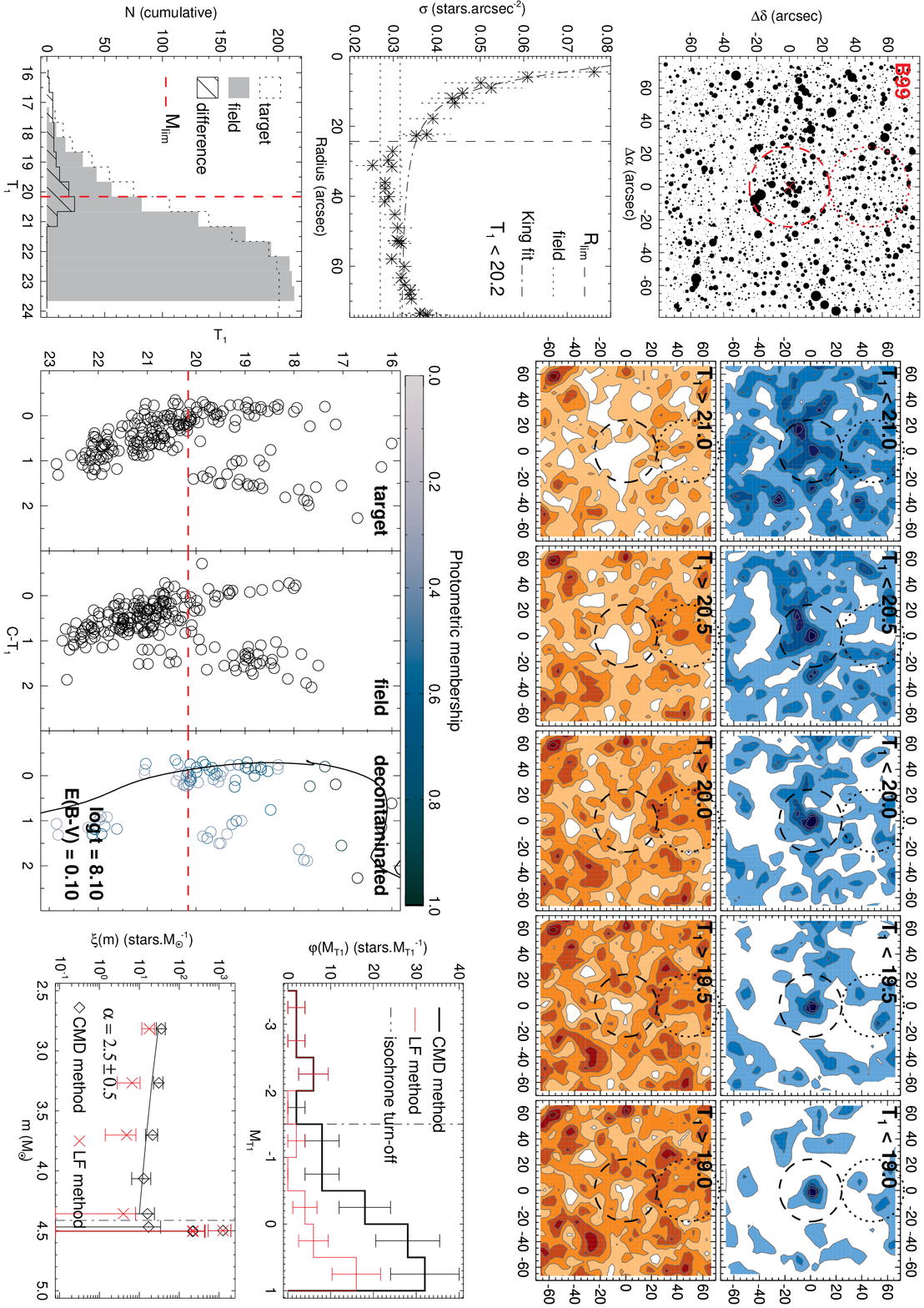}
\caption{B99 analysis charts. Panels are the same as in Fig.~\ref{onlinefig}}
\end{minipage}
\end{sideways}
\end{figure*}

\clearpage

\begin{figure*} 
\centering
\begin{sideways}
\begin{minipage}{230mm}
\includegraphics[width=16cm,angle=90]{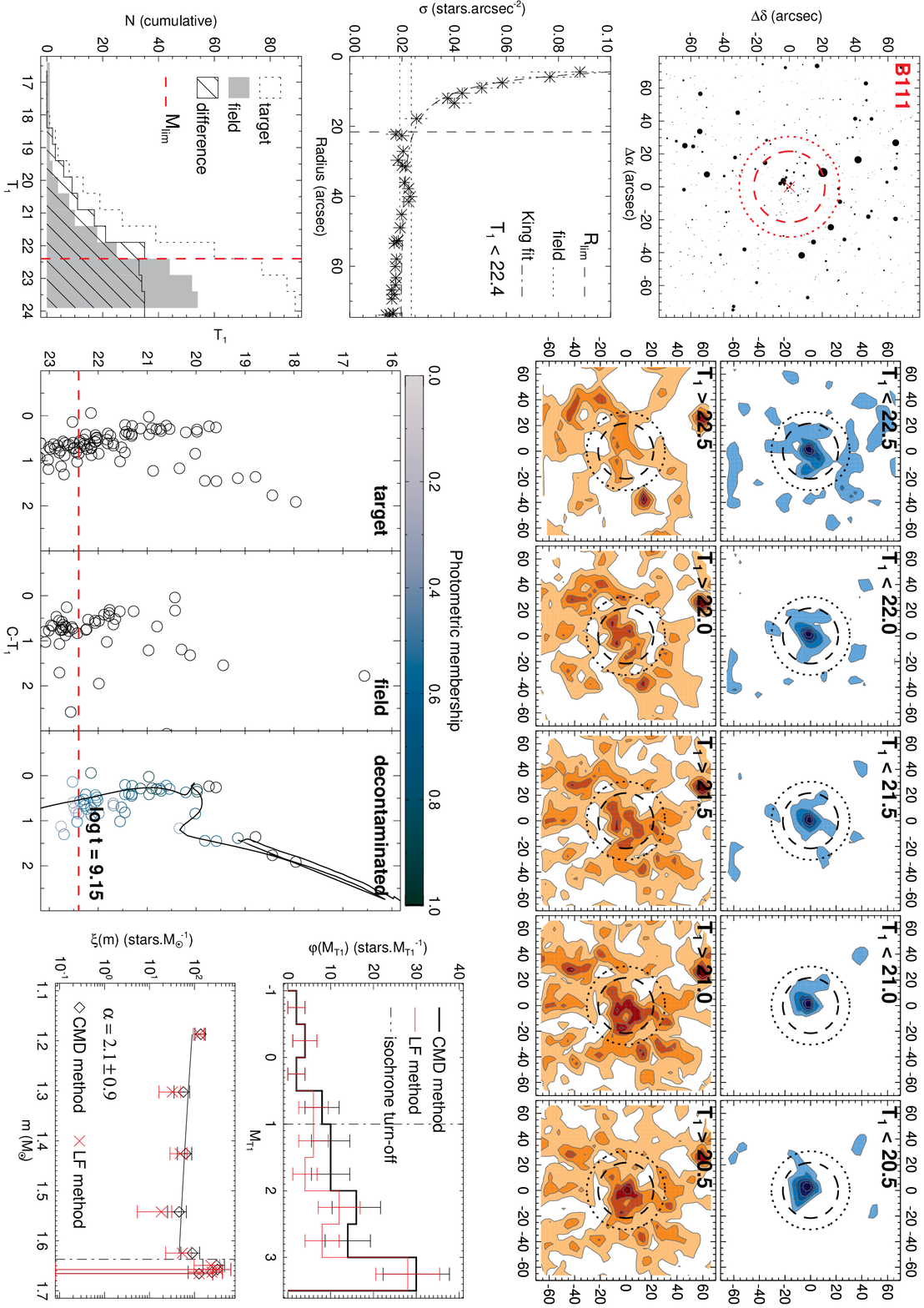}
\caption{B111 analysis charts. Panels are the same as in Fig.~\ref{onlinefig}}
\end{minipage}
\end{sideways}
\end{figure*}

\clearpage

\begin{figure*} 
\centering
\begin{sideways}
\begin{minipage}{230mm}
\includegraphics[width=16cm,angle=90]{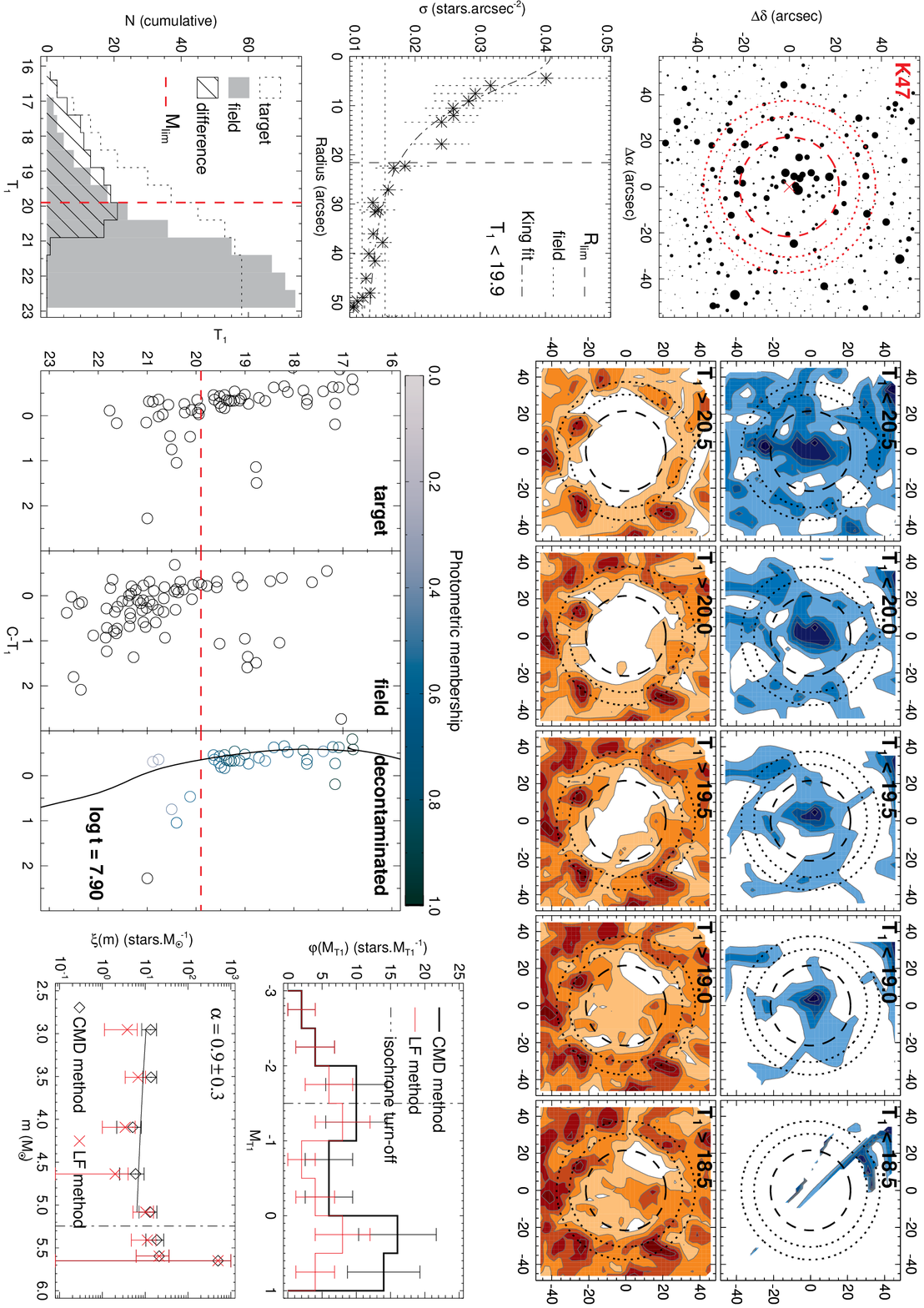}
\caption{K47 analysis charts. Panels are the same as in Fig.~\ref{onlinefig}}
\end{minipage}
\end{sideways}
\end{figure*}

\clearpage

\begin{figure*} 
\centering
\begin{sideways}
\begin{minipage}{230mm}
\includegraphics[width=16cm,angle=90]{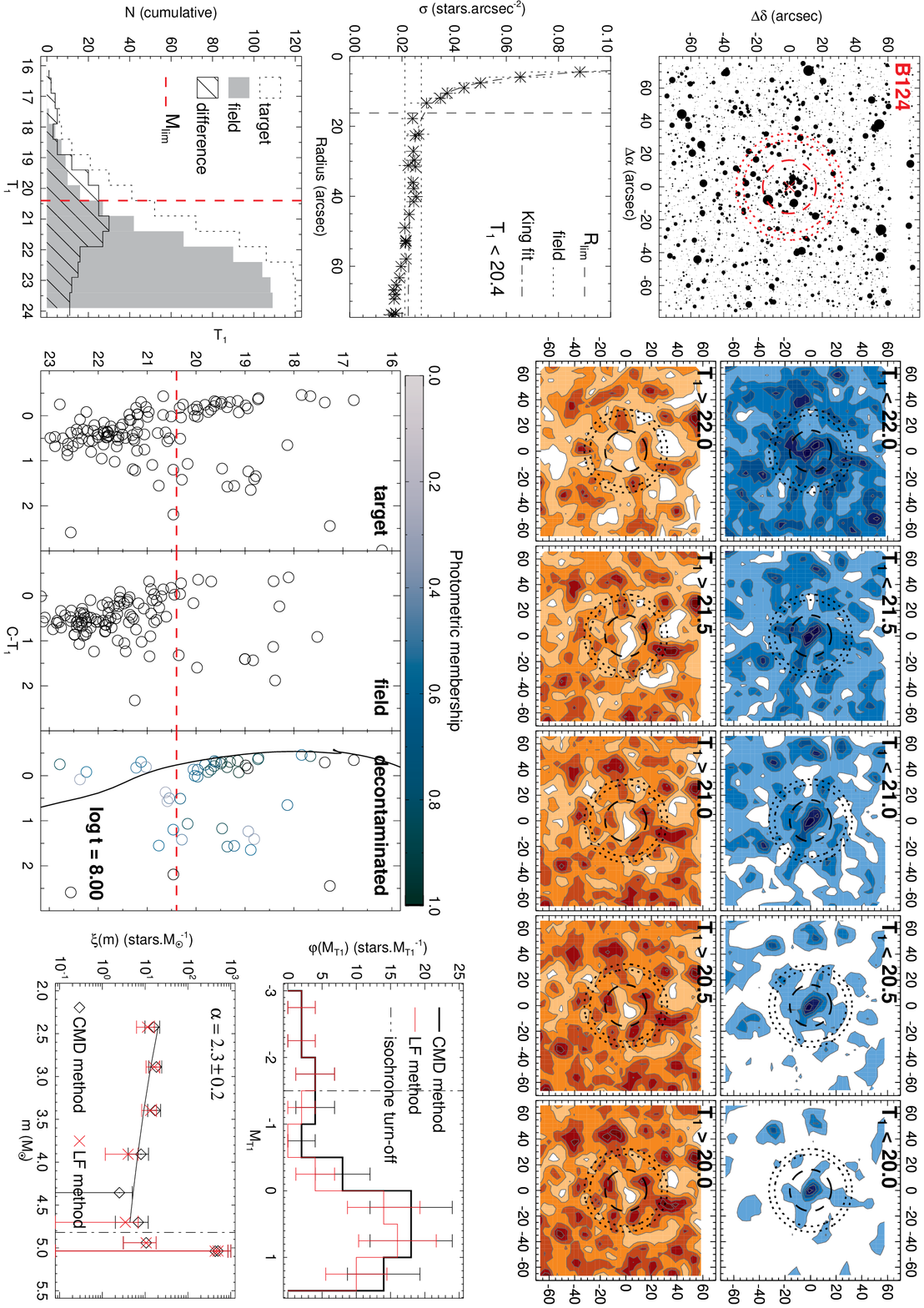}
\caption{B124 analysis charts. Panels are the same as in Fig.~\ref{onlinefig}}
\end{minipage}
\end{sideways}
\end{figure*}

\clearpage

\begin{figure*} 
\centering
\begin{sideways}
\begin{minipage}{230mm}
\includegraphics[width=16cm,angle=90]{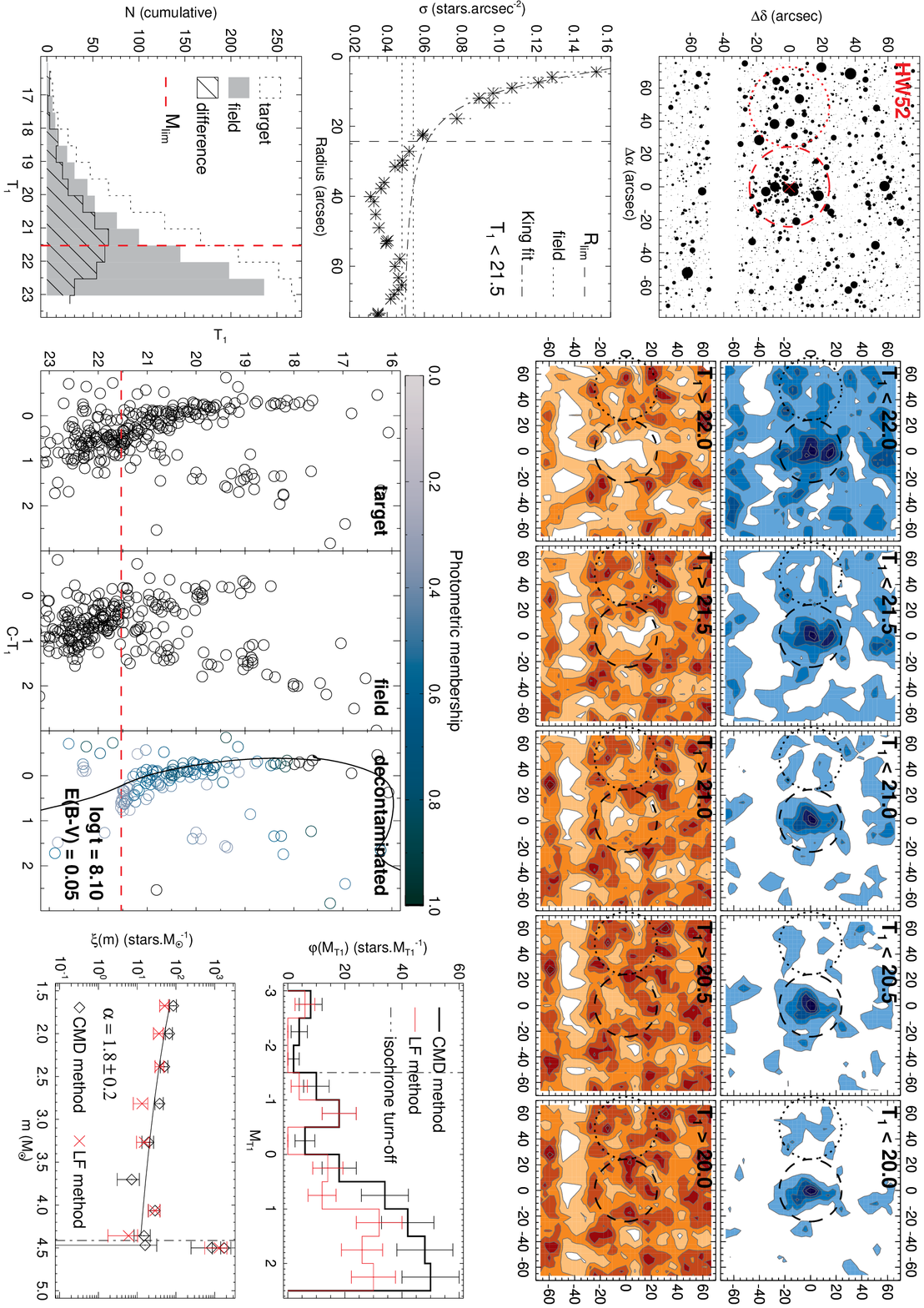}
\caption{HW52 analysis charts. Panels are the same as in Fig.~\ref{onlinefig}}
\end{minipage}
\end{sideways}
\end{figure*}

\clearpage

\begin{figure*} 
\centering
\begin{sideways}
\begin{minipage}{230mm}
\includegraphics[width=16cm,angle=90]{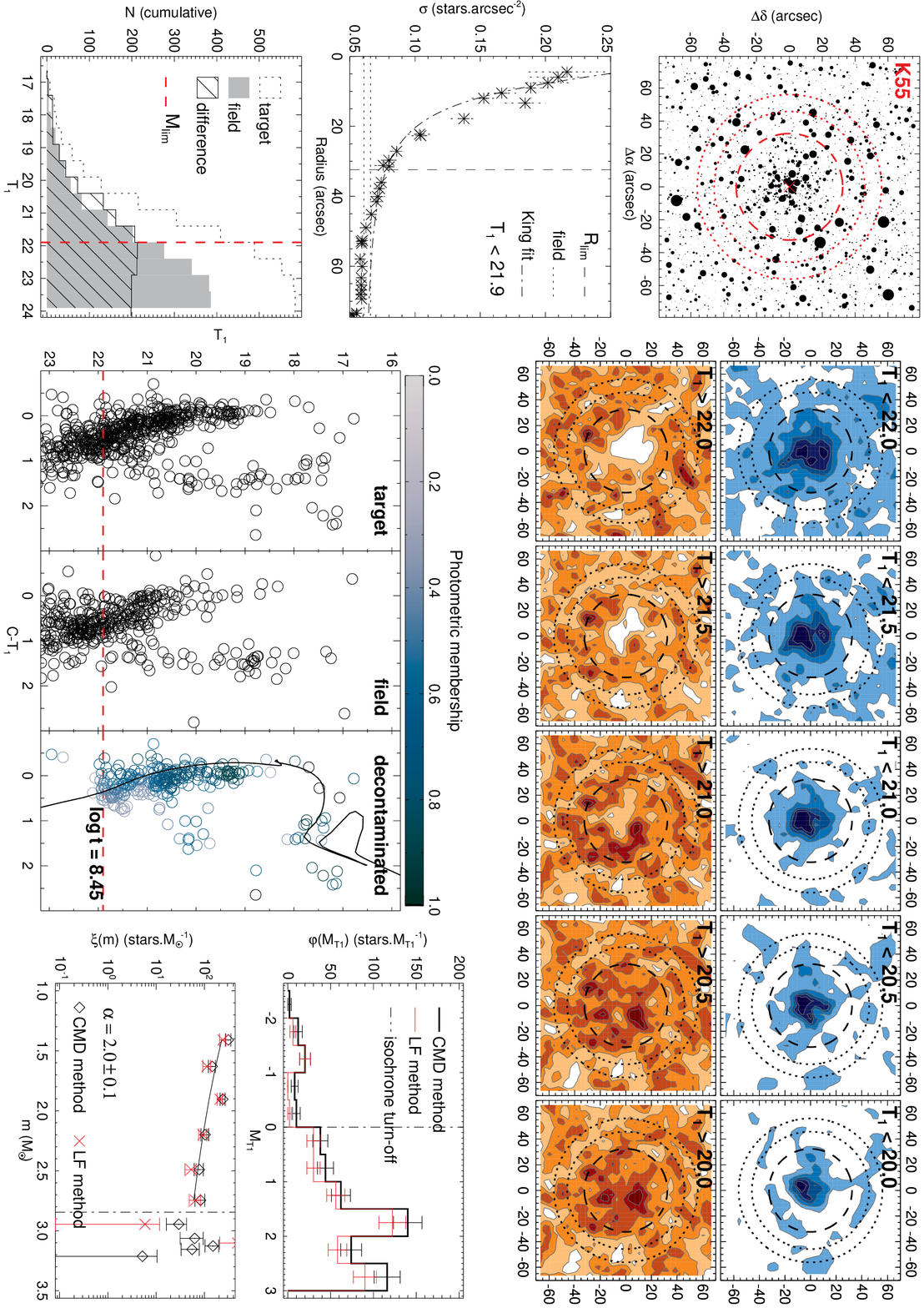}
\caption{K55 analysis charts. Panels are the same as in Fig.~\ref{onlinefig}}
\end{minipage}
\end{sideways}
\end{figure*}

\clearpage

\begin{figure*} 
\centering
\begin{sideways}
\begin{minipage}{230mm}
\includegraphics[width=16cm,angle=90]{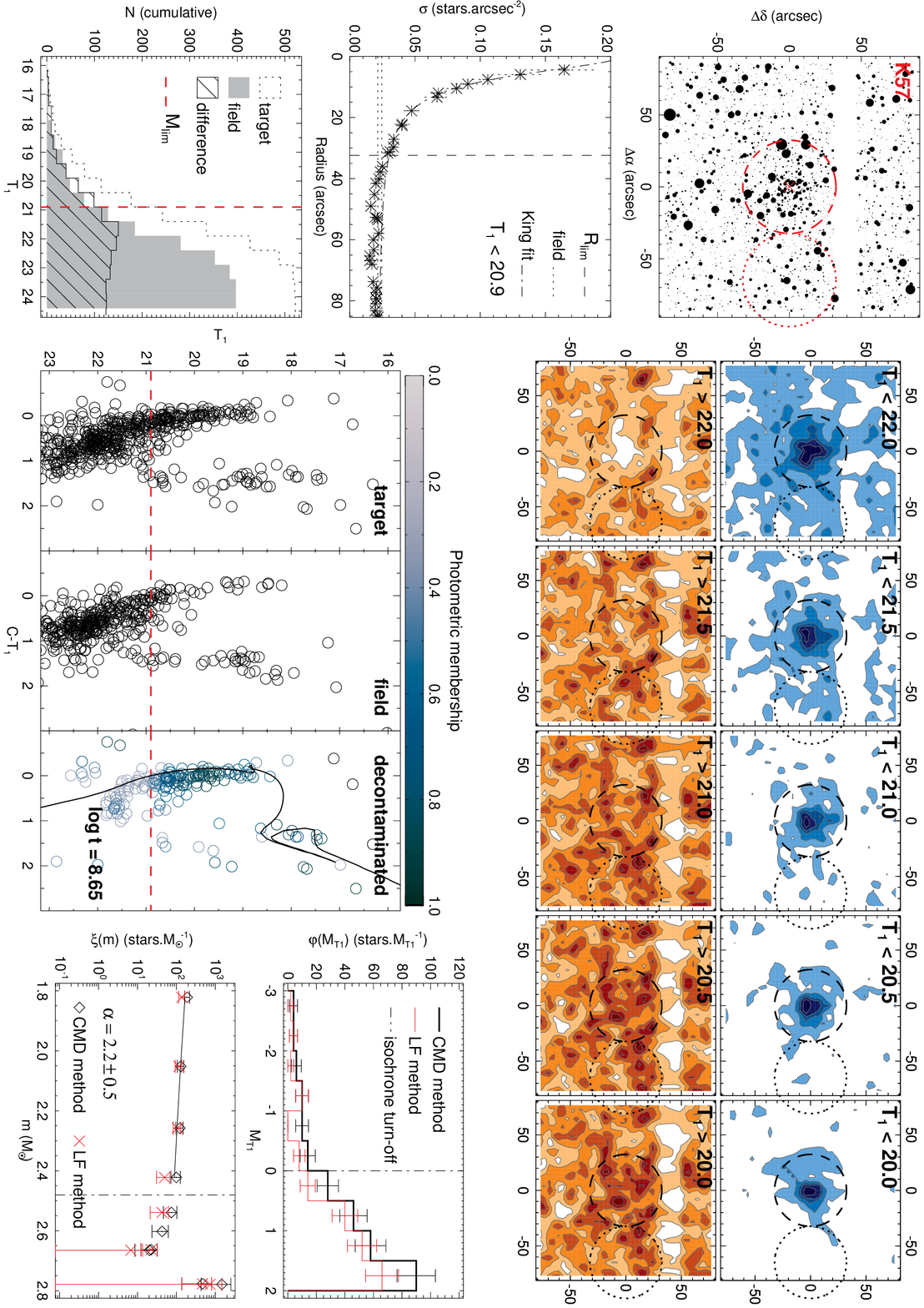}
\caption{K57 analysis charts. Panels are the same as in Fig.~\ref{onlinefig}}
\end{minipage}
\end{sideways}
\end{figure*}

\clearpage

\begin{figure*} 
\centering
\begin{sideways}
\begin{minipage}{230mm}
\includegraphics[width=16cm,angle=90]{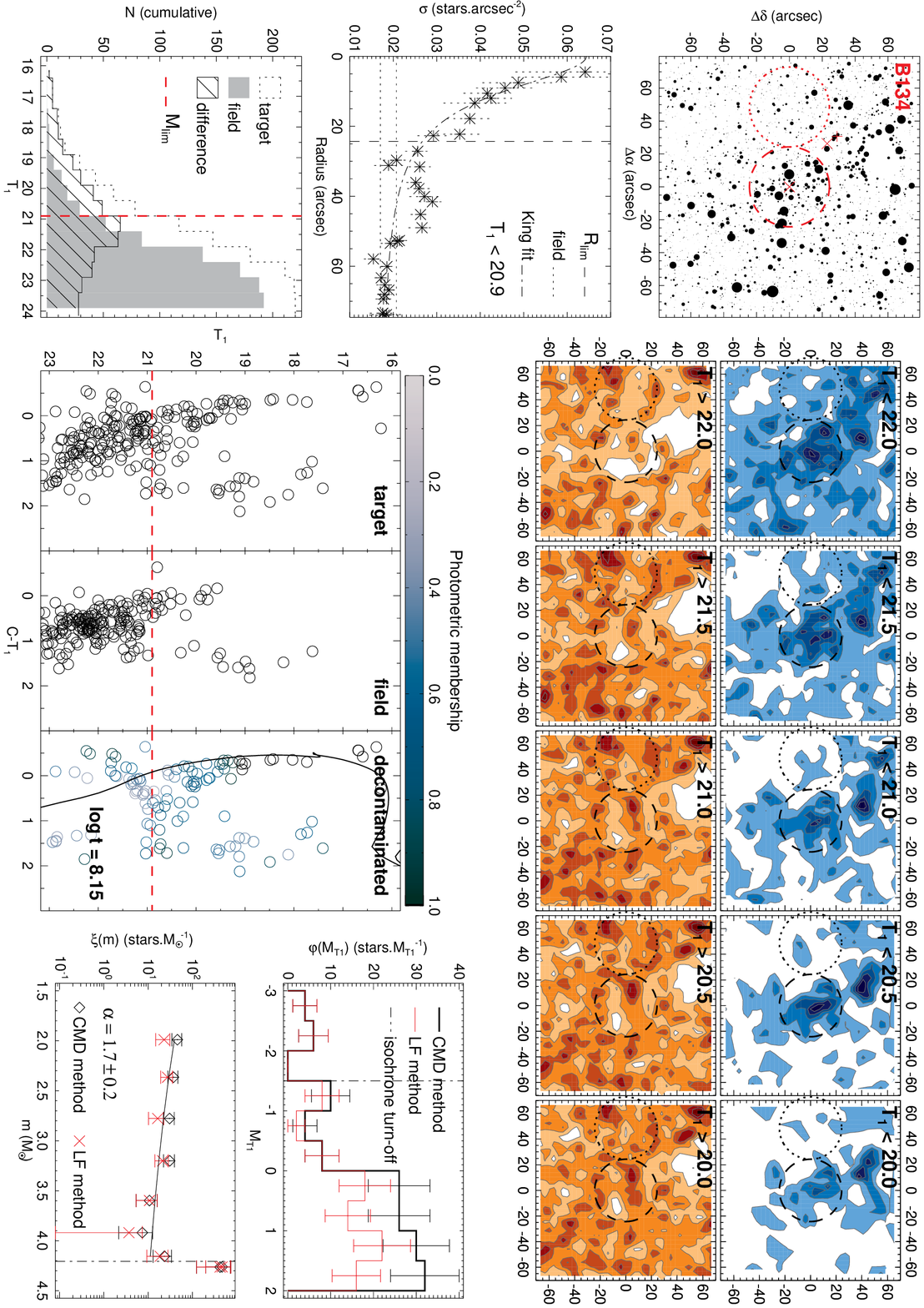}
\caption{B134 analysis charts. Panels are the same as in Fig.~\ref{onlinefig}}
\end{minipage}
\end{sideways}
\end{figure*}

\clearpage

\begin{figure*} 
\centering
\begin{sideways}
\begin{minipage}{230mm}
\includegraphics[width=16cm,angle=90]{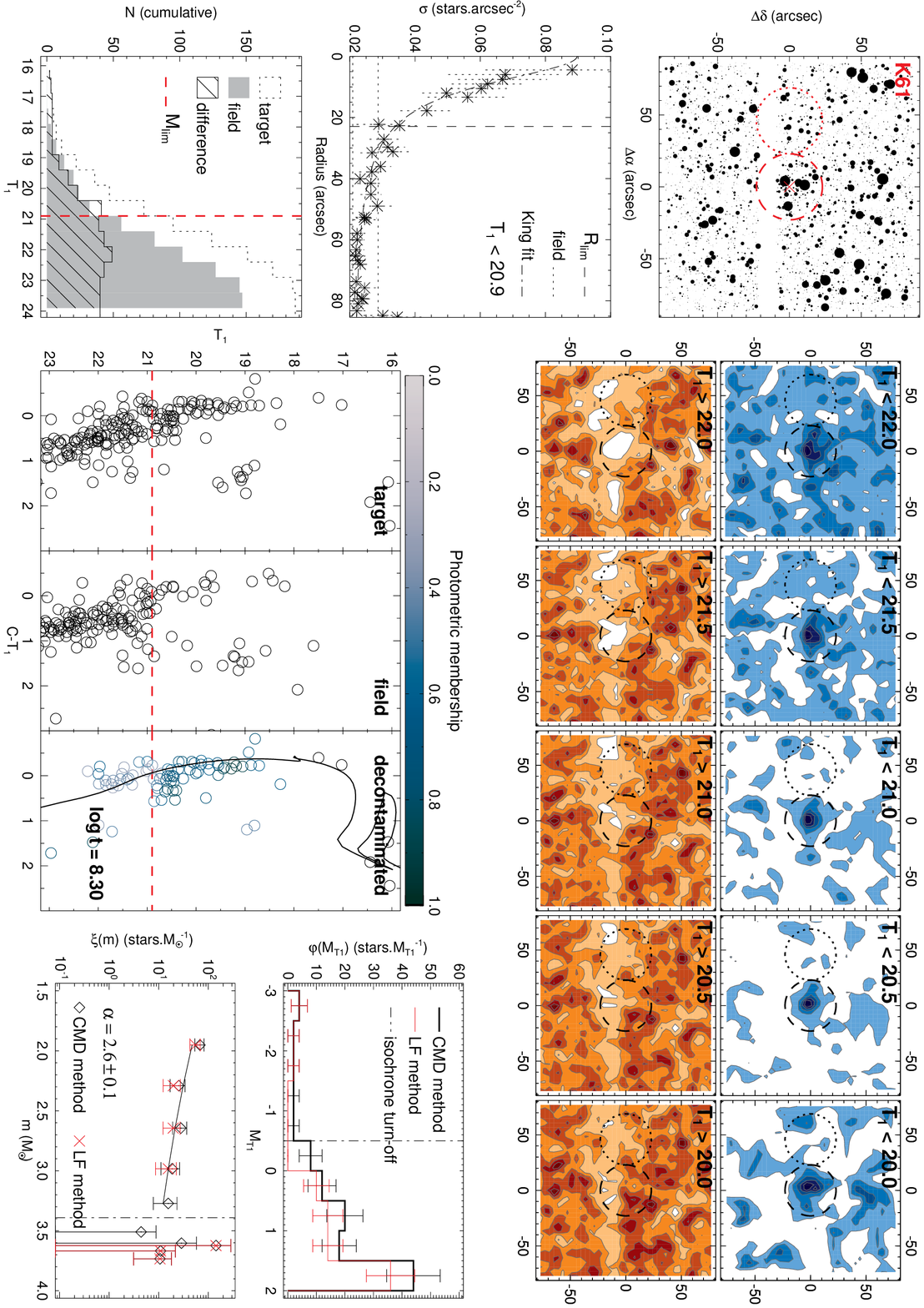}
\caption{K61 analysis charts. Panels are the same as in Fig.~\ref{onlinefig}}
\end{minipage}
\end{sideways}
\end{figure*}

\clearpage

\begin{figure*} 
\centering
\begin{sideways}
\begin{minipage}{230mm}
\includegraphics[width=16cm,angle=90]{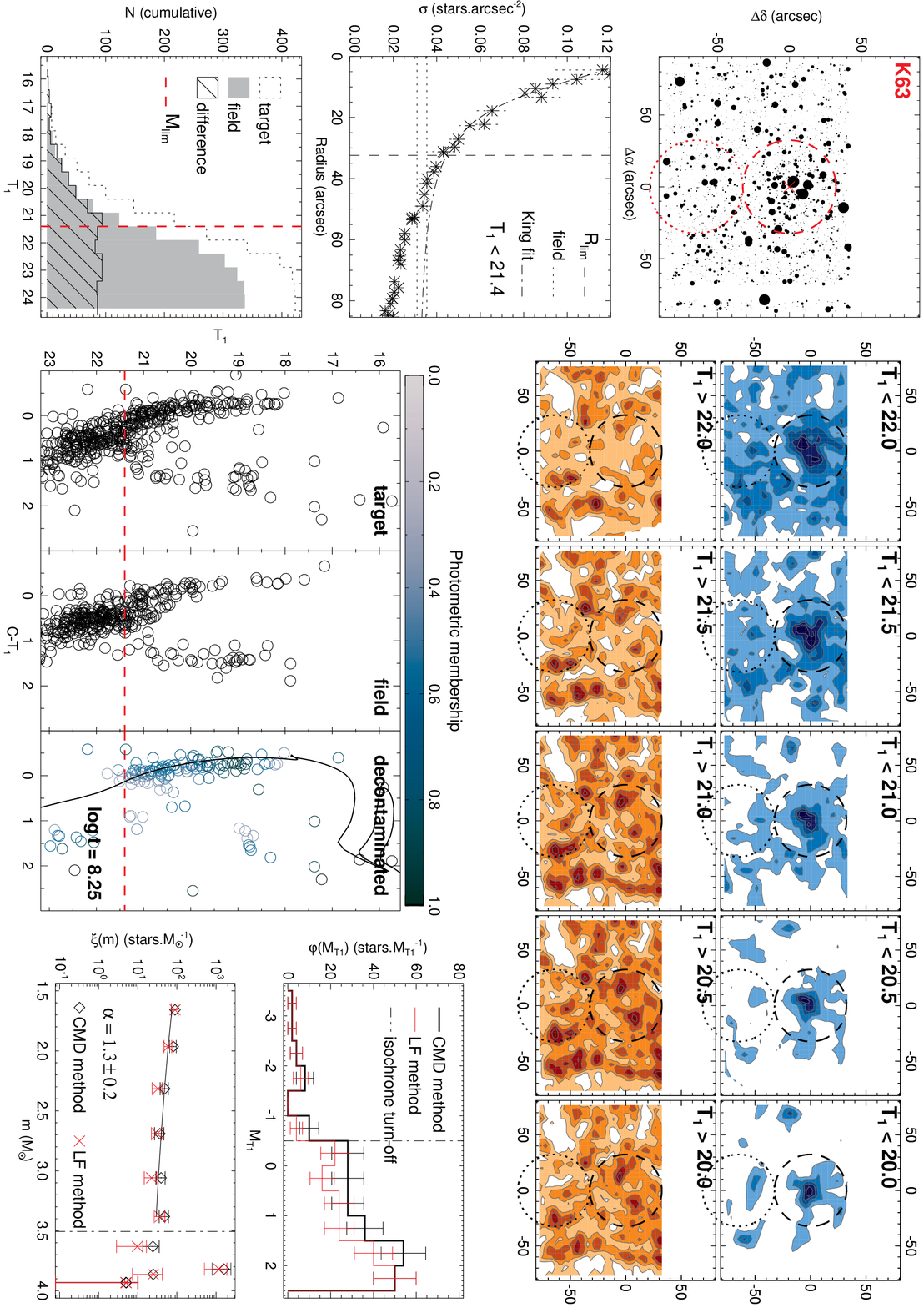}
\caption{K63 analysis charts. Panels are the same as in Fig.~\ref{onlinefig}}
\end{minipage}
\end{sideways}
\end{figure*}

\clearpage


\label{lastpage}

\end{document}